\documentclass[namedreferences,hyperref,optionalrh]{spr-sola}
\usepackage{graphicx}        
\usepackage{amssymb}        
\usepackage{color}           
\usepackage{breakurl}
\usepackage{lscape}

\chardef\us=`\_

\begin{document}

\begin{frontmatter}
\title{Solar Energetic Proton Events Observed by the High Energy Telescopes on the STEREO Spacecraft or at the Earth During the First Solar Orbit of STEREO~A (2006--2023)}

\author[addressref={aff1,aff2},corref,email={irichard@umd.edu}]{\inits{I.G}\fnm{Ian G.}~\snm{Richardson}\orcid{0000-0002-3855-3634}}
\author[addressref={aff1,aff3},email={tycho@rcn.com}]{\inits{T.T.}\fnm{Tycho T.}~\snm{von Rosenvinge}\orcid{0000-0001-7320-4141}}
\author[addressref={aff1,aff3},email={ocstcyr2@gmail.com}]{\inits{O.C.}\fnm{O. Chris}~\snm{St. Cyr}\orcid{0000-0001-8906-6097}}
\author[addressref=aff1,email={david.larioloyo@nasa.gov}]{\inits{D.}\fnm{David}~\snm{Lario}\orcid{0000-0002-3176-8704}}
\author[addressref=aff1,email={john.g.mitchell@nasa.gov}]{\inits{J.G.}\fnm{J. Grant}~\snm{Mitchell}\orcid{0000-0003-4501-5452}}
\author[addressref=aff1,email={eric.r.christian@nasa.gov}]{\inits{E.R.}\fnm{Eric R.}~\snm{Christian}\orcid{0000-0003-2134-3937}}

\address[id=aff1]{Dept. of Astronomy, University of Maryland, College Park, MD 20742, USA}
\address[id=aff2]{Heliophysics Science Division, NASA Goddard Space Flight Center, Greenbelt, MD 20771, USA}
\address[id=aff3]{Retired}

\runningauthor{I.G. Richardson et al.}
\runningtitle{STEREO HET First Orbit}

\begin{abstract}

 The twin STEREO~A and B spacecraft were launched in October 2006 into heliocentric orbits at $\sim1$~AU, advancing ahead of or lagging behind Earth, respectively, at $\sim22^\circ$/year.  The spacecraft provide in-situ observations of the solar wind and energetic particle populations, as well as remote sensing observations of solar activity and the corona.  
In particular, the High Energy Telescopes (HETs) on the STEREO spacecraft observe 0.7--4 MeV electrons and 13--100~MeV protons. This paper summarizes observations of solar energetic particle (SEP) events made by the STEREO HETs from the beginning of the mission through Solar Cycle 24 to December 2023, approaching the maximum of Solar Cycle 25 and encompassing STEREO A's first full orbit of the Sun relative to Earth, completed in August 2023; contact with STEREO~B was lost in October 2014.  Specifically, the catalog of SEP events including $\sim25$ MeV protons observed by the STEREO HETs and/or instruments on spacecraft near Earth in \cite{richardson2014} is updated to include $\sim450$ SEP events and a total of $\sim1000$ separate observations of these events from the various spacecraft locations. These extensive observations can provide unique insight into the propagation of energetic protons in the inner heliosphere and how the properties of the particle events are related to those of the associated solar eruptions. In particular, we examine the association of coronal mass ejections (CMEs) and SEP events with all 397 M and X-class solar X-ray flares in June~2010-January~2014 and demonstrate that, for these events, the occurrence of a CME accompanying a flare is required for the detection of a $\sim25$~MeV proton event. On the other hand, many flares accompanied by CMEs are not followed by detected SEP events.  The longitudinal width and intensity of the associated SEP events generally increase with the CME speed and the flare intensity. We also note evidence for a $\sim150$~day ``Rieger-like" periodicity in the SEP occurrence rate in 2020-2023 during the rising phase of Solar Cycle~25.
\end{abstract}


\end{frontmatter}

\section{Introduction}
     \label{S-Introduction} 

The twin STEREO (Solar TErrestrial RElations Observatory) A (“Ahead”) and B (“Behind”) spacecraft \citep{kaiser2008} were launched on 26 October 2006 into heliocentric orbits at approximately 1~AU, advancing ahead of, or lagging Earth in its orbit, respectively.  Separating from Earth by $\sim22^\circ$/year, in February 2011, STEREO A and B were $180^\circ$ apart above the west and east limbs of the Sun, respectively, as viewed from Earth, allowing observations of the complete solar surface to be made for the first time. STEREO~A passed directly behind the Sun as viewed from Earth in May 2015, and then approached Earth above the east limb of the Sun. In August, 2023, STEREO~A finally returned to the vicinity of Earth.
However, contact was lost with STEREO~B on October~1, 2014 when on the far side of the Sun and has not been restored.  

Each STEREO spacecraft carries a High Energy Telescope (HET) \citep{vonrosen2008} making observations of 0.7--4 MeV electrons and 13--100~MeV protons. The aim of this paper is to summarize observations of solar energetic particle (SEP) events made by the STEREO spacecraft, focusing on proton observations made by the HETs.  In particular, we update to December 2023 the catalog of SEP events of \citet{richardson2014} that include $\sim25$~MeV protons observed at one or both of the STEREO spacecraft and/or by spacecraft located near Earth, encompassing the first STEREO~A orbit of the Sun relative to Earth.  The updated catalog includes information on the associated solar phenomena (e.g., location, flare, radio, and CME associations). Some 450 individual SEP events and their solar sources are identified.  

As discussed in Section~\ref{S-Obs}, the STEREO spacecraft were widely separated both from each other and spacecraft near Earth during the rising and maximum phases of Solar Cycle~24 (see also, for example, \citet{lario2013} and \cite{richardson2014}). Therefore, they were ideally located to investigate particle acceleration and transport in longitudinally-extended SEP events, some apparently filling the inner heliosphere. STEREO~A was approaching Earth during the ascending phase of Solar Cycle 25, commencing in December~2019 when the spacecraft was $\sim80^{\circ}$ east of Earth.  When approaching closer to Earth, the observations of SEP events from STEREO~A become more similar to those from near-Earth spacecraft.  Nevertheless, the combined observations can be used, for example, to investigate smaller-scale variations in the features of SEP events that are related to local solar wind structures. An example is the October 9, 2021 event discussed by \citet{lario2022} and \citet{Palmerio2024}.  These studies also included observations from the more recently launched missions into the inner heliosphere, including Parker Solar Probe (PSP) \citep{fox2016}, Solar Orbiter (SolO) \citep{muller2020}, and Bepi-Colombo \citep{benkhoff2021}.  They clearly demonstrated the impact of a corotating interaction region \citep[CIR; e.g.,][]{Richardson2018a} on the development of an SEP event at the different spacecraft locations, as modeled by \citet{wijsen2023}.   Other recent papers discussing SEP events during the rising phase of Solar Cycle 25 observed by STEREO~A, near-Earth spacecraft and these inner heliosphere spacecraft include \citet{kollhoff2021}, who summarized observations of the first widespread SEP event of Cycle 25 in November 2020, and \citet{lario2022}, \citet{palmerio2022}, \citet{dresing2023} and \citet{khoo2024}, among others. In particular, significant SEP events with multispacecraft analyses include October 28, 2021 \citep[e.g.,][]{papaioannous2022,cohen2025}, September 5, 2022 \citep[e.g.][]{paouris2023,kouloumvakos2025}, and March 13, 2023 \citep[e.g.][]{dresing2025}. Also, \citet{farwa2025} discuss electron to proton intensity ratios during 45 SEP events in 2020--2023 observed at five locations. 

The focus of this paper is on the STEREO HET observations, and we do not attempt to provide a comprehensive review of SEP events observed by the current fleet of spacecraft. Nevertheless, in some cases, observations from the inner heliosphere spacecraft can help identify the source of an SEP event on the far side of the Sun relative to Earth or STEREO~A. Such observations considered include the ``living catalog" of SEP events observed by PSP \citep{mitchell2023}, the catalog of X-ray flares observed by the Spectrometer/Telescope for Imaging X-rays \citep[STIX,][]{Krucker2020} on board SolO (\url{https://datacenter.stix.i4ds.net}), observations from the Energetic Particle Detector (EPD) on SolO \citep{Rodriguez2020} and the catalog of multi-spacecraft SEP events in Solar Cycle 25 \citep{Dresing2024} compiled as part of the SERPENTINE (Solar EneRgetic ParticlE aNalysis plaTform for the INner hEliosphere) project (\url{https://serpentine-h2020.eu/}). 
The end of 2023 provides a natural time to update the published STEREO $\sim25$~MeV proton event list to include an additional ten years of observations beyond those summarized in \citet{richardson2014} that encompass the first solar orbit of STEREO~A with respect to Earth.         

There are several reasons why observations of SEP events at HET energies are of interest. First, it is often easier to identify SEP event onsets at such energies than at lower energies, where new solar particle events can be obscured by  particles accelerated by interplanetary shocks \citep[cf., Figure 1 of][]{richardson2017},  the effects of solar wind structures on low-energy ion intensity-time profiles, and by extended periods with elevated low-energy ion intensities. A further reason is that SEP events including $\sim25$~MeV protons have been routinely identified in observations since 1967 from the Goddard Space Flight Center (GSFC) (and other) instruments on various spacecraft and used in several studies. These include: \citet{VanHollebeke1975}, \citet{kahler1978}, \citet{cane1988}, where observations from 235 events in 1967--1985 were used to demonstrate clearly the contribution of interplanetary shocks to the distribution of SEPs in longitude, and \cite{cane2010}, who discussed 280 events in 1997--2006 during Solar Cycle 23. A summary of more than a thousand such events since 1967, including observations from STEREO, is presented in \citet{richardson2017}. Also, as noted above, this paper updates the catalog of $\sim25$~MeV proton events observed by STEREO and near-Earth spacecraft of \citet{richardson2014}.  These near-Earth observations included proton observations from the  Energetic and Relativistic Nuclei and Electron (ERNE) Sensor \citep{torsti1995} and proton and electron observations from the Electron, Proton Helium Instrument \citep[EPHIN,][]{muller1995}, both on board  the Solar and Heliospheric Observatory \citep[SOHO,][]{domingo1995}.

Proton events extending to tens of MeV are also of space weather interest, including radiation hazards to astronauts \citep[e.g.][]{cucinotta2010} and spacecraft systems \citep[e.g.,][]{iucci05}.  Observations from the HETs have contributed to the development of the SEPSTER (SEP predictions based on STEReo observations) \citep{richardson2018} and SEPSTER2D \citep{bruno2021} empirical SEP prediction models \citep[see also Sections 3.26 and 3.27 in][for summaries of these models]{Whitman2023} that are providing near real-time predictions to the Coordinated Community Modeling Center (CCMC) SEP Scoreboard (\url{https://sep.ccmc.gsfc.nasa.gov/intensity/}).

\citet{richardson2014} provided a comprehensive summary of the properties of over 200 individual SEP events up to December 2013, only 10 months before the loss of contact with STEREO B, and in particular focused on 25 widespread SEP events seen at both STEREO spacecraft and at Earth. An extended set of 43 such events is discussed by \citet{vonrosen2015}, including examples after December 2013.  However, the STEREO spacecraft were then converging on the far side of the Sun from Earth prior to the loss of contact with STEREO~B and thus less favorably configured to provide a widespread view of SEP events.  Therefore, in this paper, we do not extend the analysis of widespread events of \citet{richardson2014} and \citet{vonrosen2015}. Among the studies referring to the \citet{richardson2014} catalog (including the updated version discussed in this paper), recent modeling by \citet{strauss2023} suggests that the differences in electron and proton arrival times at the different spacecraft locations reported by \citet{richardson2014} might be accounted for by diffusive particle transport from a single particle source, while \citet{posner2024} argue that the dependence on the solar event longitude of proton energy dispersion during the onsets of the most widespread SEP events is inconsistent with a source associated with an expanding CME-driven shock. \citet{Posner2025} note that all the 19 ``ground level enhancements" observed on the surface of Mars by the Radiation Assessment Detector on the Mars Science Laboratory \citep[MSL/RAD][]{Hassler2012} from August 2012 to 2024 were associated with widespread SEP $\sim25$~MeV proton events, including events reported in this paper. \citet{Richardson2023} used the catalog of \cite{richardson2014} to identify SEP events up to 2013 associated with CMEs observed in the low corona by the Mauna Loa Solar Observatory (MLSO) Mk3/4 Coronameters, while around 27 SEP events associated with CMEs observed by the MLSO K-coronagraph (K-Cor) in 2013-2022 have been identified using the extended catalog described here \citep{Burkepile2025, stcyr2025}. 

The outline of this paper is as follows: Section~\ref{S-Obs} summarizes HET SEP observations during the STEREO mission up to the end of 2023. The expanded SEP event catalog is introduced in Section~\ref{S-catalog} and is included in Appendix~A. The inter-calibration of the STEREO~A HET and SOHO instruments at the beginning of the mission and end of the first STEREO orbit is discussed in Section~\ref{calib}.   We then summarize examples of statistical studies of the relationship between SEP events and the properties of their solar sources using this data set in Section~\ref{S-dist}. We also consider where the highly energetic SEP events observed by the Payload for AntiMatter Exploration and Light-nuclei Astrophysics (PAMELA) reported by \cite{bruno2018} and the events associated with long duration gamma ray flares \citep{bruno2023} lie in the longitudinal distribution of $\sim25$~MeV proton events.   In Section~\ref{S-stat}, we investigate the CME/flare/SEP association for a large sample of 397 M/X flares. Section~\ref{s-lomb} discusses evidence of periodicity in the SEP occurrence rate during the rise phase of Cycle 25.  
The results are summarized in Section~\ref{S-Summary}.  Finally several individual SEP events are discussed in Appendices B and C.  

\begin{figure}
    \centering
    \includegraphics[width=1\linewidth]{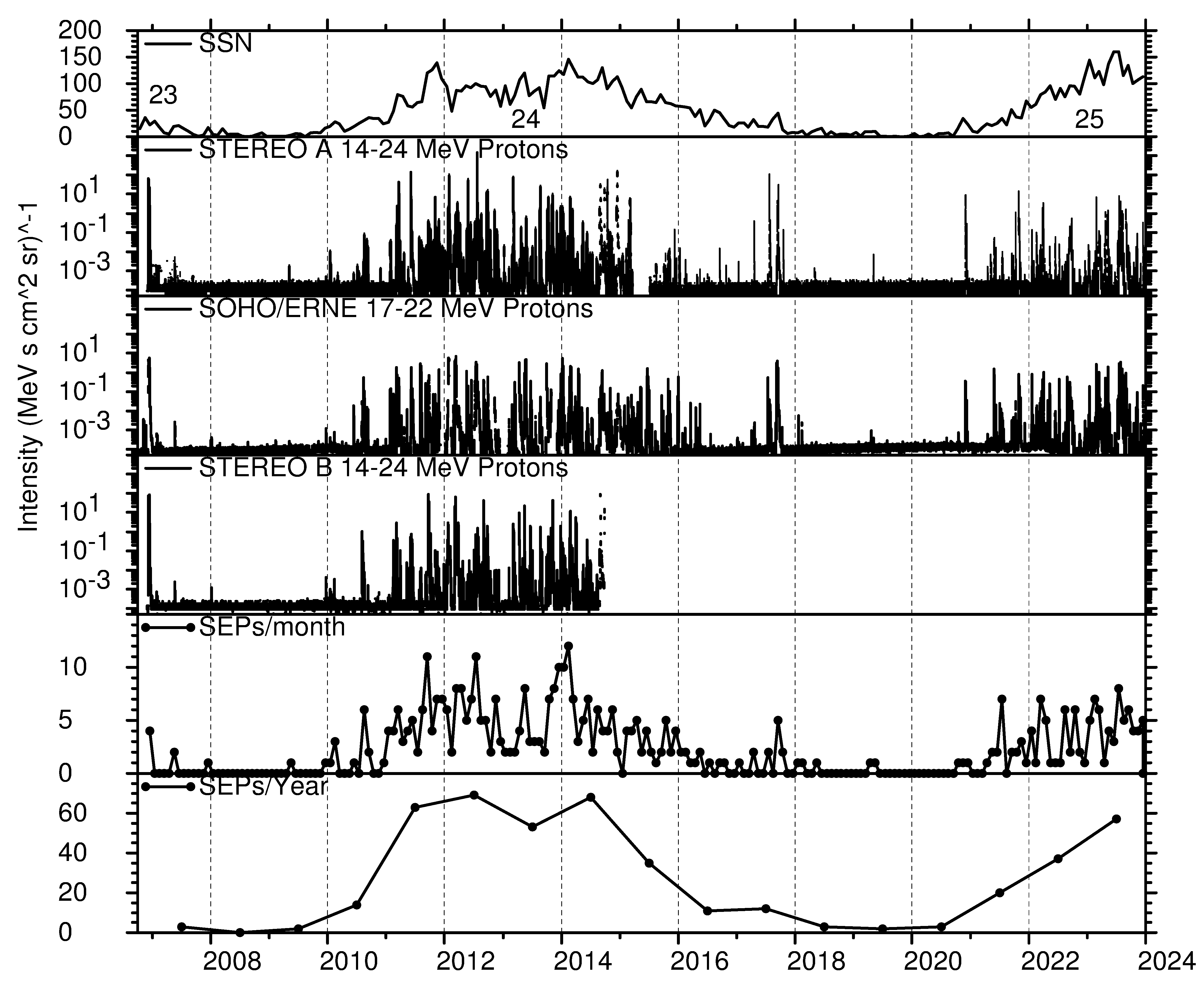}
    \caption{Summary of STEREO A (panel~2) and B (panel 4) 14-24 MeV proton intensities and SOHO ERNE 17-22~MeV proton intensities (panel 3) from STEREO launch in October 2006 to December 2023, just beyond the end of the first solar orbit (relative to Earth) of STEREO~A in August 2023.  Contact with STEREO B was lost on October 1 2014, when the STEREO spacecraft were both $\sim180^\circ$ heliolongitude from Earth. The top panel shows the monthly sunspot number, extending from the late declining phase of Solar Cycle~23 to the ascending phase of Solar Cycle 25. The bottom two panels show the number of individual $\sim25$~MeV proton events per month and per year cataloged in Tables~1 to \ref{tab11}. Note that these SEP rates are influenced by the loss of STEREO~B observations and STEREO~A approaching Earth in Cycle 25 (cf., Figure~\ref{fig:stloc}). }
    \label{fig:3SC}
\end{figure}

\section{STEREO and Near--Earth SEP Observations in 2006--2023}
\label{S-Obs} 
\begin{figure}
    \centerline{
    \includegraphics[width=0.5\linewidth]{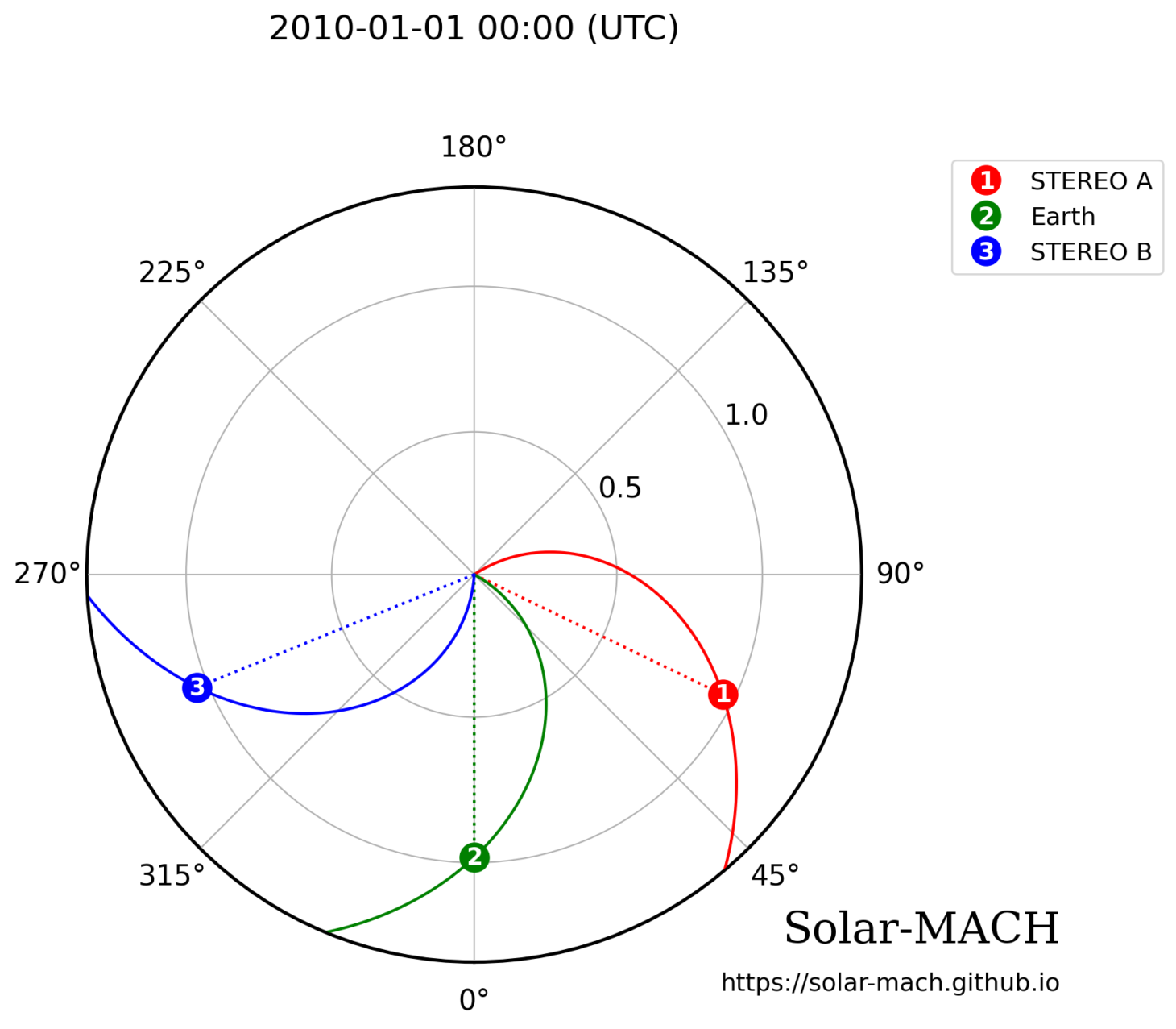}
    \includegraphics[width=0.5\linewidth]{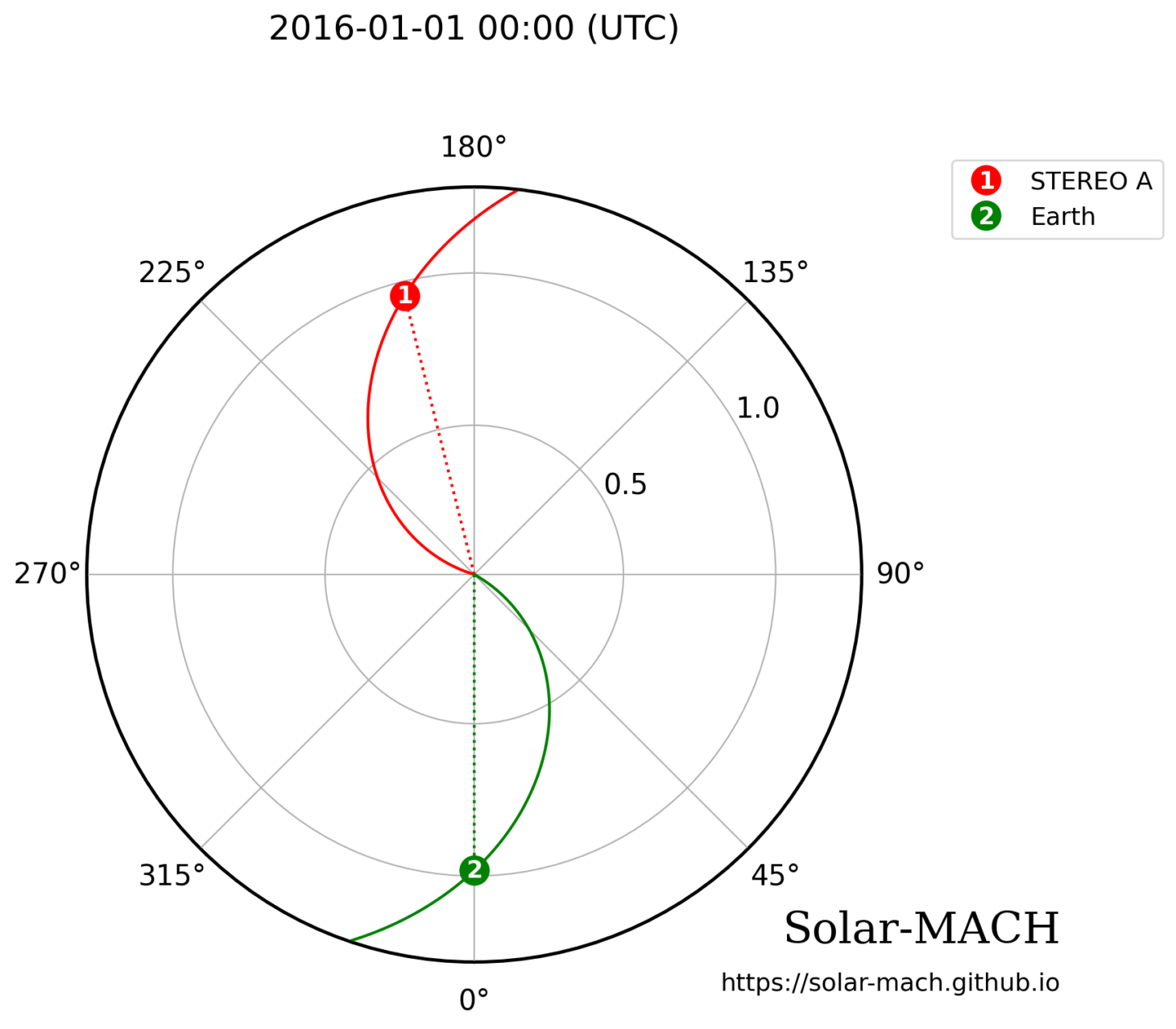}
    }
    \centerline{
    \includegraphics[width=0.5\linewidth]{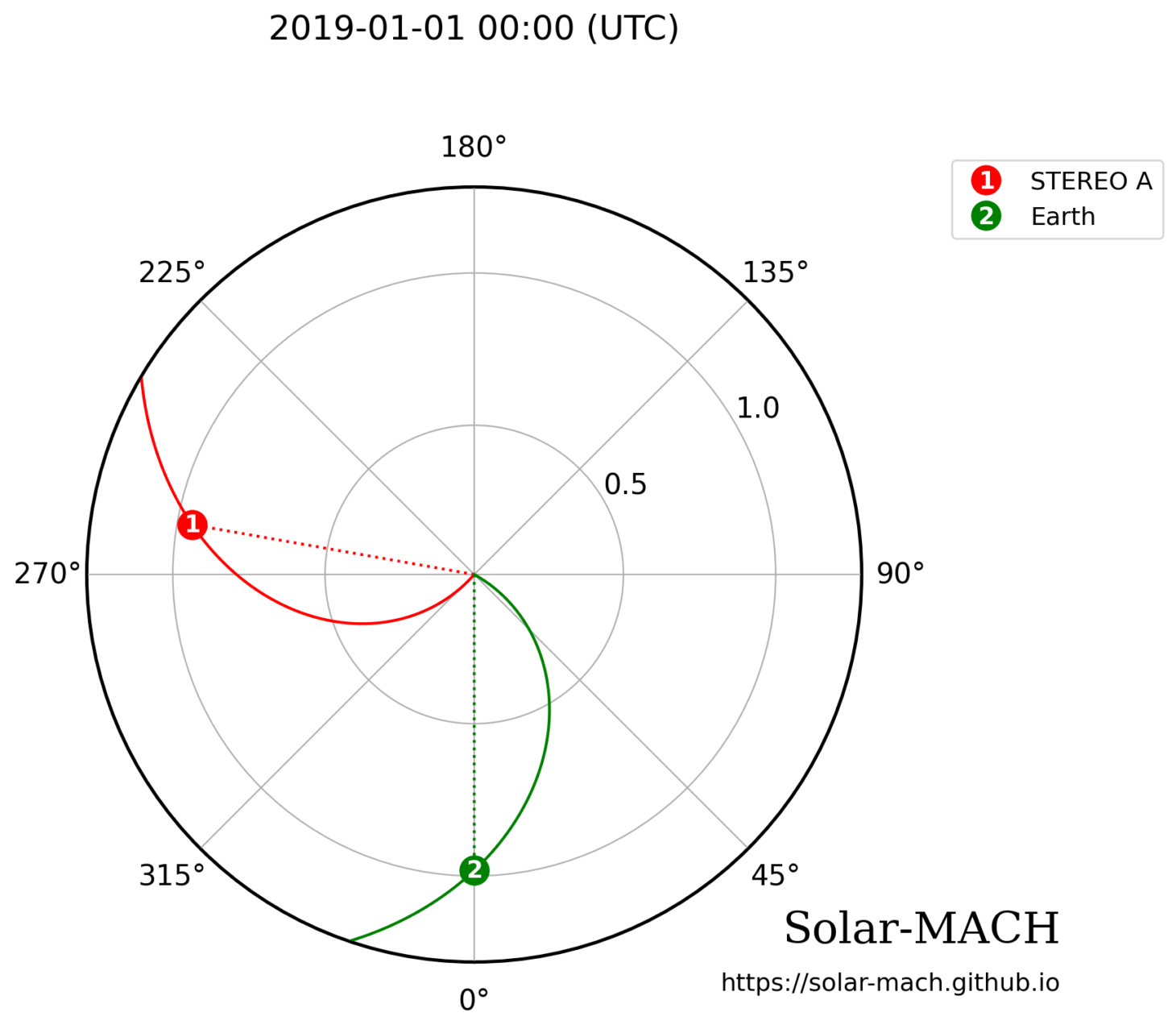}
    \includegraphics[width=0.5\linewidth]{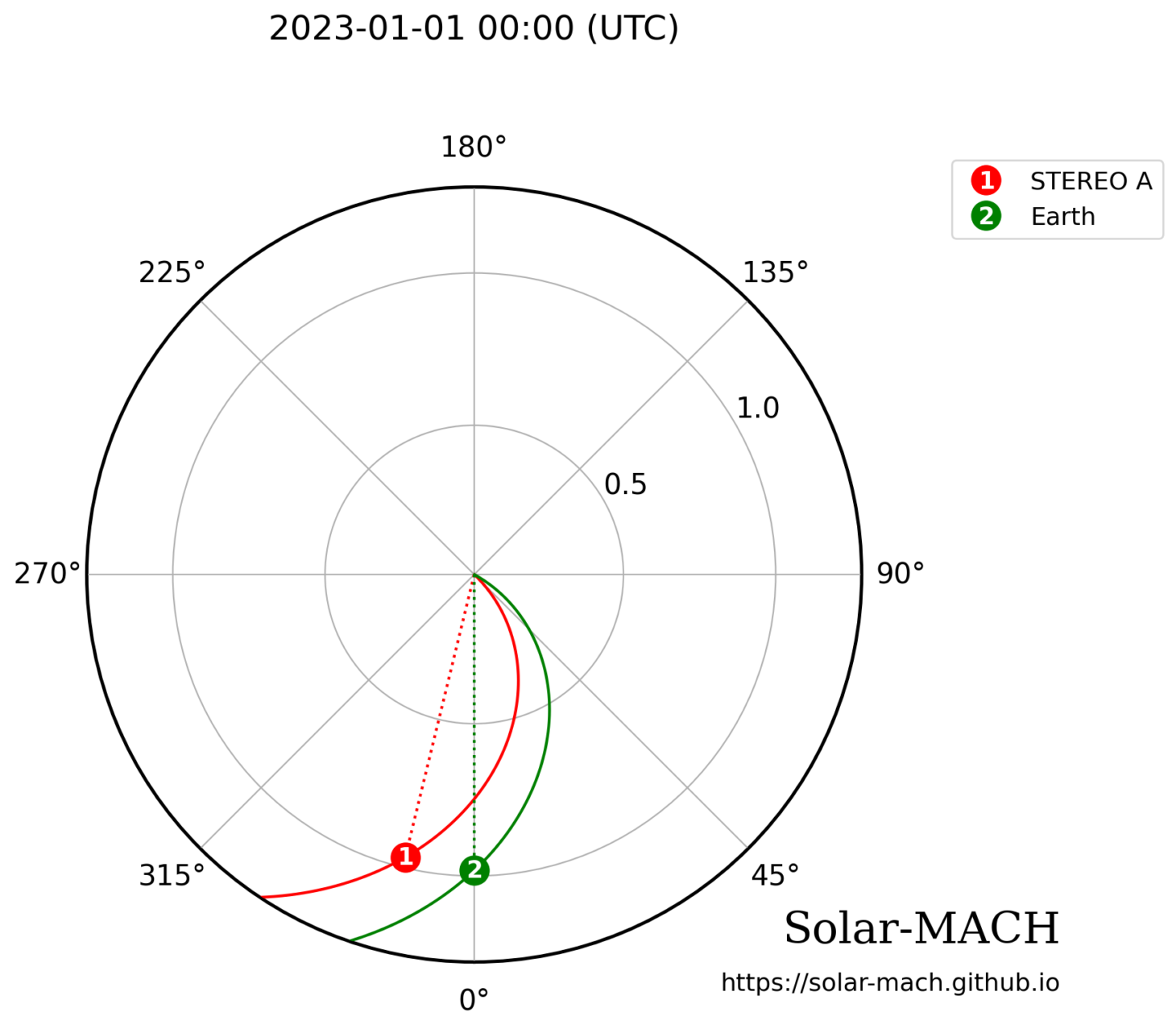}
    }
    \caption{Locations of STEREO~A (1, red) and B (3, blue) relative to Earth (2, green) on January 1 of 2010, 2016, 2019 and 2023. Contact was lost with STEREO~B in October 2014. Nominal Parker spiral magnetic field lines (assuming a solar wind speed of 400 km/s) passing each location are also shown. Figures produced using Solar-MACH \citep[Solar MAgnetic Connection Haus,][]{giesler2023}.}
    \label{fig:stloc}
\end{figure}

Figure~\ref{fig:3SC} shows an overview of the proton intensities at 14--24~MeV measured by the STEREO A and B HETs and, in a similar energy range (17--22~MeV)\footnote{The ERNE data used are level~2 ``export data" served by the NASA Space Physics Data Facility Virtual Energetic Particle Observatory (VEPO, \url{https://spdf.gsfc.nasa.gov/research/vepo/}). However, VEPO provides the approximate on-board energy channel ranges, and these have been corrected to the revised energy ranges given in \url{https://export.srl.utu.fi/export_data_description.txt}.} , by the ERNE instrument on SOHO located at L1 upstream of Earth (note that ERNE saturates in the largest events in Figure~\ref{fig:3SC}). The observations cover the period from STEREO launch (October 2006) to December 2023, shortly after STEREO~A returned to near the Earth in August 2023; Figure~\ref{fig:stloc} shows the STEREO spacecraft locations relative to Earth on January~1 of 2010, 2016, 2019 and 2023. The period in Figure~\ref{fig:3SC} extends from the late declining phase of Solar Cycle 23 to the ascending phase of Cycle 25, as shown by the monthly-averaged sunspot number (from the World Data Center for the Sunspot Index and Long-term Solar Observations (WDC-SILSO), \url{https://www.sidc.be/SILSO/datafiles}) in the top panel.  The occurrence of SEP events in Figure~\ref{fig:3SC} clearly follows the $\sim11$~year solar activity cycle. The final large SEP events of Solar Cycle~23, in December 2006, occurred just after STEREO launch \citep[e.g.,][]{vonrosen2009}.  Only isolated particle enhancements were observed during the next $\sim3$ years in the extended solar minimum between Solar Cycles 23 and 24. Then, following brief intervals of activity in early and mid-2010, the SEP rate finally increased in early 2011, more than two years into Solar Cycle 24 (beginning in December, 2008).  By this time, the STEREO spacecraft were already above the limbs of the Sun when viewed from Earth, allowing observations of SEP events at widely-separated locations \citep[e.g.,][]{richardson2014}.  A notable feature of Solar Cycle 24 in Figure~\ref{fig:3SC} is the temporary reduction in the occurrence of SEP events in late 2012-2013, evident in the monthly and yearly numbers of individual SEP events shown in the bottom two panels (the identification of these events is discussed below in Section~\ref{S-catalog} and Appendix~A).  Such a temporary decrease in the rate of energetic solar events, associated with the ``Gnevyshev Gap" \citep[e.g.,][]{Gnevyshev1967, Gnevyshev1977, Storini2003}, is a frequent feature near the maxima of solar activity cycles.     
Loss of contact with STEREO~B occurred on October 1, 2014, so there are no STEREO~B data after this time.

Occasional proton events occurred during the late decay phase of Solar Cycle 24. Most notably, a burst of activity in July-October, 2017, similar to that in December 2006 near the end of Cycle~23, produced some of the largest SEP events of Cycle 24. In particular, the SEP events on 4-10 September, 2017 have been discussed for example by \citet{bruno2019}, \citet{mishev2018},  \citet{mav2018}, \citet{matthia2018} and \citet{jiggens2019}.   Rare small events occurred during the solar activity minimum between Solar Cycles~24 and 25, when STEREO~A was returning towards Earth above the east limb of the Sun (Figure~\ref{fig:stloc}); an example is highlighted in Appendix~B. SEP activity then increased following the onset of Cycle 25.  As noted above, the first widespread event of Cycle 25, on November 29, 2020, was discussed by \citet{kollhoff2021}. At this time, STEREO~A was 70$^\circ$ east of Earth, and the event was also detected by near-Earth and inner heliosphere spacecraft. Figure~\ref{fig:stloc} shows that STEREO~A continued to approach Earth during the rising phase of Cycle~25, finally passing the longitude of Earth in August 2023.

\section{SEP Catalog}
\label{S-catalog}

As mentioned above, \citet{richardson2014} cataloged the $\sim25$~MeV proton events observed at the STEREO spacecraft and/or at Earth during the first seven years of the STEREO mission up to December 2013 and summarized their properties, including the associated solar phenomena (flares, radio emission and CMEs).  Tables~1 to \ref{tab11} in Appendix~A update this catalog to December 2023, slightly beyond the end of the first STEREO~A orbit in August 2023. For completeness, and because parameters for a few events have been updated or corrected, we have included the events identified by \citet{richardson2014}. The catalog is also available at the Harvard Dataverse \citep{richardson2024} and will be further updated as data availability permits. “$\sim25$~MeV” as a description of the proton events aligns with that used in previous similar catalogs such as \citet{cane2010}.  In addition, the intensities are based on observations from various spacecraft instruments around this energy that cover different energy ranges, as discussed below and in Appendix~A.

\begin{figure}
    \centering
    \includegraphics[width=0.9\linewidth]{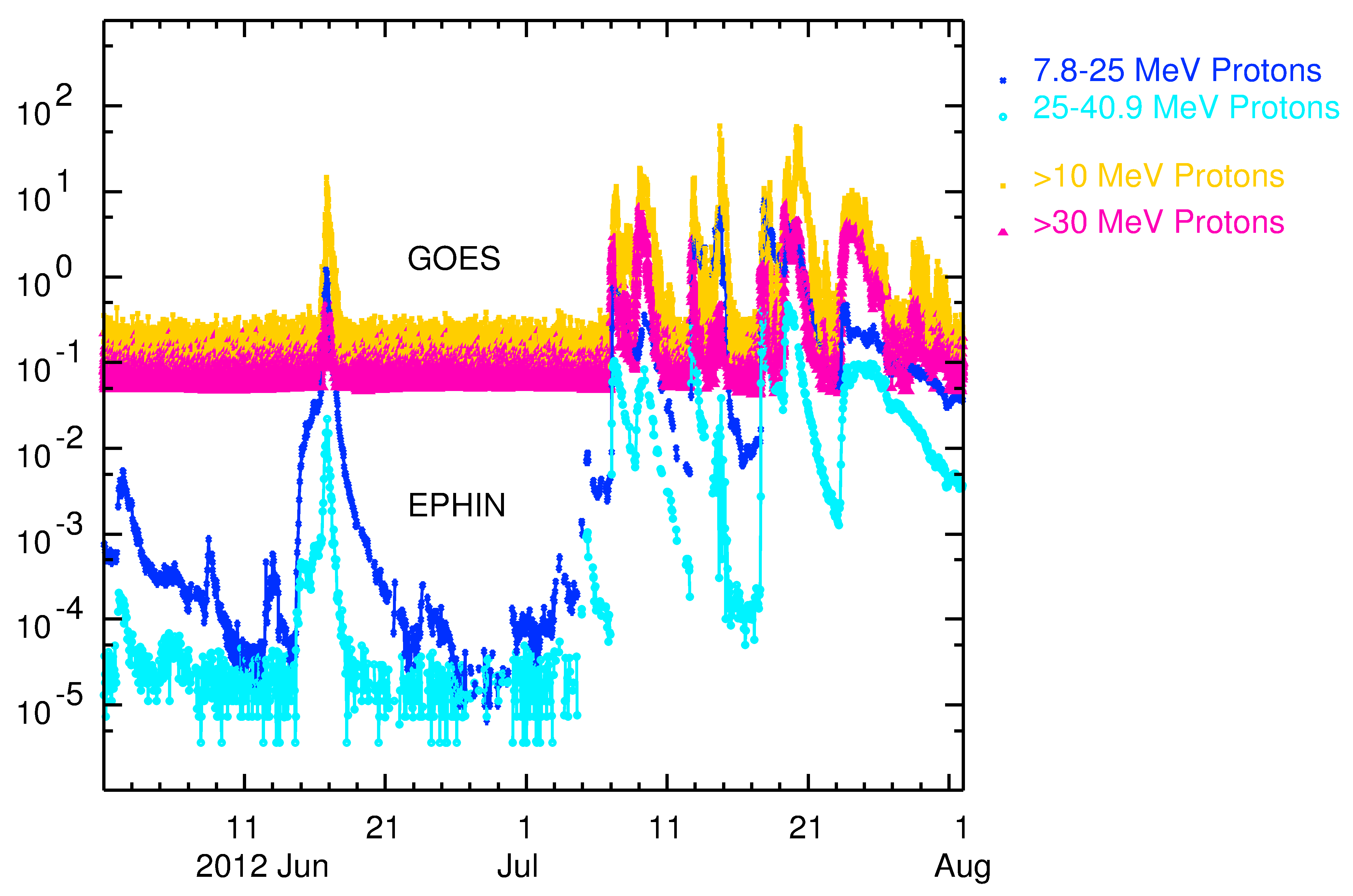}
    \caption{Comparison of GOES $>10$ and $>30$~MeV proton fluxes in (s~cm$^2$~sr)$^{-1}$ (= proton flux unit, pfu) and 7.8-25~MeV and 25-40.9~MeV proton intensities (in (MeV~s~cm$^2$~sr)$^{-1}$) from SOHO/EPHIN for an interval in June-July, 2012, illustrating how the high background in the GOES fluxes obscures all but the largest SEP events that are observed by EPHIN. Using the standard/operational SEP event definition of $>10$~pfu at $>10$~MeV further limits the number of ``SEP events", and these are unrepresentative of the large dynamic range of SEP event intensities evident in the EPHIN observations. From \citet{Richardson2023}. }
    \label{fig:goesephin}
\end{figure}

The SEP events are predominantly identified from examination of $\sim1$~hour-averaged (and also higher cadence) electron and proton observations from the STEREO HETs and the ERNE and EPHIN instruments on SOHO in orbit around the L1 point upstream of Earth. In a few cases (e.g., to identify SEP onsets during occasional SOHO data gaps), observations from the Advanced Composition Explorer \citep[ACE,][]{stone1998} and Wind \citep{wilson2021} spacecraft, also at L1, as well as the GOES (Geostationary Operational Environmental Satellite) spacecraft in geosynchronous Earth orbit, have also been considered.  

Although integral proton flux data from the GOES spacecraft are widely used for near-Earth SEP studies and compiling event catalogs \citep[e.g.,][]{rotti2022}, as well as operations \citep[e.g.,][]{Bain2021}, we do not focus on GOES data here for the following reasons, also discussed by \citet{Richardson2023}: Figure~\ref{fig:goesephin} compares GOES $>10$~MeV and $>30$~MeV proton fluxes with 7.8--25~MeV and 25--41~MeV proton intensities from EPHIN during a representative period in June--July, 2012.  It illustrates how the high background in the GOES observations obscures many small SEP events that are evident in EPHIN, which has a much lower background, similar to that in the HETs (see Figure~\ref{fig:3SC}). Although unimportant for space weather or operations, such events form part of the SEP population, and, as discussed further below, may include both small events that nonetheless have clear solar sources, such as that discussed in Appendix~B, and particles from large events originating at distant locations on the Sun that are important for understanding particle acceleration and transport.  Furthermore, the standard National Oceanic and Atmospheric Administration (NOAA) list of solar proton events affecting the Earth environment (\url{ftp://ftp.swpc.noaa.gov/pub/indices/SPE.txt}) requires a threshold of 10~(s~cm$^2$~sr)$^{-1}$ (= proton flux unit, pfu) in the GOES $>10$~MeV proton flux to be exceeded. Many studies use this criterion, which corresponds to the lowest level (S1) of the NOAA Space Weather Prediction Center Solar Radiation Storm Scale (\url{https://www.swpc.noaa.gov/noaa%2Dscales%2Dexplanation}), to determine whether an SEP event or ``no event" is present. However,  Figure~\ref{fig:goesephin} clearly illustrates that these events are literally the ``tip of the iceberg" of the range of SEP event size.    Although this criterion may be suitable for identifying events of space weather interest \citep[e.g., ][]{Bain2021} including spacecraft operations, and is a focus of many efforts to predict SEP events \citep[e.g.,][]{Whitman2023}, considering just a limited sample of relatively large events (which may also suffer from the ``Big Flare Syndrome" \citep{kahler1982}) is extremely restrictive for studies aimed at understanding the nature of SEP events and their solar sources. In addition, SEP event onsets in GOES-based SEP event lists are often based on the time when the 10~pfu threshold is crossed, which may be considerably delayed from the time of the related solar event, especially when the intensity is slowly rising. Here, we select any SEP event that is evident at proton energies of {\bf $\sim25$} MeV above the low instrumental backgrounds of the HETs and SOHO instruments while also recognizing that there are many additional SEP events that do not reach such energies that are worthy of study.  Furthermore, we acknowledge that the GOES instruments, designed for operations, do have the advantage that they are less likely to saturate in exceptionally intense SEP events.

Details of the parameters included in the proton event catalog are given in Appendix~A.  These include the peak proton intensities at $\sim25$~MeV during the ``first peak" of the SEP event at STEREO A/B and near the Earth (i.e., not including any later ``energetic storm particle (ESP)" intensity peak associated with interplanetary shock passage),  the longitudes of the associated solar event/flare relative to Earth and the STEREO spacecraft, the GOES soft X-ray flare peak intensity (for frontside or near limb events with respect to Earth), type II and III radio emissions observed by instruments on Wind and the STEREO spacecraft and the associated CME speed and width (mainly from the CDAW (Coordinated Data Analysis Workshop) SOHO Large Angle and Spectrometric Coronagraph Experiment (LASCO) CME catalog, \url{https://cdaw.gsfc.nasa.gov/CME_list/}) for each SEP event, where available. As in the \citet{richardson2014} catalog extending to December, 2013, the updated catalog again shows that essentially every $\sim25$~MeV proton event can be associated with an identifiable solar eruption, a CME observed by LASCO (51\%  are identified as LASCO ``halo CMEs" in the CDAW CME catalog indicating that the CME front surrounded the C2 occultor, though not necessarily symmetrically), and also with type III radio emissions observed by Wind and/or the STEREO spacecraft. Where information is available, 48\% of these events were accompanied by type II emissions observed by these spacecraft. Three ``problematic" events, where the solar event associations are more challenging to discern are discussed in Appendix~C.

Several other SEP catalogs exist that are complementary to our catalog. These include:
\begin{itemize}
    \item The STEREO SEP event list available at \url{https://stereo-ssc.nascom.nasa.gov/data/ins_data/impact/level3/}. This includes the subset of events meeting the criterion that the flux of 13-100 MeV protons from HET measurements is greater than 10 pfu, mimicking the NOAA Solar Proton Event list based on GOES $>10$~MeV proton data. Event intensity-time plots are also provided;
    
    \item The catalog of 55--80 MeV solar proton events extending through Solar Cycles 23 and 24 of \citet{paassilta2017};
    \item The catalog of $>55$ MeV wide-longitude solar proton events observed by SOHO, ACE, and the STEREOs at $\sim 1$~AU during 2009 - 2016 compiled by \citet{paassilta2018};
    \item \citet{kuhl2017} discuss 42 SEP events with protons above 500 MeV in 1995--2015 measured with SOHO/EPHIN;
    \item The SEPServer (\url{https://sepserver.eu/}) event catalog \citep{vainio2013,papaioannou2014};
    \item The multi-spacecraft solar energetic particle event catalog for Cycle 25 produced by the SERPENTINE project \citep[][\url{https://data.serpentine-h2020.eu/catalogs/sep-sc25/}]{Dresing2024};
    \item The Wind/EPACT(Energetic Particles Acceleration, Composition, Transport) Proton Event Catalog (1996--2016) compiled by \citet{miteva2018};
     \item \citet{pande2018} include 35 SEP events observed by GOES or SOHO/ERNE associated with $\ge$ M flares in 2010-2014;
      \item The ``Integrated Geostationary Solar Energetic Particle Events Catalog" of \citet{rotti2022} based on GOES data;
      \item The ``Catalogues of Solar Proton Events in the 20-25 Cycles of Solar Activity" compiled by Moscow State University (\url{https://swx.sinp.msu.ru/apps/sep_events_cat/}; and
      \item The ``Catalog of Solar Proton Events in the 24th Cycle of Solar Activity (2009–2019)" by \citet{logachev2022}.
\end{itemize}

\section{HET Calibration Check in December 2006 and August 2023.}
\label{calib}
As discussed by \citet{richardson2014}, the occurrence of the last large SEP events of Solar Cycle 23 in December, 2006 \citep{vonrosen2009}, around two months after STEREO launch when the STEREO spacecraft were still close to Earth, allowed the HET calibration to be checked relative to instruments on near-Earth spacecraft over a large dynamic range in particle intensity. This timing was extremely fortuitous since (Figure~\ref{fig:3SC}) comparable SEP events were only again observed after an interval of several years when the STEREO spacecraft were already widely separated from the Earth.  While the proton responses of the HETs and instruments on near-Earth spacecraft were found to be similar, an unexpected factor of $\sim14$ reduction in the 0.7-4~MeV electron intensity measured by HET relative to these other instruments was noted, as also discovered independently by \citet{lario2013}.  

\begin{figure}
    \centering
    \includegraphics[width=1.0\linewidth]{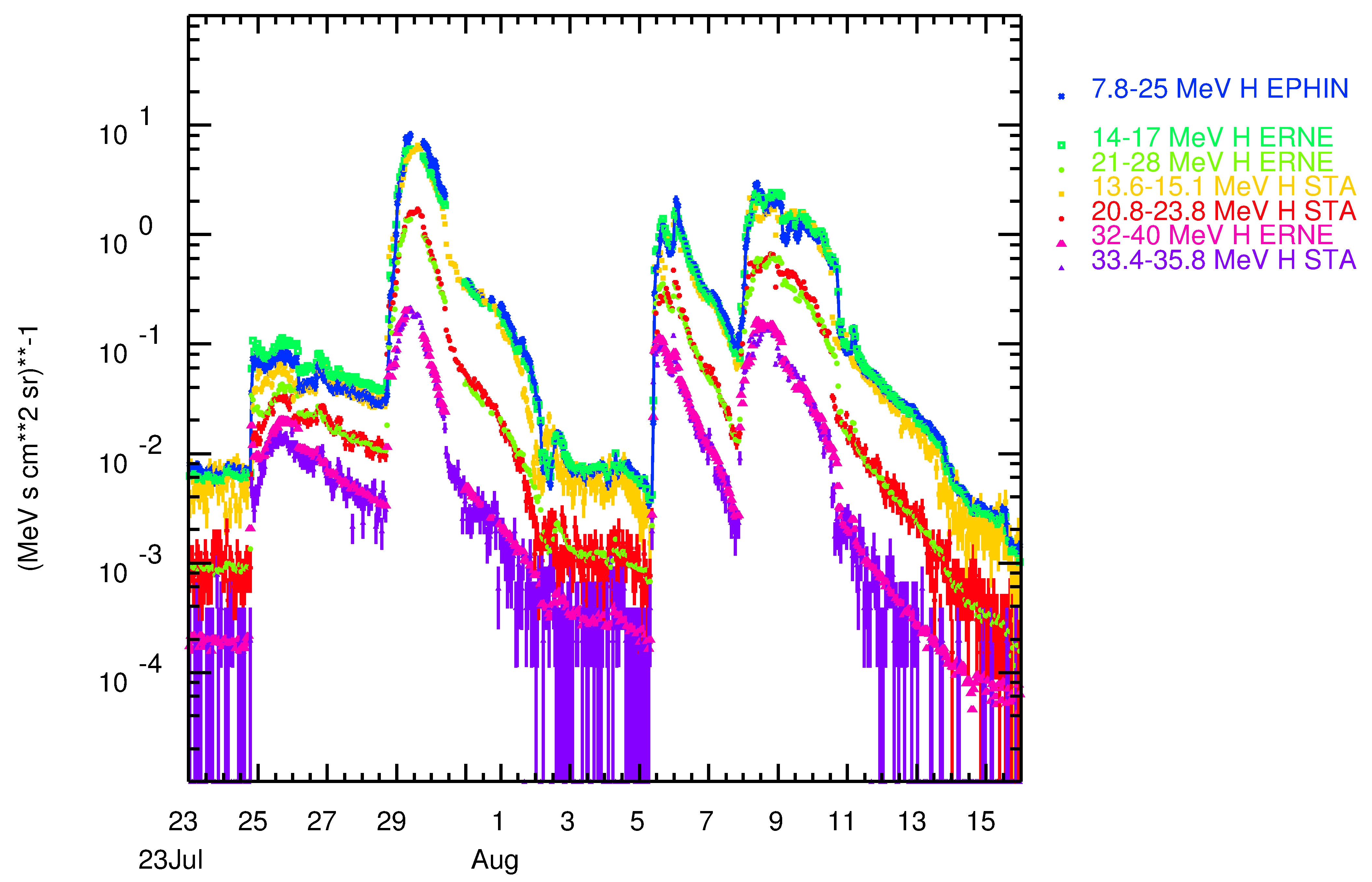}
    \caption{Comparison of hourly-averaged proton intensities observed by SOHO/EPHIN (7.8-25 MeV), SOHO/ERNE (14-17~MeV, 21-28~MeV and 32-40 MeV) and STEREO~A HET (13.6-15.1, 20.8-23.8 MeV and 33.4-35.8~MeV) during a sequence of SEP events between July 23 and August 15, 2023 when STEREO~A moved from $1.9^\circ$ east of Earth to $0.3^\circ$ west. The intensities are consistent in channels of similar energy ranges. }
    \label{fig:hetsohointcomp}
\end{figure}

STEREO~A returned to the vicinity of Earth during the rising phase of Solar Cycle~25 and hence observations of SEP events are available to check whether the HET and SOHO instrument responses remained consistent. Several such events are illustrated in Figure~\ref{fig:hetsohointcomp}, which compares one-hour averaged proton intensities from SOHO/EPHIN, SOHO/ERNE and STEREO~A HET on July 23-August 15, 2023 when STEREO~A moved from $1.9^\circ$ east of Earth to $0.3^\circ$ west. The observations show that similar intensities were measured in comparable energy channels of the STEREO A HET and SOHO instruments during these events.

\begin{figure}
    \centering
    \includegraphics[width=1\linewidth]{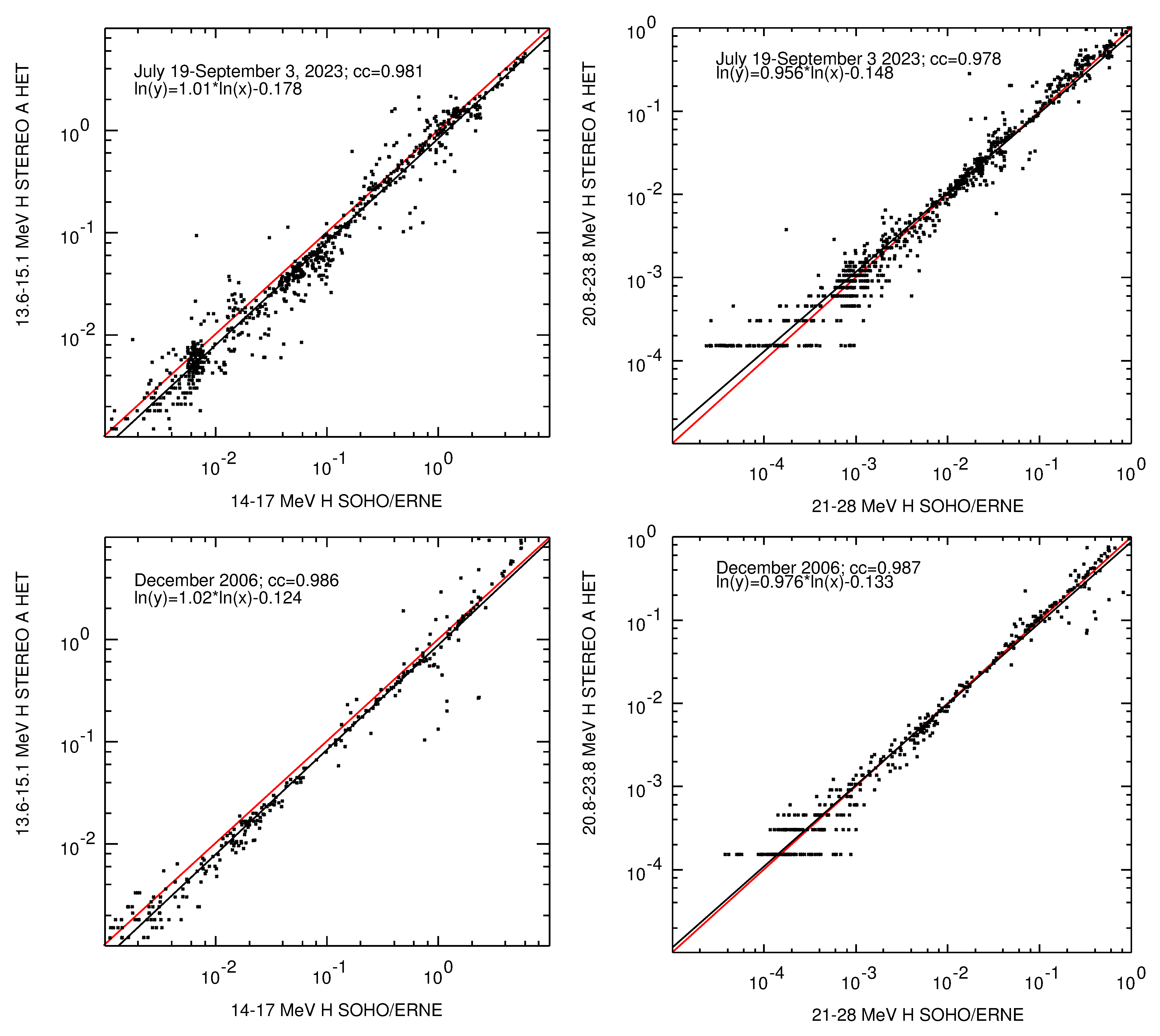}
    \caption{The top panels shows the correlation of SOHO/ERNE and STEREO~A~HET hourly-averaged proton intensities ((MeV s cm$^2$ sr)$^{-1}$) in two energy ranges during July~19-September~3, 2023 when STEREO~A moved from $2.3^\circ$ east of Earth to $2.0^\circ$ west.  The red lines indicate equal intensities. The bottom panels show essentially identical correlations in December 2006 near the start of the STEREO mission. The energy channels for SOHO/ERNE (STEREO~A HET) are 13--17 (13.6--15.1)~MeV in the left panels and 21--28 (20.8--23.8) MeV in the right panels. The ``quantization" in the HET data in the right panels is due to low particle counts.}
    \label{fig:staernechans}
\end{figure}

Another way of comparing the observations in these instruments is to investigate the correlation between the intensities measured in similar energy channels, as in \citet{richardson2014}. As an example, the top panels of Figure~\ref{fig:staernechans} compare the proton intensities measured by STEREO~A HET (SOHO/ERNE) at 13.6-15.1~MeV (14-17~MeV) and 20.8-23.8~MeV (21-28~MeV) using simultaneous hourly-averaged data from July~19 (avoiding a large SEP event commencing late on July~17 in which ERNE saturated) to September~3, 2023. During this interval, STEREO~A moved from $2.3^\circ$ east to $2.0^\circ$ west of Earth.  Black lines indicate the least squares fits to the data, which are close to the lines of equality (red). The bottom panels show the essentially identical correlations using the same energy channels for December 2006, when, as noted above, several SEP events occurred while the STEREO spacecraft were still close to Earth shortly after launch \citep[e.g.,][]{vonrosen2009,richardson2014}. 
These results again indicate that, at least at proton energies around $\sim20$~MeV, the HET and ERNE proton inter-calibration at the end of the first STEREO~A orbit is consistent with that in the early stages of the mission.  

\begin{figure}
    \centering
    \includegraphics[width=0.75\linewidth]{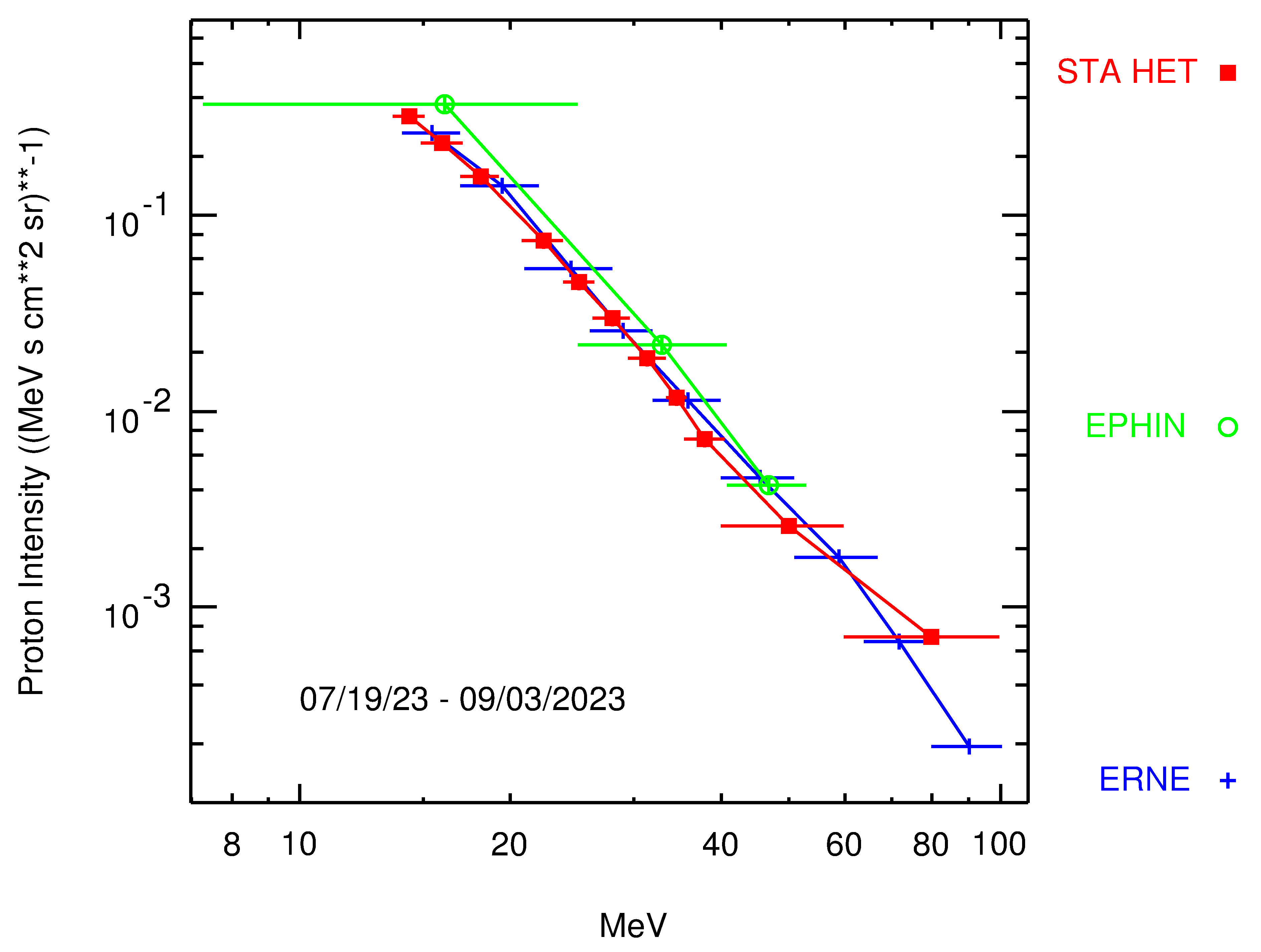}
    \caption{Comparison of proton spectra at $\sim13$-100~MeV from SOHO EPHIN (green circles) and ERNE (blue crosses), and STEREO~A HET (red squares) for July~19-September~3, 2023, showing excellent agreement between the STEREO~A HET and the SOHO instruments. (Data generated by the VEPO with the ERNE energy channels corrected.) }
    \label{fig:sohostaspectra}
\end{figure}

Another comparison may be made using particle spectra. Figure~\ref{fig:sohostaspectra} compares proton spectra at $\sim13$-100~MeV from SOHO/EPHIN, SOHO/ERNE and the STEREO~A HET for the period in Figure~\ref{fig:staernechans}.  Again the proton response of the different instruments appears to remain in excellent agreement.  Based on these comparisons, and assuming that they represent the instrument responses throughout the first STEREO A orbit, and that the STEREO B HET also behaved similarly until loss of spacecraft contact, we conclude that it is not necessary to correct the proton intensities in Tables~1 to \ref{tab11} in Appendix~A for any drift in the calibration of HET relative to the SOHO instruments during the first STEREO~A orbit.  

We do not consider the electron response here since the electron response of the EPHIN instrument has changed since the December 2006 events (B. Heber, private communication, 2023).  Therefore, the equivalent comparison cannot be made for the 2023 period and requires further analysis that is beyond the scope of this paper. We note however, that \citet{farwa2025} (see their Appendix~B) estimated a factor of 10 difference in the electron calibrations of the SolO HET (an instrument unrelated to the STEREO HETs) and STEREO~A HET during the decay phases of several SEP events, which is reasonably consistent with the factor of $\sim14$ difference between the SOHO-STEREO HET electron calibrations at the beginning of the STEREO mission.  

\begin{figure}
    \centering
    \includegraphics[width=1\linewidth]{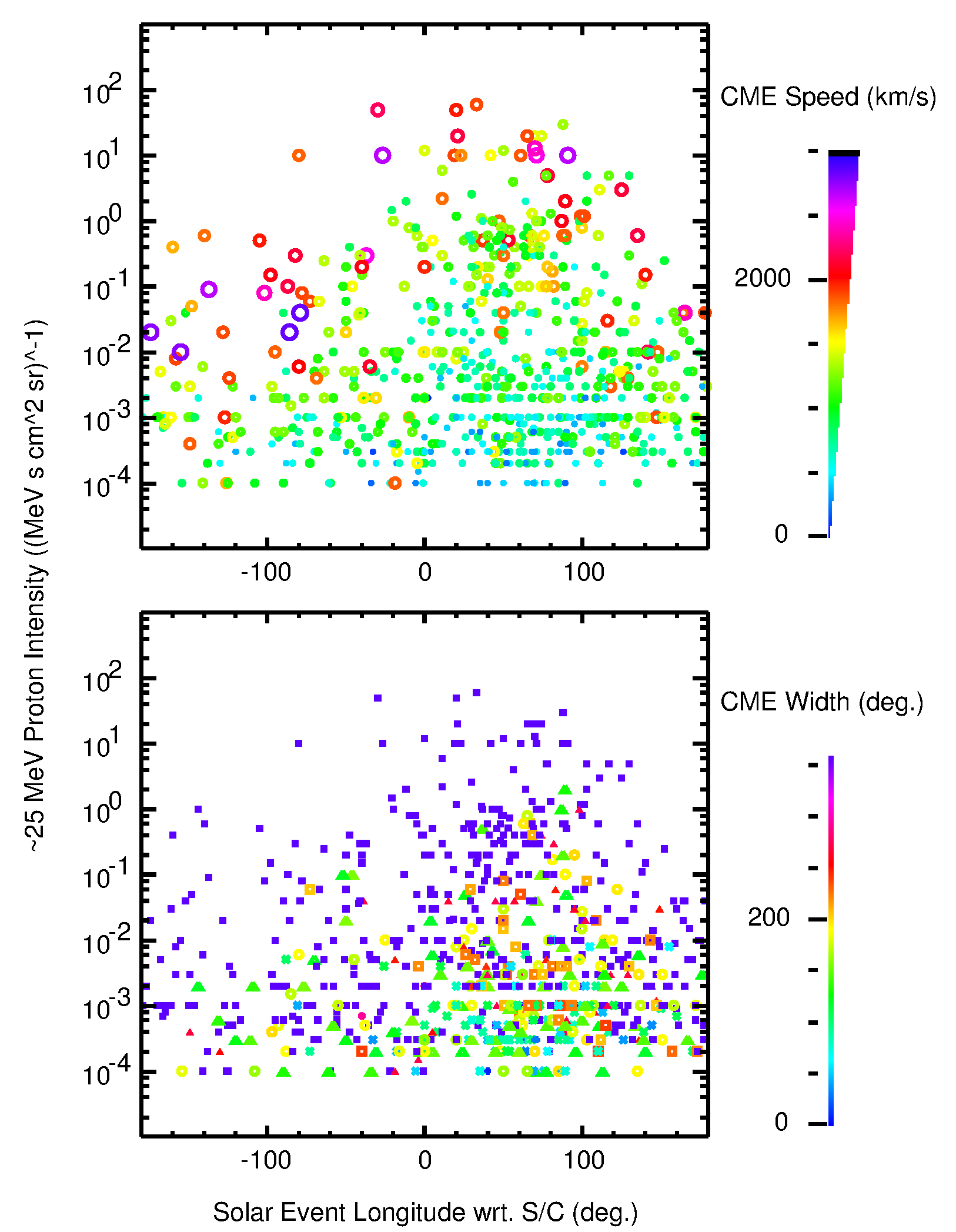}
    \caption{$\sim25$~MeV proton intensity vs. solar event longitude relative to the observing spacecraft for the events in Tables 1-11. Colors/symbols indicate the speed of the associated CME in the top panel and the width of the CME in the bottom panel (blue squares indicate halo (360$^\circ$ width) CMEs). CME parameters are from Tables~1--11 in Appendix~A and, with a few exceptions, from the CDAW LASCO CME catalog.}
    \label{fig:IvslongvsCME}
\end{figure}

\section{Comparison of SEP Events With the Properties of the Associated Solar Events}
\label{S-dist}
\citet{richardson2014} discussed several properties of the SEP events and the associated solar events observed up to December 2013, and typically the conclusions from that study also apply to the expanded set of events. One difference is that, as noted above, the events in \citet{richardson2014} were mainly observed when the STEREO spacecraft and Earth were approximately evenly-spaced in longitude around the Sun. Hence, for example, the number of locations at which an SEP event was detected gave a reasonable idea of how widespread an event was. With the loss of contact with STEREO B, and STEREO A making an orbit of the Sun relative to Earth, instead we consider the observations in Tables~1 to \ref{tab11} as a large ensemble of SEP observations made over a wide range of heliolongitudes relative to the location of the related solar event, a single SEP event providing one to three observations depending on the number of spacecraft locations at which the SEP event was detected. Here, we summarize some of the basic properties of this ensemble of SEP events and their associated solar events.  We also point out some biases that are inherent in these observations.

Figure~\ref{fig:IvslongvsCME} summarizes, in the top panel, $\sim25$~MeV proton intensities vs. the solar event longitude relative to the observing spacecraft when an SEP event was detected (i.e., we omit cases where observations are available from a particular spacecraft location, but no SEP event was detected, or there was a high background or a data gap). This figure again emphasizes the result from \citet{richardson2014} and  \citet{richardson2017} that SEPs may be detected at all longitudes relative to a solar eruption, though the largest, and also frequent small events, tend to be observed when the eruption is on the western hemisphere (0-90$^\circ$ longitude) relative to the observing spacecraft, and hence likely to be well-connected to the spacecraft by the nominal spiral interplanetary magnetic field (IMF).  The color and size of the symbols indicate the speed of the associated CME. The fastest CMEs ($\sim2000$~km/s and above), indicated by the larger red and purple symbols, tend to be associated with the largest SEP events at all longitudes.  However, somewhat slower CMEs are also apparently associated with SEP events detected far from the spacecraft, suggesting that an exceptionally fast CME in the mid-corona is not a requirement for SEPs to be detected far from the associated solar event. On the other hand, the SEPs associated with the slowest CMEs (small blue symbols) tend to be weak and predominantly detected within $\sim60^\circ$ of the nominal magnetic connection at $\sim$W60$^\circ$. 

We should emphasize that the low speeds for many of these SEP-associated CMEs are not due to, for example, projection effects from using the CDAW plane of the sky speeds. In particular, \citet{richardson2015} show many clear examples of $\sim25$~MeV proton events associated with slow CMEs using CME parameters from several catalogs, including three-dimensional speeds from the DONKI (Database Of Notifications, Knowledge, Information) database (\url{https://kauai.ccmc.gsfc.nasa.gov/DONKI/}), and CME parameters inferred using coronagraph observations from spacecraft in quadrature to the solar event, reducing projection effects.  We also note that MLSO K-Cor observations suggest that some SEP-associated CMEs do have higher speeds in the low corona than in the middle corona observed by LASCO \citep{stcyr2025}. 

Although these results may not necessarily implicate particle acceleration at the expanding CME-driven shock as the sole mechanism for distributing protons in widespread events, they do indicate that the intrinsically largest SEP events, typically associated with fast CMEs, tend to be detected at large longitudinal distances from the solar event, as was also noted by \cite{richardson2014}.  This study, as well as several others \citep[e.g.,][]{lario2013, cohen2017,richardson2017, paassilta2018, bruno2021}, found that the longitudinal intensity dependence in individual SEP events may be expressed as a Gaussian with a standard deviation of $\sim40^\circ$ relative to the IMF footpoint, and qualitatively, the ``Gaussian-like" dependence is evident in Figure~\ref{fig:IvslongvsCME}. 

\begin{figure}
    \centering
    \includegraphics[width=1\linewidth]{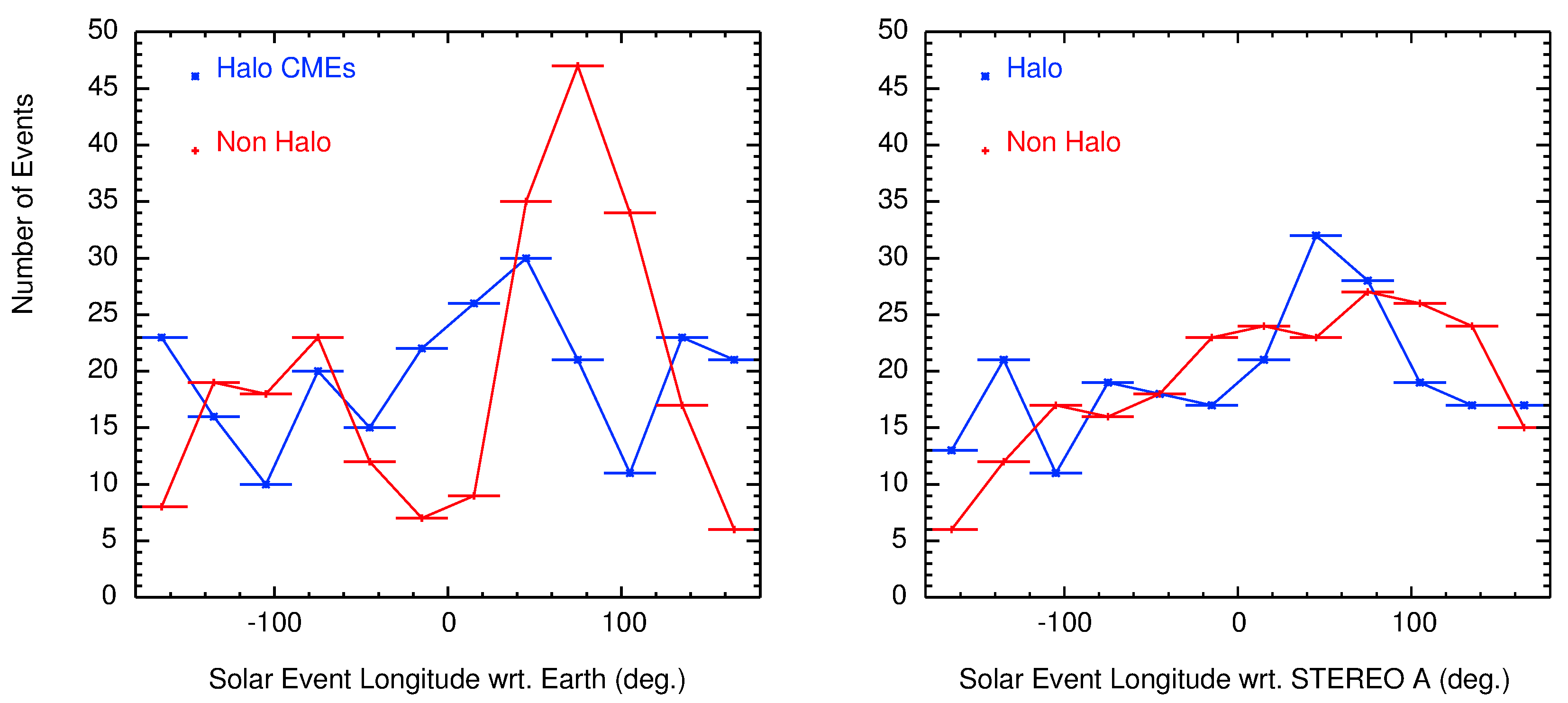}
    \caption{Left: Number of SEP-associated CDAW catalog ``halo" (blue graph) and ``non-halo" (red graph) SOHO/LASCO CMEs for the events in this study vs. the longitude of the associated solar event relative to Earth. The solar events associated with halo CMEs are approximately evenly distributed in longitude whereas non-halo CMEs tend to be associated with solar events located around the east and west limbs relative to Earth (longitudes $\sim \pm90^{\circ}$).  This pattern for the non-halo CMEs is removed when the event longitude relative to STEREO~A is used (right panel) due to the changing spacecraft location relative to Earth (and similarly for STEREO~B, not shown).  The biases in the left panel due to the use of CME widths from the CDAW LASCO catalog are included in the bottom panel of Figure~\ref{fig:IvslongvsCME}. }
    \label{fig:nhal}
\end{figure}

In the bottom panel of Figure~\ref{fig:IvslongvsCME}, the symbol color indicates the CME width from the CDAW SOHO/LASCO catalog. As noted above, a large fraction of the SEP-associated CMEs are ``halo" (360$^\circ$ width) CMEs in this catalog (though not in other catalogs, e.g., \citet{richardson2015}) and these are associated with SEPs detected at all longitudes relative to the solar event. The narrower CMEs tend to be associated with SEPs from western events that are reasonably well connected with the observer. However, there is a bias introduced into this figure because the LASCO CME widths are projected against the plane of the sky. In particular, a halo designation in the CDAW catalog indicates that some signature of the CME surrounds the LASCO C2 occultor, but the CME is frequently asymmetric and associated with a solar event away from central meridian. The blue graph in the left panel of Figure~\ref{fig:nhal} shows that the locations of the related solar events for the SEP-associated halo CMEs in Tables~1 to \ref{tab11} of Appendix~A are approximately evenly distributed in longitude relative to Earth. This clearly demonstrates that these halo CMEs are not generally directed towards or away from the Earth (as is often erroneously assumed for CDAW ``halo" CMEs, as discussed by \citet{stcyr2005}) but are approximately equally likely to have any direction relative to the Sun-Earth line.  This result is consistent with \citet{kwon2014} who noted that SEP-associated CMEs tend to be halo-like even when observed well away from the direction of propagation. On the other hand, the solar events associated with non-halo CMEs (red graph in the left panel) tend to cluster near the limbs relative to Earth (longitudes $\sim \pm90^{\circ}$), although with a western bias presumably because of the required association with SEP events and the influence of the spiral IMF, and are rarely located near central meridian. This distribution suggests that CMEs are more likely to be classified as non-halo, with a width less than 360$^\circ$, when originating towards the limbs, where projection effects on the width are reduced. In the right panel of Figure~\ref{fig:nhal}, the distribution of non-halo CMEs is more uniform when plotted vs. the solar event location relative to STEREO~A because the spacecraft was  separated from Earth and SOHO (a similar plot could be made for STEREO~B). Since the bottom panel of Figure~\ref{fig:IvslongvsCME} includes results from both STEREO and near-Earth spacecraft, the different longitudinal distributions of halo and non-halo CMEs at each spacecraft and the bias in CME widths from the CDAW catalog will be incorporated into this figure.

\begin{figure}
    \centering
    \includegraphics[width=1\linewidth]{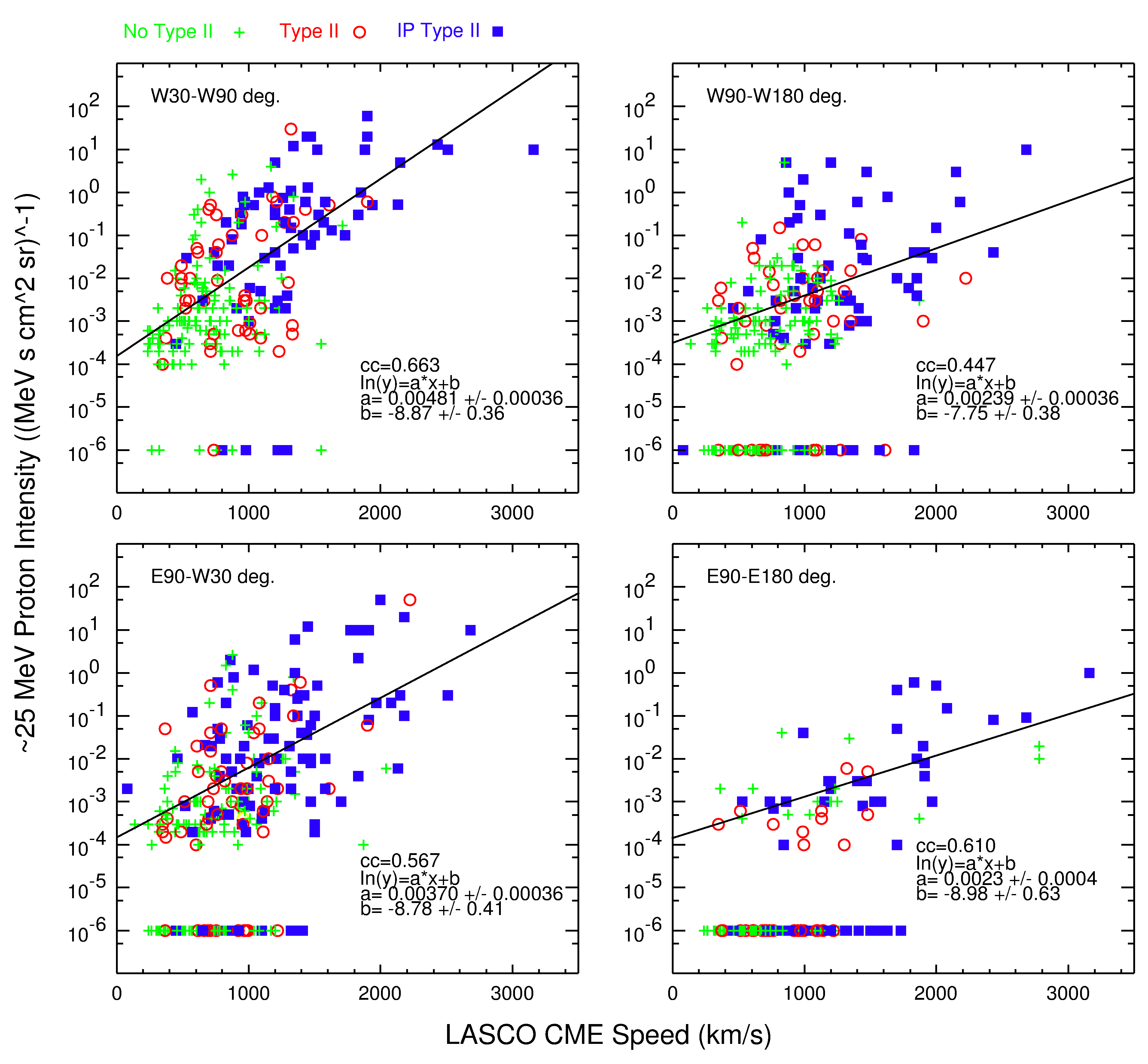}
    \caption{$\sim25$~MeV proton intensity vs. LASCO CME speed (from the CDAW catalog) for events in different longitude ranges relative to the observing spacecraft. The symbols indicate whether type II (red open circles), IP type II radio emission, extending below 1~MHz  (solid blue squares) or no emission (green crosses) observed by spacecraft is reported (events where this information is not available are excluded from the figure). Cases where no SEP event was detected at the observing spacecraft or there was a high background from previous events are shown at an intensity of $10^{-6}$. These are excluded from the least squares fits and correlations shown. }
    \label{fig:IvcmetypeII}
\end{figure}

Figure~\ref{fig:IvcmetypeII} summarizes the relationship between the LASCO CME speed and the $\sim25$~ MeV proton intensity in various longitude ranges relative to the observing spacecraft. The symbols indicate whether the WAVES or SWAVES (STEREO WAVES) instruments on the  Wind or STEREO spacecraft, respectively, detected interplanetary (IP) type II radio emission extending below 1~MHz \citep[e.g., ][solid blue squares]{canee2005}, type II emission that did not reach 1~MHz (red open circles), or no type II emission was detected (green crosses), as reported in the CDAW Type II catalog (\url{https://cdaw.gsfc.nasa.gov/CME_list/radio/waves_type2.html}) and indicated in Tables~1 to \ref{tab11}. Cases where no SEP event was detected at the observing spacecraft, or there was a high background from previous events, are shown at an intensity of $10^{-6}$. These are excluded from the least squares fits and correlations shown. The top left panel shows a reasonable correlation (correlation coefficient = 0.663) between proton intensity and CME speed for well-connected events (W30-W90$^\circ$), as reported in many previous studies \citep[e.g.,][]{kahler1978, reames1999,richardson2014}.  As also noted by \citet{richardson2014} for a smaller sample of events, SEP events without type II emissions tend to be weaker and associated with slower CMEs than those with type II emissions, with the larger events typically associated with IP type II emission. Similar patterns are evident in events originating in other longitude ranges though the correlations with CME speed become weaker.  The top right panel of Figure~\ref{fig:IvcmetypeII} shows results for behind-the-west-limb events (all locations are relative to the observing spacecraft), the bottom right for behind-the-east-limb events, and the bottom left, front side events located between the east limb and W30$^\circ$. Thus, at all longitudes, the SEP intensity shows some correlation with the CME speed.

   \begin{figure}
         \centering
         \includegraphics[width=0.6\linewidth]{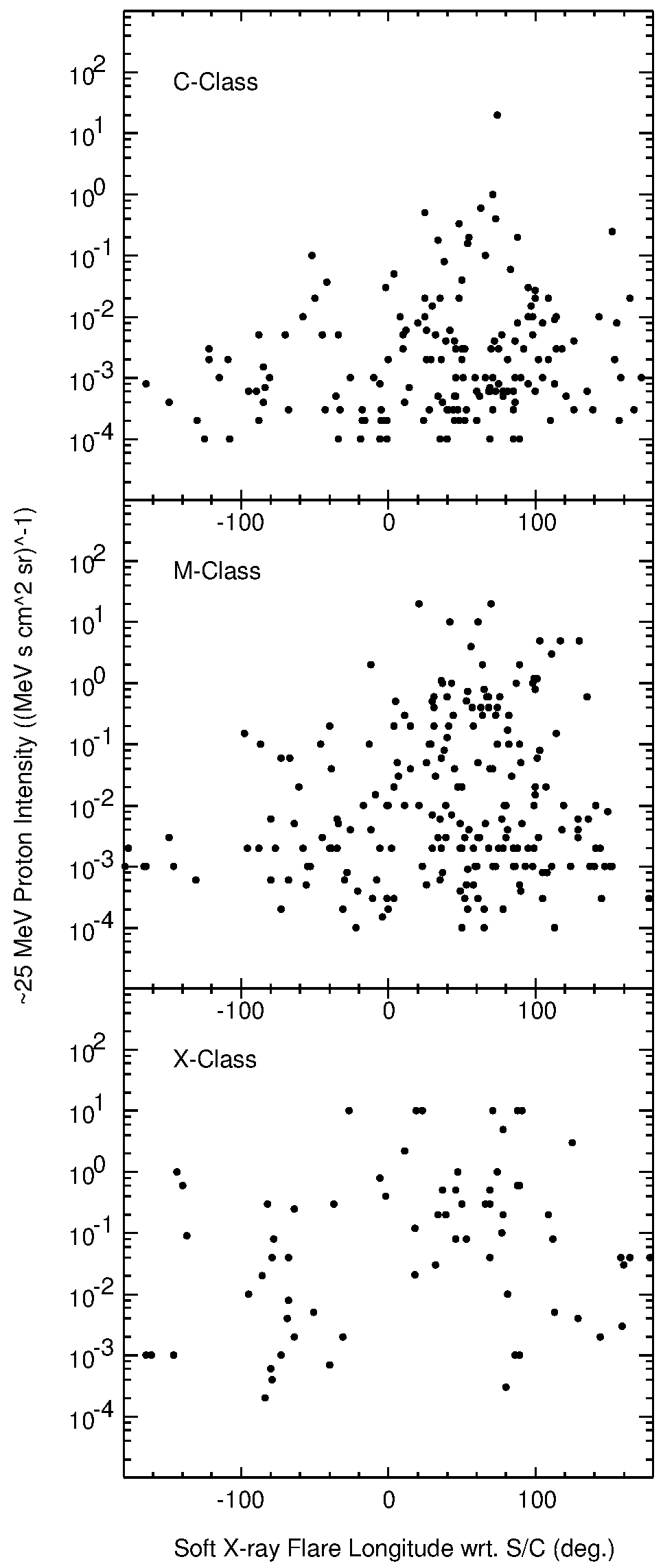}
         \caption{SEP proton intensities vs. the longitude of the associated GOES soft X-ray flare relative to the SEP-observing spacecraft for cases where a C, M, or X (and higher)-class flare is reported. Flare intensities for events after December 18, 2017 have been multiplied by 0.7 to make them consistent with earlier flare intensities. Twelve events associated with flares observed by SolO/STIX are also included based on the estimated equivalent GOES flare size. }
         \label{fig:IflareXray}
     \end{figure}  

Figure~\ref{fig:IflareXray} summarizes the $\sim25$~MeV proton intensity vs. the longitude of the associated soft X-ray flare relative to the SEP-observing spacecraft reported in Tables~1 to \ref{tab11} of Appendix~A. (The flare longitude here is to be distinguished from the solar event location relative to the observing spacecraft used in other figures which does not require the observation of an X-ray flare.) The X-ray flare observations are from GOES except for twelve cases from SolO/STIX. The GOES X-ray intensities have been corrected for instrumental changes and correspond to pre-GOES-R (pre-December 18, 2017) intensities, as discussed further in Appendix~A and indicated in the figure caption. The three panels show cases where SEP events are associated with C-class, M-class or X- (and higher) class flares.  Again, there is a clear western hemisphere bias in the number of proton events and the largest events associated with each class of flares.  For SEP events associated with western C and M-class flares, there is a wide spread in intensities, and overall, even C-class flares may be associated with SEP events that extend far from the flare location. Considering SEP events associated with X-class flares, the weak SEP events that are associated with smaller flares are generally absent, and the SEP events extend to distant longitudes. Note that these results require the detection of an SEP event, but as discussed below and in Section~\ref{S-stat}, there are significant numbers of X-ray flares that are not associated with detected SEP events.

\begin{figure}
    \centering
    \includegraphics[width=0.75\linewidth]{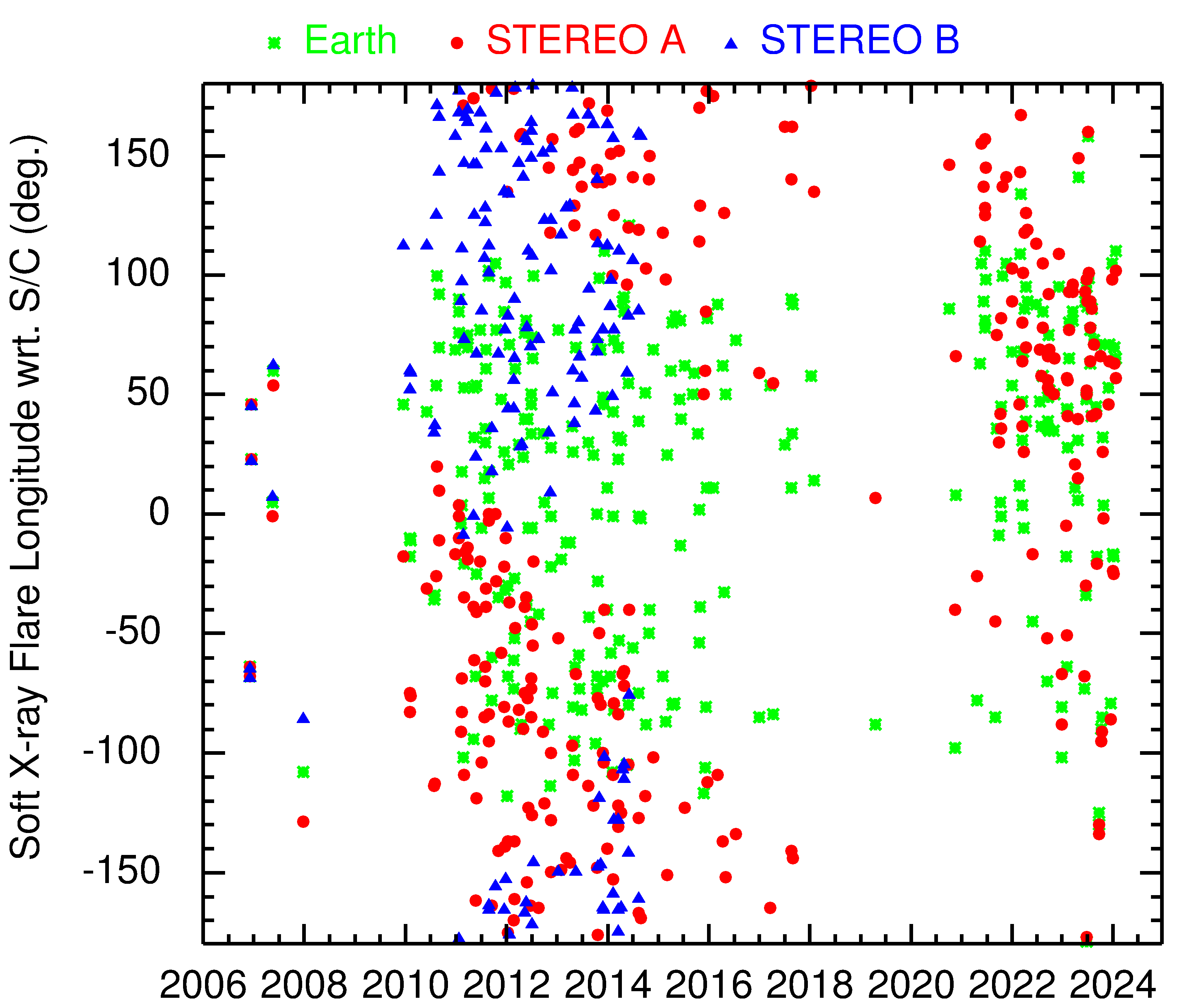}
    \caption{Locations of SEP-associated GOES soft X-ray flares in 2006-2023 relative to Earth (green squares), STEREO A (red circles) or STEREO B (blue triangles), illustrating the bias towards western events (positive longitudes) in this sample arising from the loss of contact with STEREO~B, the solar cycle variation in the occurrence of SEP-associated flares, and the unfavorable (eastern/far western) locations of these flares relative to STEREO~A in Cycle~24.  This bias is reflected in Figure~\ref{fig:IflareXray}. Twelve flares observed by SolO/STIX are also included in the figure. }
    \label{fig:sxrt}
\end{figure}

\begin{figure}
    \centering
    \includegraphics[width=0.75\linewidth]{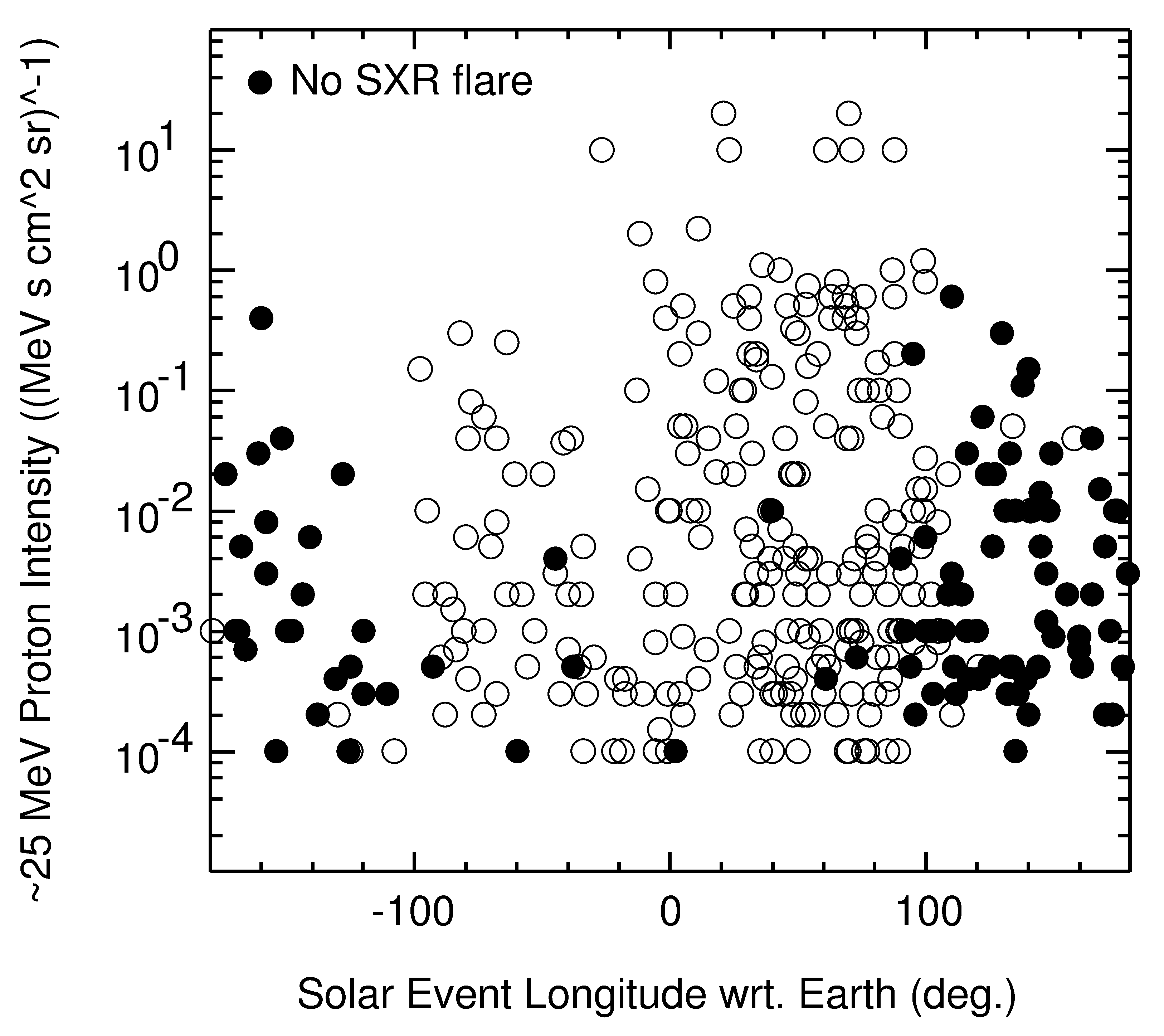}
    \caption{$\sim25$~MeV proton intensity observed at Earth vs. solar event longitude relative to Earth. Filled circles indicate the 27\% of these SEP events at Earth that are not associated with GOES soft X-ray (SXR) flares (typically these events originate on the far side) or, in a few cases, the association is uncertain. Several SEP events associated with far side X-ray flares observed by SolO/STIX are also shown. }
    \label{fig:noflare}
\end{figure}

Figure~\ref{fig:IflareXray} also includes biases due to uneven coverage in the X-ray flare longitude with respect to the observing spacecraft. Figure~\ref{fig:sxrt} shows the locations of the SEP-associated soft X-ray flares observed by GOES or SolO/STIX relative to Earth or the STEREO spacecraft in 2006 to 2023. (In the following discussion, we neglect the few SolO events.) The GOES flares are only observed when on the front side or just over the limbs relative to Earth (green squares in Figure~\ref{fig:sxrt}) and so lie predominantly in the longitude range +90 to -90$^\circ$. Considering STEREO A (red circles) moving ahead of Earth, most of the GOES flares in Solar Cycle 24 were progressively further to the east of the spacecraft (negative longitudes), then switched to behind the west limb ($>90^\circ$) towards the end of Solar Cycle 24 as STEREO~A passed behind the Sun and began to return towards Earth (cf., Figures~\ref{fig:3SC} and \ref{fig:stloc}). Therefore overall, STEREO~A was relatively poorly positioned to observe SEPs from X-ray flares observed at Earth in Solar Cycle~24. Then in Solar Cycle 25, the front side flares were predominantly western (longitude $>0^\circ$) relative to STEREO~A as the spacecraft approached Earth. Considering STEREO~B (blue triangles), lagging behind Earth, the SEP-associated GOES soft X-ray flares were predominantly western relative to the spacecraft during Solar Cycle 24, progressing to far behind the west or east limbs just before loss of contact.   
The overall result of these various factors is a deficiency of cases where the GOES X-ray flare is east of the spacecraft location, and this accentuates the western bias evident in Figure~\ref{fig:IflareXray} beyond that expected because of preferential magnetic connection to the western hemisphere.

Considering just the SEP events detected at Earth, Figure~\ref{fig:noflare} shows the proton intensity vs. solar event longitude for events associated with GOES soft X-ray flares (open circles) and with no clearly-associated flare (solid circles). (Again, several SEP events at Earth associated with far side flares observed by SolO/STIX are included in Figure~\ref{fig:noflare} but not considered in this discussion.)  This figure emphasizes that a significant fraction of the $\sim25$~MeV proton events detected at Earth are not associated with GOES soft X-ray flares, which has implications for SEP prediction schemes \citep[e.g.,][and references therein]{Whitman2023} requiring direct observations of the related solar activity either at X-ray or other wavelengths. For this sample of 331 SEP events detected at Earth, 91 (27\%) have no clearly-associated soft X-ray flare, with 19\% associated with solar events behind the west limb at longitudes $\ge90^\circ$ and 6.3\% associated with solar events behind the east limb (longitudes $<90^\circ$), consistent with the results of \citet{richardson2014} (see their Figure~11) and \citet{cane2010}. The few front side events with no associated flare are cases where the flare association is uncertain or the SEP event appears to be associated with an eruption (often of a filament) that is not accompanied by significant soft X-ray emission.   

\begin{figure}
    \centering
    \includegraphics[width=1\linewidth]{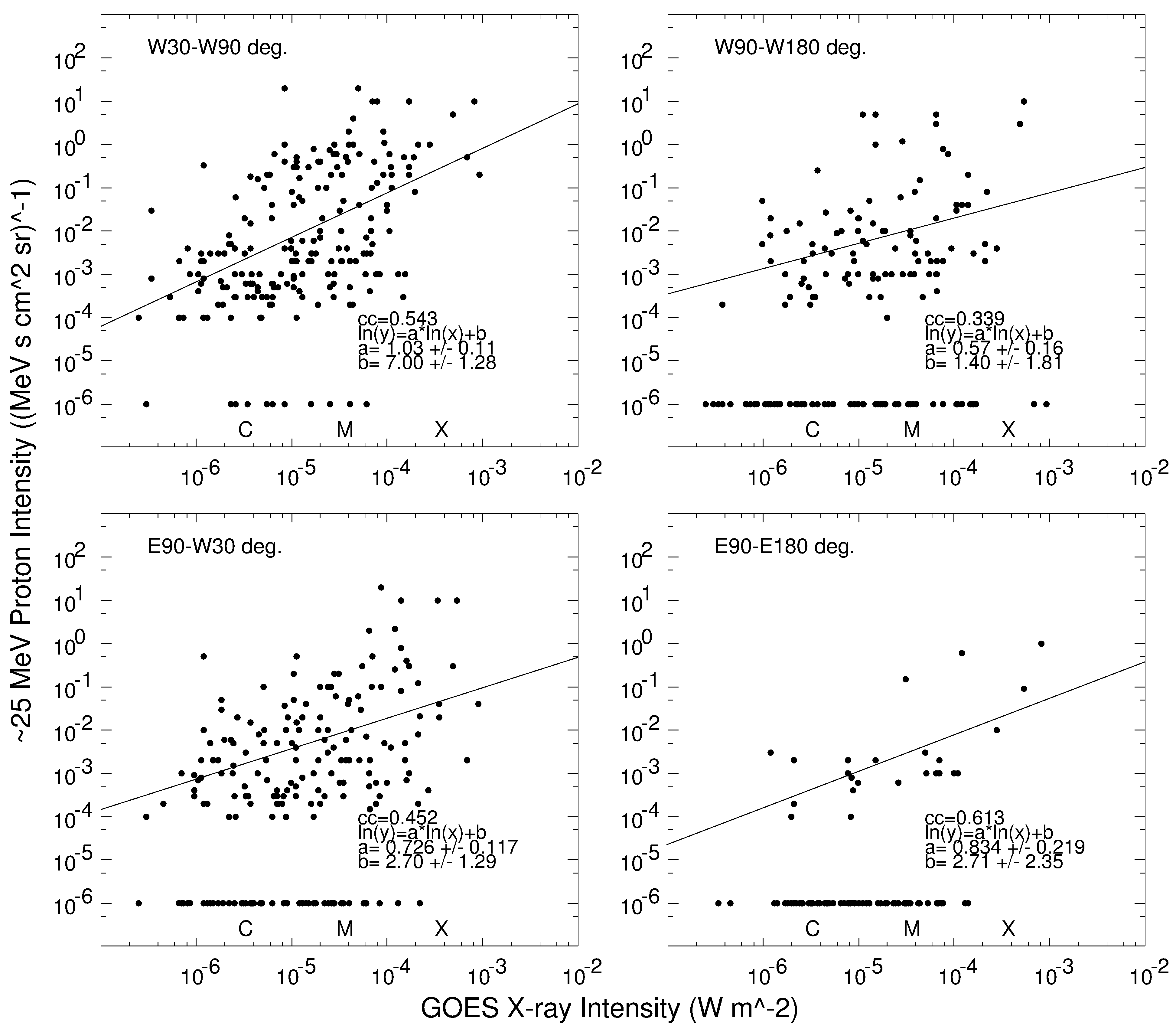}
    \caption{SEP intensity vs. GOES X-ray intensity for different ranges of solar event longitude with respect to the observing spacecraft. Cases when no SEP event was detected or there was a high background at a particular spacecraft location are shown at an intensity of $10^{-6}$ and are not included in the calculation of the least squares fits and correlations. X-ray intensities after December 18, 2017 have been multiplied by 0.7 to make them consistent with earlier event intensities (see Appendix~A for details). }
    \label{fig:ixraylon}
\end{figure}

Figure~\ref{fig:ixraylon} summarizes the $\sim25$~MeV proton intensity vs. the GOES soft X-ray flare intensity for the same solar event longitude ranges relative to the observing spacecraft as in Figure~\ref{fig:IvcmetypeII}. Again, the GOES X-ray intensities have been corrected for instrumental changes and correspond to pre-GOES-R (pre-December 18, 2017) intensities, as discussed further in Appendix~A and indicated in the figure caption.  Note that this figure includes SEP observations from the STEREO spacecraft and hence the X-ray flares are located over a more extended longitude range relative to the observing spacecraft than for the near-Earth observations in Figure~\ref{fig:noflare}, as previously discussed in relation to Figure~\ref{fig:sxrt}. For well-connected events at W30-90$^\circ$ in the top left panel, there is a modest (cc=0.543) correlation between the SEP and flare intensities, as has been noted in some other studies \citep[e.g.,][]{richardson2017}. (Again, the correlation does not include cases where no SEP event was detected, or there was a high background, at a particular spacecraft, shown at an intensity of $10^{-6}$.) While the upper envelope of the distribution indicates a clear correlation between the X-ray flare and SEP event size, there are also weaker SEP events for a given flare intensity, including relatively weak events that are associated with high M and low X-class flares.  Correlations between the X-ray and SEP intensities are also present for other longitude ranges, though these are weaker except for the small number of cases when the flare is behind the east limb relative to the observing spacecraft (bottom right panel of Figure~\ref{fig:ixraylon}). Thus, even when the flare is well behind the east limb of the Sun relative to the spacecraft, the SEP intensity may show some dependence on the X-ray flare intensity, as is also evident in Figure~\ref{fig:IflareXray}. 

Correlations such as those in Figures~\ref{fig:IvcmetypeII} and \ref{fig:ixraylon} might be used to argue for the CME- or flare-acceleration of SEPs, respectively. For this group of $\sim25$~MeV proton events, the correlation coefficients are generally larger in each longitude range for the SEP intensity vs CME speed in Figure~\ref{fig:IvcmetypeII}. This is to be  expected since, as already noted, relatively weak SEP events may be associated with relatively strong flares, but not with especially fast CMEs.  On the other hand, the correlations are not comparable. In particular, the SEP intensity-X-ray correlations are double-logarithmic and cover a larger dynamic range.  Therefore, a simple comparison with the values of the log-linear intensity-CME speed correlation coefficients may not provide meaningful insight into whether the observations favor SEP acceleration by either flares or CMEs. We also note that \citet{lario2014} reported ``triangular" proton intensity vs. CME speed distributions when plotted using a double-logarithmic scale and GOES proton data, similar to those found for the X-ray intensity in Figure~\ref{fig:ixraylon}, and emphasized the positive correlations between SEP intensity and the CME speed (and also the X-ray intensity) along the upper envelope of the distributions. They also suggested that the SEP intensities are influenced by solar wind structures, including at the observing spacecraft, an aspect that we do not consider here.      

\begin{figure}
    \centering
    \includegraphics[width=1.0\linewidth]{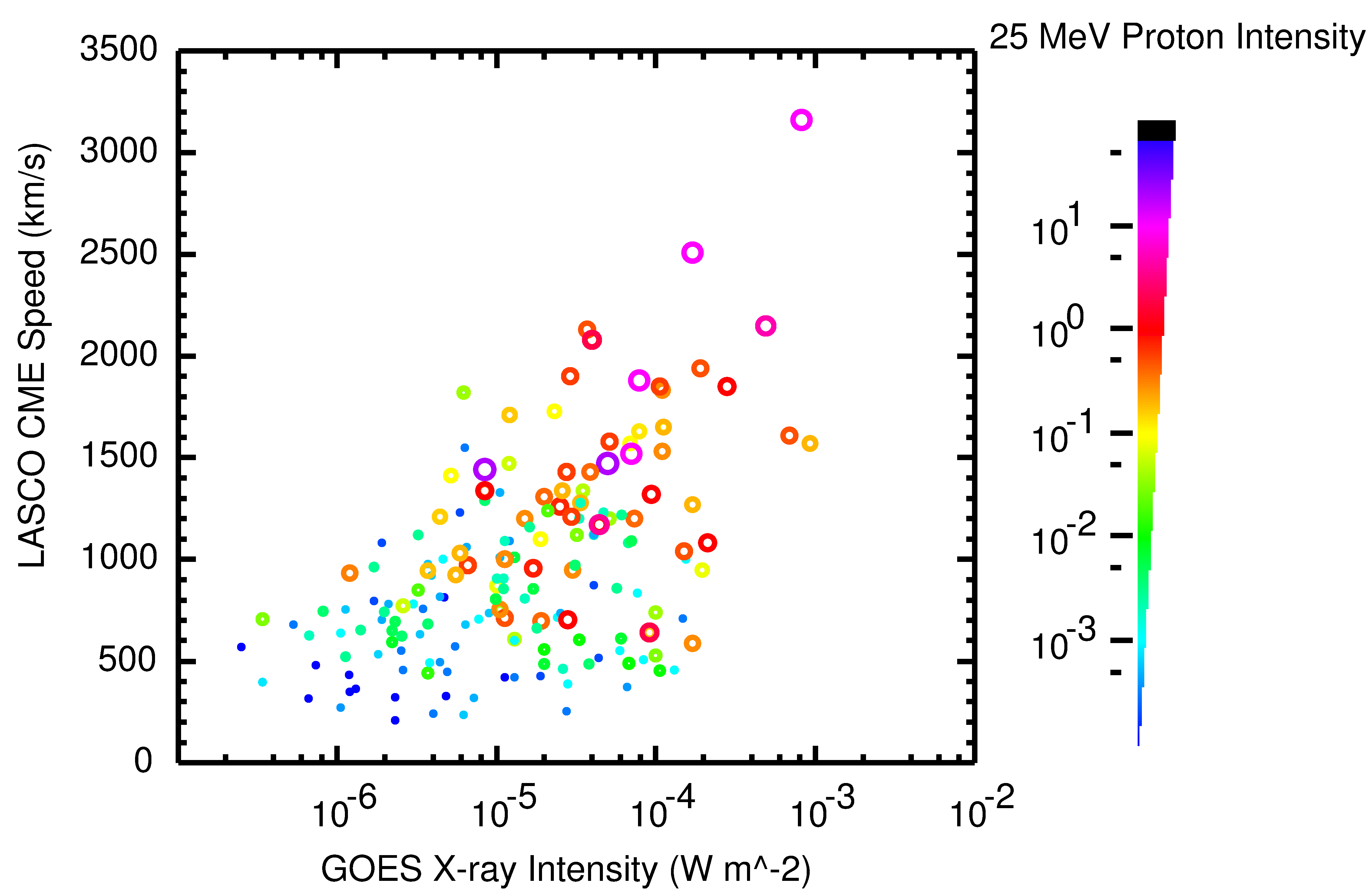}
    \caption{LASCO CME speed vs. GOES X-ray intensity for 200 SEP-associated solar events at W30-90$^\circ$ relative to the observing spacecraft.  The symbol size and color indicates the intensity of $\sim25$~MeV protons. X-ray intensities after December 18, 2017 have been multiplied by 0.7 to be consistent with earlier event intensities.}
    \label{fig:cmexrayint}
\end{figure}

\begin{figure}
    \centering
    \includegraphics[width=0.85\linewidth]{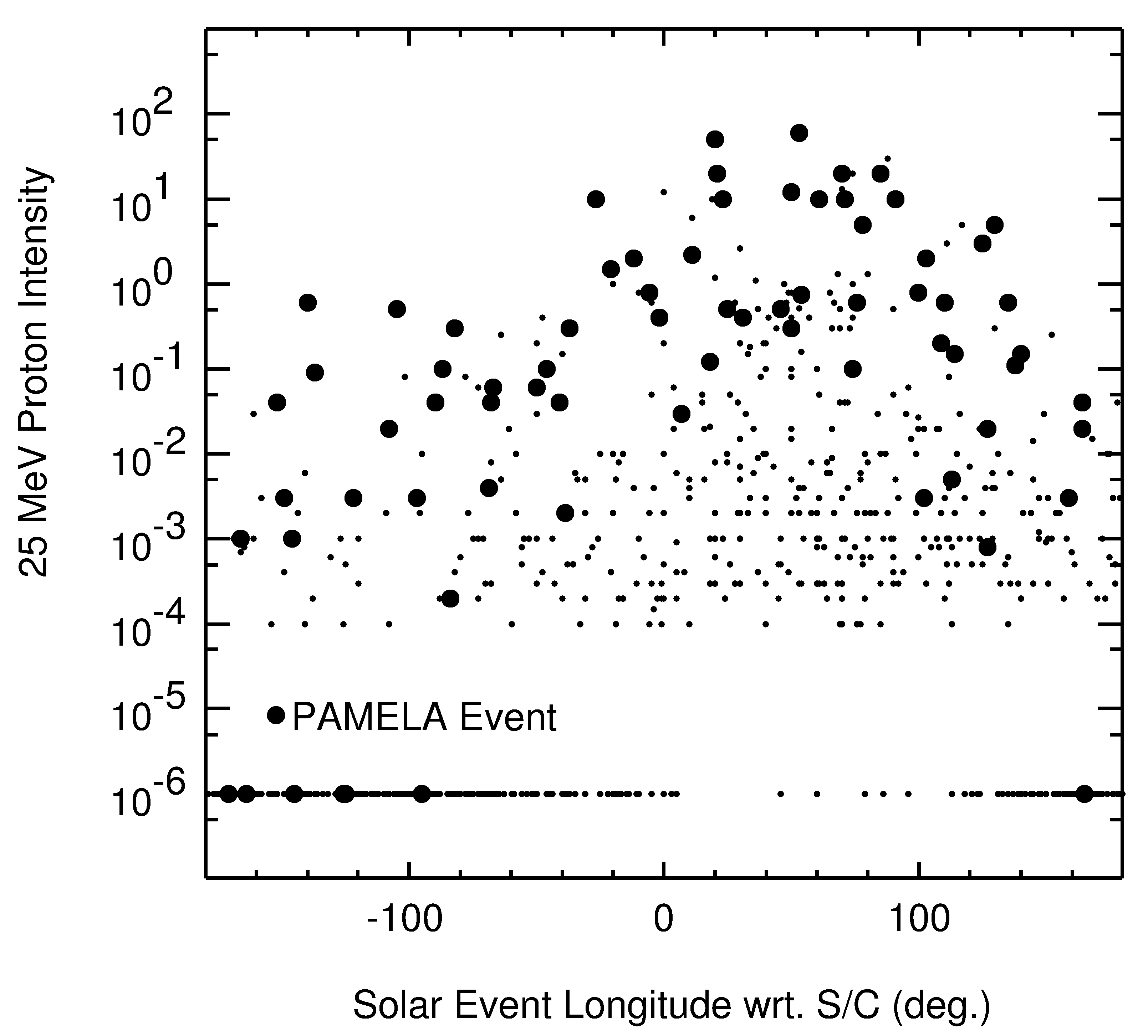}
    \caption{25 MeV proton intensity ((MeV s cm$^2$ sr)$^{-1}$) vs. solar event longitude relative to the observing spacecraft in December 2006-September 2014 with observations related to the high energy events detected by PAMELA at 80~MeV-few GeV during the same period \citep{bruno2018} indicated by large circles. Observations without a detectable 25~MeV proton event are plotted at an intensity of $10^{-6}$. Where such a data point is indicated as a PAMELA event, this means that, although the SEP event was observed by PAMELA, it was not detected at 25~MeV by a spacecraft widely separated from the related solar event.    }
    \label{fig:pamela}
\end{figure}

Figure~\ref{fig:cmexrayint} summarizes the relationship between the combination of X-ray flare intensity and CME speed, and the 25 MeV proton intensity, for 200 events with well-connected solar flares at W30-90$^\circ$ relative to the observing spacecraft.  The symbol size and color indicate the proton intensity. Consistent with the previous discussion, the larger SEP events (intensities above $\sim10^{-1}$) tend to be associated with CME speeds above $\sim1000$~km/s and M class and larger X-ray flares. There is also clearly an overall correlation between CME speed and X-ray intensity, as is well established \citep[e.g.,][]{vrsnak2005}.

\citet{bruno2018} reported the spectra of 30 high-energy SEP events in December 2006-September 2014 detected by the Payload
for Antimatter Matter Exploration and Light-nuclei Astrophysics (PAMELA) instrument on the Resurs-DK1 satellite at energies of 80~MeV – few GeV.  Figure~\ref{fig:pamela} shows, for SEP events during this period, the $\sim25$~MeV proton intensity observed by STEREO A, B, or at Earth vs. the longitude of the related solar event relative to the spacecraft. The large filled circles indicate $\sim25$~MeV proton observations made by STEREO A, B and SOHO of the PAMELA events of \citet{bruno2018}. At $\sim25$~MeV, the SEP events that were observed by PAMELA at higher energies generally extended to all longitudes and were among the highest intensity events observed at any longitude, although in a few cases, the PAMELA events were not detected at 25~MeV by distant spacecraft.  This is consistent with the conclusion of \citet{bruno2018} that SEP events observed at the highest energies, including ground level enhancements observed by neutron monitors, are just the most intense events in a population of SEPs with similar spectra and are not a separate/special class of events.  There are also other events that have intensities at $\sim25$~MeV that are comparable to those of the PAMELA events, suggesting that their spectra were softer and did not extend into the PAMELA energy range.  

\begin{figure}
    \centering
    \includegraphics[width=0.85\linewidth]{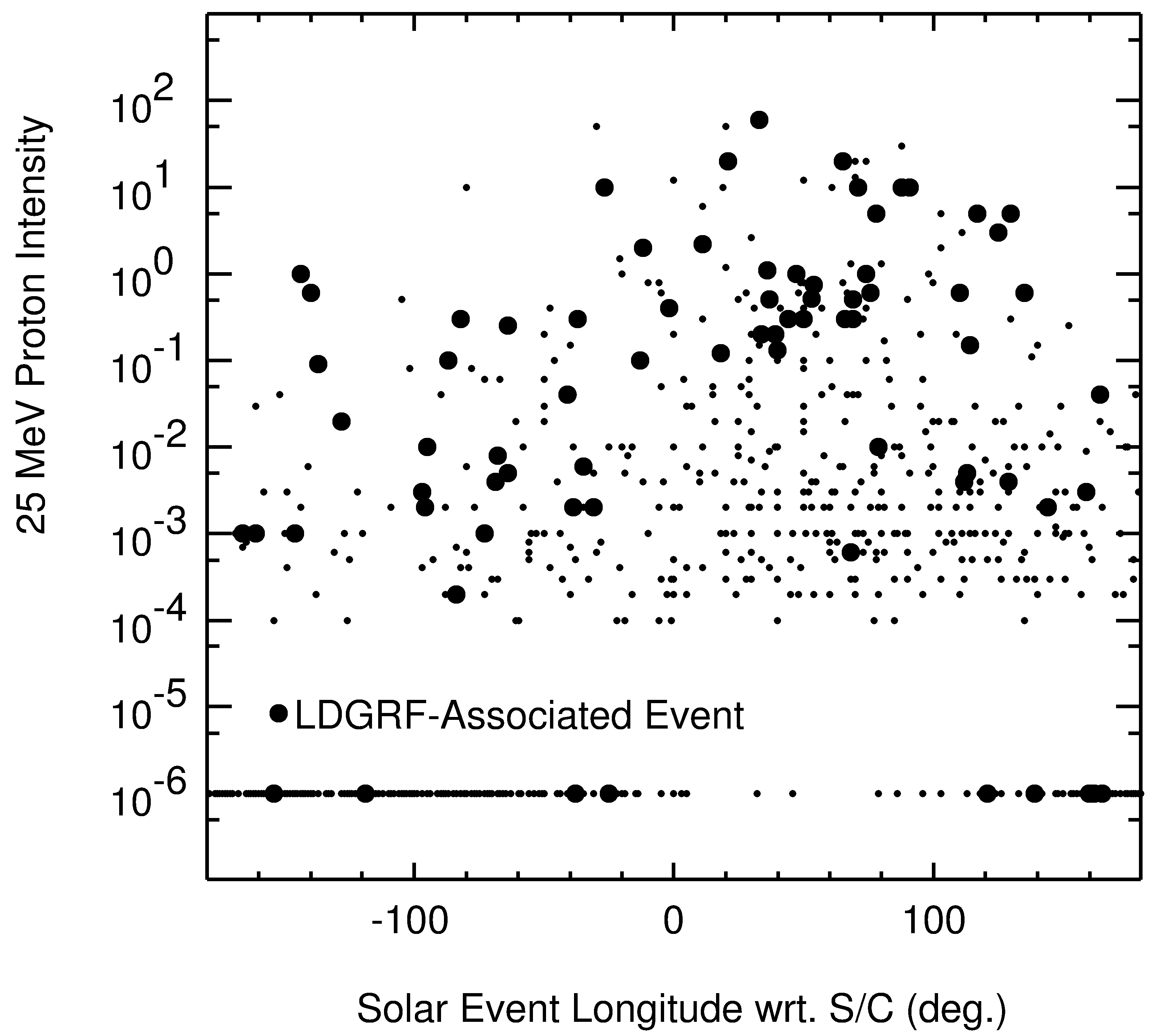}
    \caption{25 MeV proton intensity ((MeV s cm$^2$ sr)$^{-1}$) vs. solar event longitude relative to the observing spacecraft in March 2011-September 2017. Large circles indicate observations of the events associated with the long duration gamma ray flares (LDGRFs) observed by FERMI/LAT during this period summarized in Table~3 of \citet{bruno2023}. Observations without a detectable SEP event are plotted at an intensity of $10^{-6}$. }
    \label{fig:LDGRF}
\end{figure}

Figure~\ref{fig:LDGRF} indicates, in a similar format, those SEP events that are related to the long duration gamma ray flares (LDGRFs) observed by the Fermi/LAT (Large Area Telecope) and compiled from several studies in Table~3 of \citet{bruno2023}. The SEP events shown cover the period of March 2011-September 2017 that encompasses these LDGRF observations.  Again, the LDGRF events tend to be associated with larger SEP events at $\sim25$~MeV, but there are exceptions. There are also many comparable SEP events that are not associated with reported LDGRFs. Given the lower proton energy that we are considering, it is unclear whether these observations directly contribute to the debate as to whether LDGRFs are caused by high energy ($>300$~MeV) protons backstreaming onto the photosphere from CME-driven shocks (\citet{bruno2023} present arguments against this scenario) or by processes in the corona such as the acceleration of particles trapped in large coronal loops \citep[e.g.,][]{Ryan1991}. Recently, \cite{bruno2025} have discussed an LDGRF on July~16, 2024 associated with a slow CME and weak SEP event extending only to a few tens of MeV where such coronal loops were present, that might favor the latter scenario.

\section{Association of M/X Flares With SEP Events}
\label{S-stat}
The results summarized above consider cases where an SEP event was detected at least at one location (the STEREO spacecraft or near Earth) following a solar event typically associated with a CME and evidence of flaring. However, there are many flares and CMEs that are not accompanied by observed SEP events, even when observations are made at multiple locations. For example, \citet{richardson2018} noted that 85\% of the 334 CMEs in the DONKI database during October 2011-July 2012, near the peak of Solar Cycle~24, were not associated with a $\sim25$~MeV proton event observed at the widely-spaced STEREO or near-Earth spacecraft.   This low SEP association rate poses challenges for SEP prediction schemes based on CME observations, as discussed by \citet{Whitman2023} and also by \citet{richardson2018}.  They demonstrated how various filtering schemes (for example, making SEP predictions only for CMEs exceeding a specified speed threshold) may be applied to reduce the ``false prediction" rate for the CME-triggered SEPSTER model. However, \citet{lario2020} noted that even ``fast and wide" CMEs occasionally may not be associated with SEP events, and considered the factors that might lead to the absence of SEPs. An even larger imbalance with SEP events exists for solar flares. Here, we start with a large sample of soft X-ray flares and use a subset of the SEP events identified in Tables~1 to \ref{tab11} in Appendix~A to investigate the relationship between SEP events, eruptive solar events, flares and CMEs, specifically including solar events that are not associated with SEP events.

\begin{figure}
    \centering
    \includegraphics[width=0.81\linewidth]{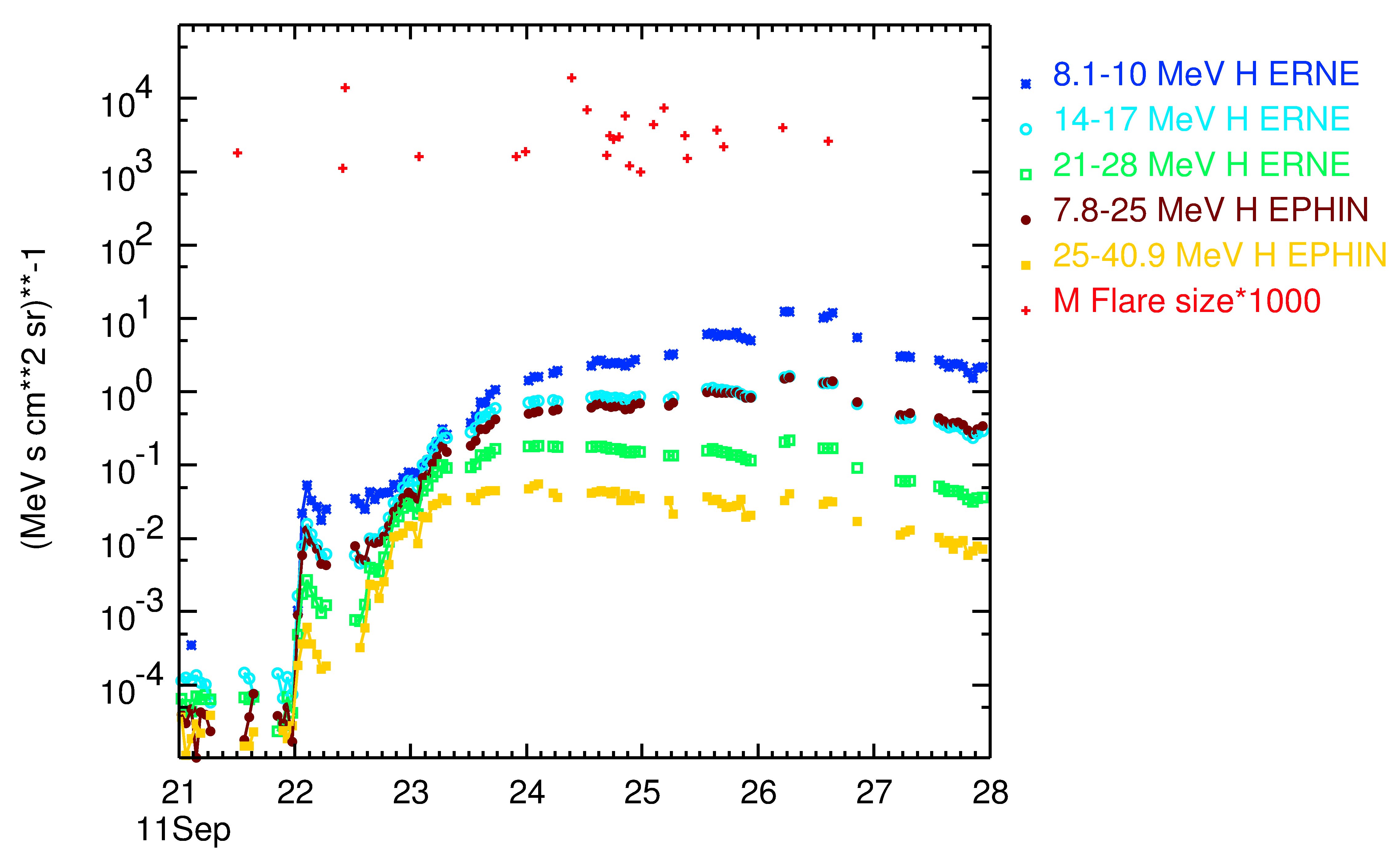}
    \includegraphics[width=0.9\linewidth]{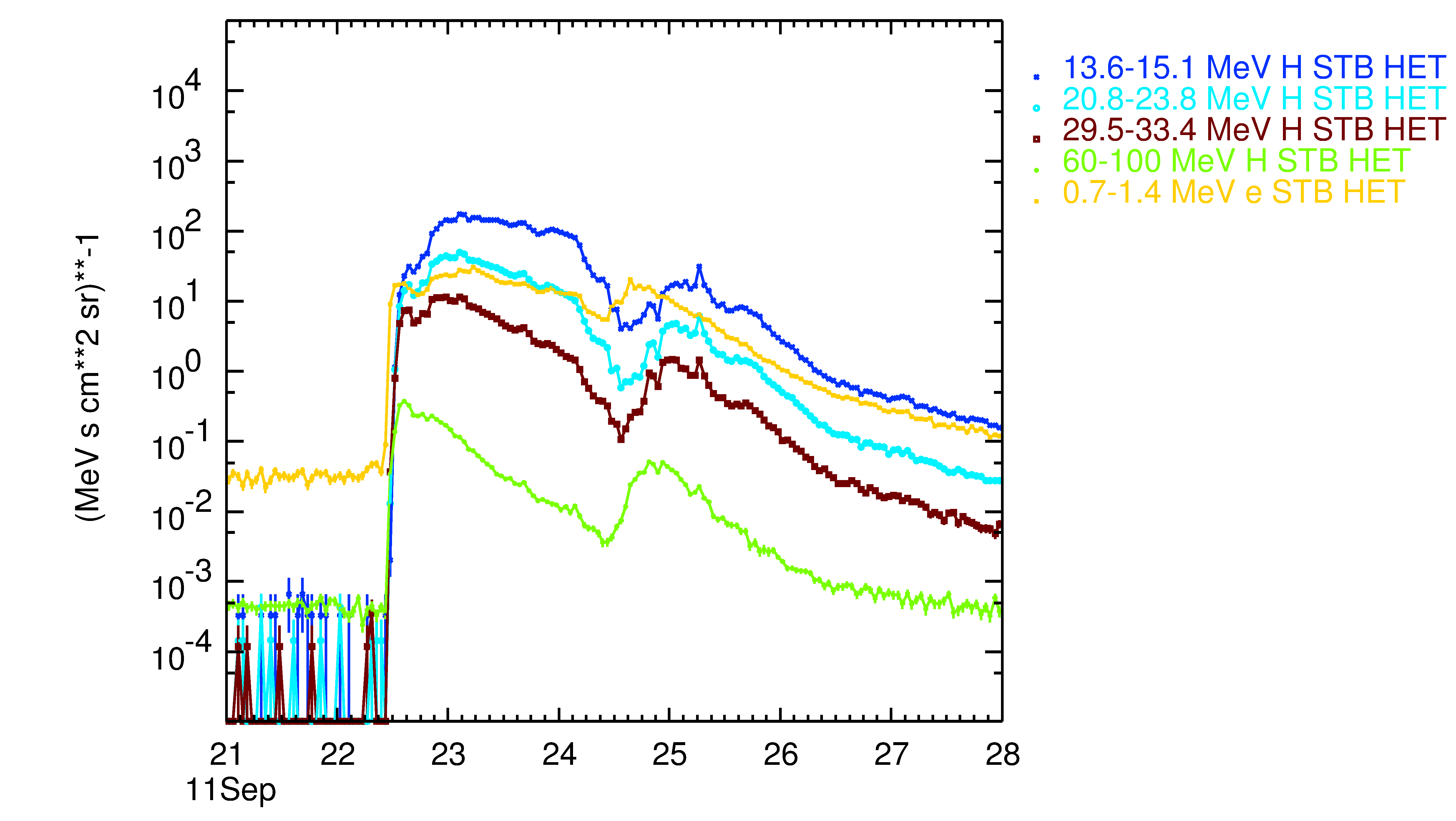}
     \includegraphics[width=0.6\linewidth]{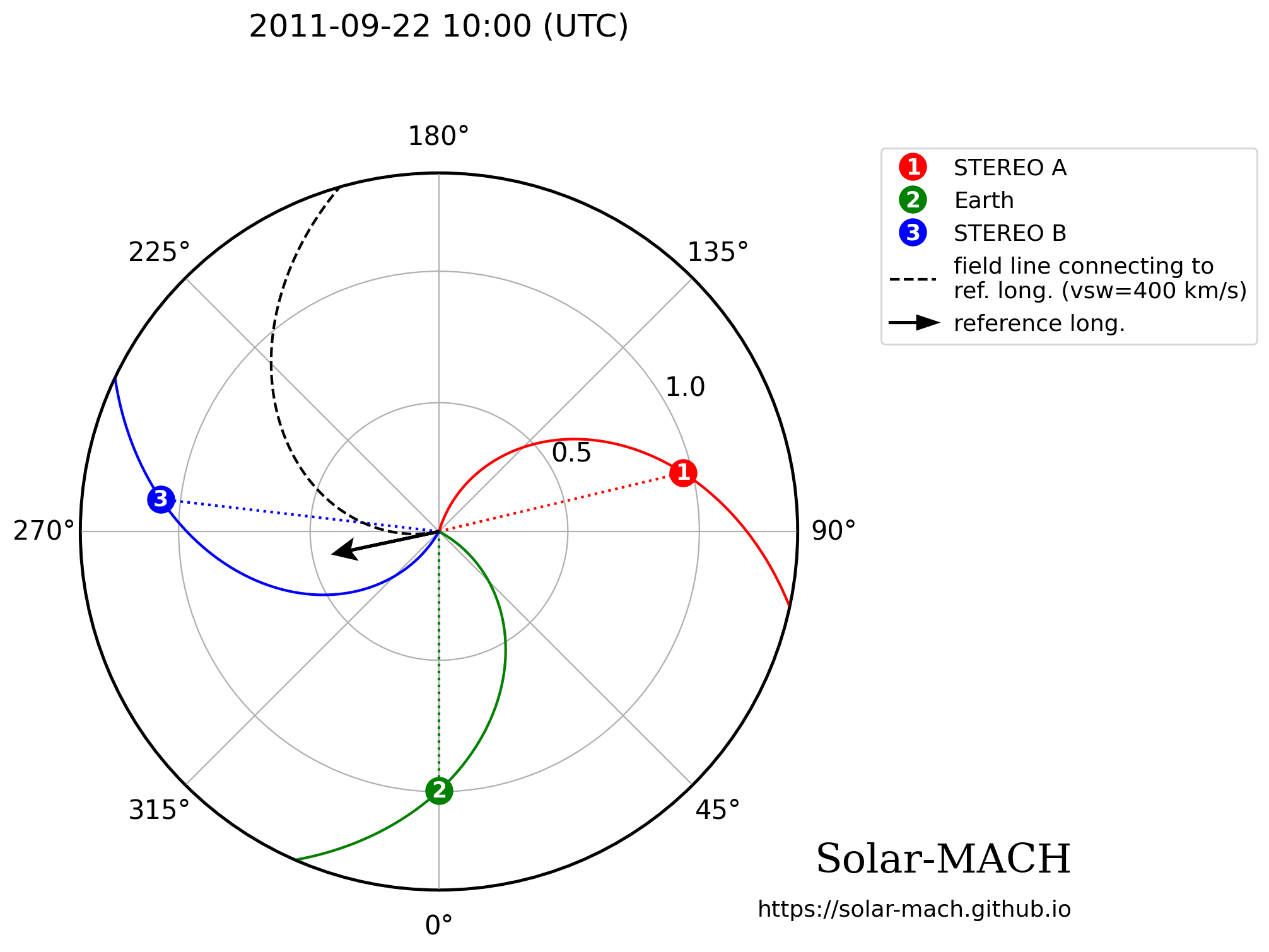}
    \caption{Top: An extended-duration widespread SEP event commencing on September 22, 2011 at $\sim10$~UT (proton data are from SOHO ERNE and EPHIN), associated with an X1.4 flare at E78$^\circ$. During the SEP event, a cluster of 20 M/X class flares was observed; the flare intensities (red crosses) are plotted at 1000 times the M intensity. A small preceding SEP event commencing at $\sim22$~UT on September 21, is also evident, originating from a source behind the west limb at W109$^\circ$, and hence with no associated flare.  Center: STEREO B HET proton and electron observations for the same period, showing the prompt onset of the September~22 SEP event and a second SEP event associated with an X1.9 flare on September 24, at E60$^\circ$ relative to Earth and W37$^\circ$ relative to STEREO~B. Note that at STEREO~B, the profile of the first event at lower energies is also modulated by an ICME-driven shock at 09:06~UT on this day (\url{https://stereo-ssc.nascom.nasa.gov/data/ins_data/impact/level3/STEREO_Level3_Shock.pdf}). Bottom: The spacecraft configuration at the time of the flare on September 22 at the location indicated by the black arrow. }
    \label{fig:flaresep}
\end{figure}

To examine the relationship between flares, CMEs and $\sim25$~MeV SEP proton events, we have considered all the 397 X or M class soft X-ray flares observed by the GOES spacecraft from June 2010 (shortly after the launch of the Solar Dynamics Observatory \citep[SDO,][]{pesnell2012}) to January 2014, encompassing much of the peak of Solar Cycle 24 (Figure~\ref{fig:3SC}). During this interval, STEREO~A(B) moved from $\sim70^\circ$ to $\sim150^\circ$ west (east) of Earth.  These flares have also formed the basis of a series of papers on the ``global energetics" of solar flares \citep{Aschwanden2014, Aschwanden2015, Aschwanden2016, Aschwanden2016a,Aschwanden2017}. In \citet{Aschwanden2017}, the total energy in SEPs (1.3-68 $\times 10^{30}$~erg) was estimated for ten solar flares where the associated SEPs were observed at both STEREO spacecraft and near the Earth, and for which the associated particles could be isolated from those related to other solar events. The total SEP energy was then estimated using the SEP spectra at each location and assumptions about the spatial distribution of the SEPs. The limited number of energy estimates, even when starting from such a large sample of X-ray flares, reflects the challenges in identifying multi-spacecraft SEP events suitable for this analysis and, as will be discussed below, the paucity of flares associated with detectable SEP events. As a first step in the event identification for \citet{Aschwanden2017}, a survey of the association of SEP events with these flares was made, and forms the basis of the following discussion.  As is evident from Tables~1 to \ref{tab11} in Appendix~A and Figure~\ref{fig:IflareXray}, SEP events can be associated with $<$M class flares, but just considering X/M class flares gives a more manageable number of events and focuses on the stronger flares. The interval considered ends before the loss of contact with STEREO~B.

One frequent challenge when associating X-ray flares and SEP events is that an unrelated SEP event may already be in progress when a flare occurs.  In the absence of any signature in the particle intensity-time profile, it may be uncertain whether the flare was not associated with an SEP event, or whether an SEP event was present but obscured by the ongoing event. This can be a significant problem since multiple flares may occur during the duration of a large SEP event which may extend for several days, and flares also tend to occur in clusters at times of high solar activity.  Such a situation is shown in the top panel of Figure~\ref{fig:flaresep}, where around 20 M/X flares (indicated by the red crosses near the top of the panel) occurred during the first $\sim4$~days of the SEP event observed by SOHO ERNE and EPHIN commencing at $\sim10$~UT on September 22, 2011 associated with an X1.4 flare at E78$^\circ$. (This was preceded by a small SEP event late on the previous day from behind the west limb with no associated X-ray flare; see Table~\ref{tab2}.) Fourteen of these flares were from Active Region 1302, also the origin of the X1.4 flare and SEP event. Thus, around 5\% of our sample of 397 X/M flares occurred just in the period in Figure~\ref{fig:flaresep}. We are unable to determine whether any other flare was associated with an SEP event based on observations at Earth. However, the center panel of Figure~\ref{fig:flaresep} shows STEREO~B HET observations during the same period (the bottom panel shows the spacecraft configuration). As noted in Table~\ref{tab2} and evident in this figure, an additional SEP event associated with an X1.9 flare on September 24 at E60$^\circ$ was observed by STEREO~B; the flare was at W37$^\circ$ relative to STEREO~B above the east limb. This period illustrates how multi-spacecraft observations can help to determine whether a flare was associated with an SEP event, although SEP events that are narrow in longitude might still be overlooked. As discussed above, the requirement here for a GOES flare to be observed tends to bias the detection of SEP events in Solar Cycle~24 towards Earth and STEREO~B, which was favorably placed to detect SEPs associated with eastern hemisphere GOES flares. On the other hand, STEREO~A was poorly connected to front side flares and only detected the more widespread SEP events such as the September~22, 2011 event (STEREO~A observations for this event are not shown in Figure~\ref{fig:flaresep}, but see Table~2).   

  Considering SEP observations at Earth and both STEREO spacecraft for the 397 M/X flares, these together provide potentially 1191 (i.e., $3\times397$) {\it independent} SEP observations.  Of these observations, in 602 cases (51\%), we could not determine whether an SEP event was detected because of an ongoing SEP event at the observing spacecraft or other factors, such as a data gap. For 462 cases (39\%), SEP observations are available but there is no evidence of a $\sim$25 MeV proton event above the prevailing low background. For the remaining 127 cases (11\%), an SEP event was observed that was clearly associated with the X-ray flare.  Considering just these last two groups with suitable SEP observations, a $\sim25$~MeV proton event was only detected in  21.6\% (127/589) of the observations.  Thus, even when considering SEP events above the low instrumental threshold and relatively large M/X flares, an SEP ($\sim25$~MeV proton) event was detected in only around a quarter of STEREO or near-Earth observations without ongoing events or data gaps. 

Considering just SEP observations at Earth for the 397 M/X flares, in 235 (59\%) of cases, an ongoing event or data gap was present at the time of the X-ray flare. For another 107 flares (27\%), SEP observations are available but no SEP ($\sim25$~MeV proton) event is evident, while an SEP event was detected for 55 (14\%) of the flares.  Again, combining the last two groups, an SEP ($\sim25$~MeV proton) event was observed at Earth in 55 out of 162 (34\%) cases that were not affected by a high background or data gaps.   Since a front side flare is required here, this helps to explain the higher rate of SEP detection (34\% vs. 22\%) when considering just near-Earth SEP observations than if observations from the STEREO spacecraft, well-separated from Earth, are also included (see Figure~\ref{fig:sxrt}).  As a comparison, the NOAA SEP list based on GOES data includes 20 $>10$~MeV proton events exceeding 10~pfu that are associated with M/X flares during the same period, around one third of the 55 $\sim25$~MeV proton events at Earth identified here.

\begin{figure}
    \centering
    \includegraphics[width=1\linewidth]{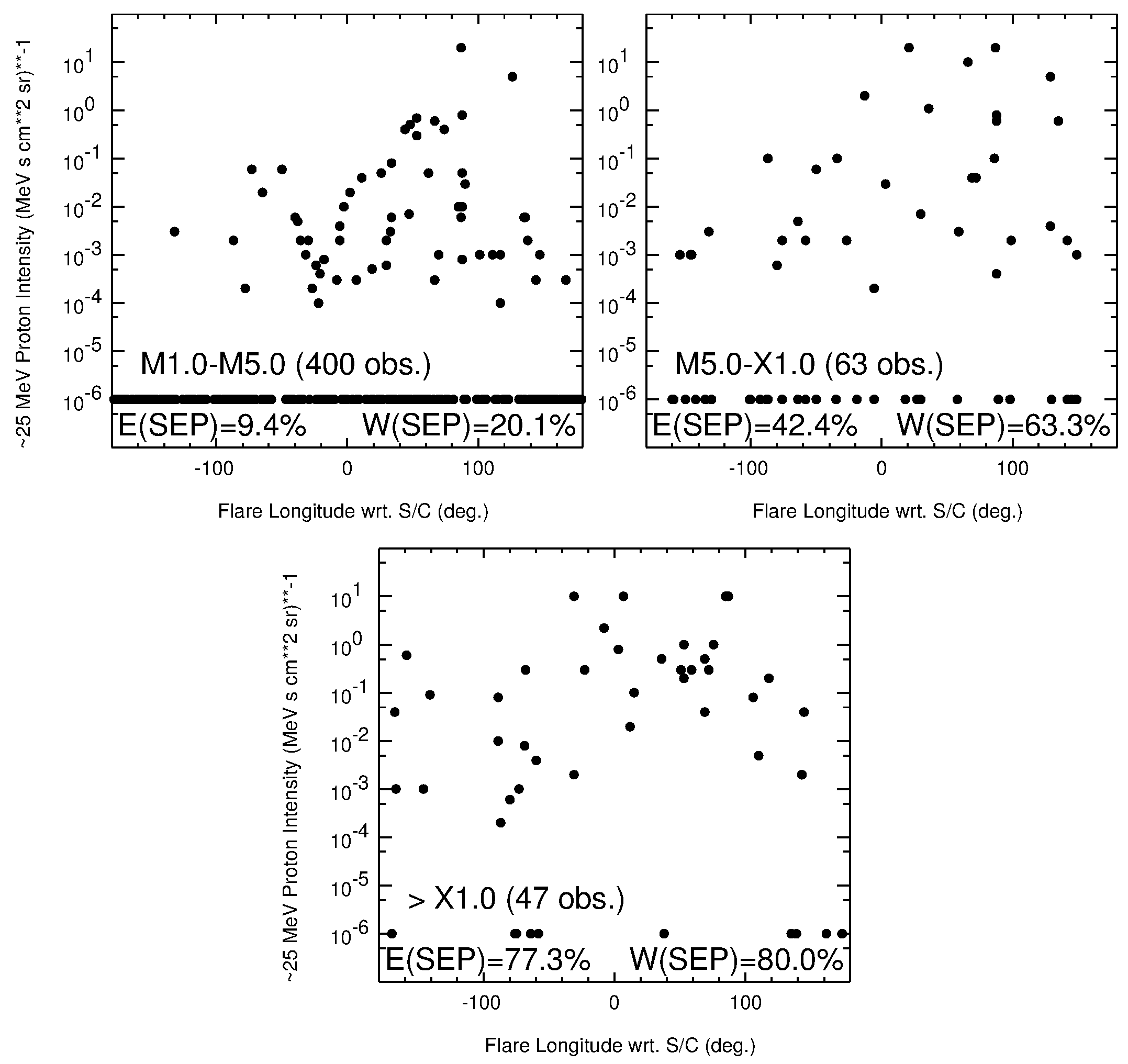}
    \caption{SEP $\sim25$~MeV proton intensity vs. flare longitude with respect to the observing spacecraft for (top left) M1.0-M5.0, (top right) M5.0-X1.0, and bottom $>$X1.0 flares between June 2010 to January 2014 with suitable SEP observations (e.g., no ongoing events or data gaps) at the STEREO spacecraft or at Earth (502 observations in total). When there is no SEP event detected at the observing spacecraft, an intensity of $10^{-6}$ is plotted.  The probability of an SEP event being detected (indicated for observations when the flare is east (E(SEP)) or west (W(SEP)) of the observing spacecraft) and over a wider longitude range, clearly increases with flare size.}
    \label{fig:flarelongint}
\end{figure}

In Figure~\ref{fig:flarelongint}, we have combined observations from the STEREO~A/B and SOHO spacecraft to compare the dependence of the 25~MeV proton intensity on the flare size and longitude with respect to the observing spacecraft.  Thus, a flare may be plotted up to three times if suitable SEP observations (i.e., without data gaps or large ongoing events) are available at the different spacecraft locations. In total, 502 SEP observations without data gaps or ongoing SEP events are available, considering measurements from the STEREO and near-Earth spacecraft.  Cases where no SEP event is observed at a particular spacecraft are plotted at an intensity of $10^{-6}$. 

In the top left panel, of the 400 SEP measurements associated with M1.0-5.0 flares, an SEP event was observed in only 59 cases (14.8\%).  There is a clear western (longitude $>0^\circ$) bias in the number of events and their intensity, indicating the strong influence of the spiral interplanetary magnetic field on the detection of these events. (Since observations from both STEREO spacecraft are available, the western bias introduced in the larger event sample due to the loss of contact with STEREO~B (cf., Figure~\ref{fig:sxrt}) is not a factor here.)   Of the SEP measurements made when the M1.0-5.0 flares were to the west of the observing spacecraft, SEP events were detected in 20.1\% of cases, compared with 9.4\% of the SEP measurements made when the flare was east of the spacecraft.  For M5.0-X1.0 flares (top right panel), SEP events were detected in  33 out of 63 (52.4\%) of the SEP measurements, again with a clear western bias in the largest events -- SEP events were detected in 63.3\% of the measurements made when M5.0-X1.0 flares were to the west of the observing spacecraft and 42.4\% of the cases when the flare was to the east. For $>$X1.0 flares (bottom panel), of 47~SEP observations, an SEP event was present in a majority of cases (37, 78.7\%). The SEP event detections were widely distributed in longitude relative to the flare and tend to be relatively intense compared to those associated with weaker flares. SEP events were detected in 80\% of cases where the $>$X1.0 flare was west of the spacecraft and 77.3\% of the cases when the flare was to the east.  Thus, the probability of detecting a $\sim25$~MeV proton event following an X-class flare is around 80\% for both eastern and western flares, and these SEP events are likely to be detectable at all longitudes relative to the flare.

\begin{figure}
    \centering
    \includegraphics[width=0.7\linewidth]{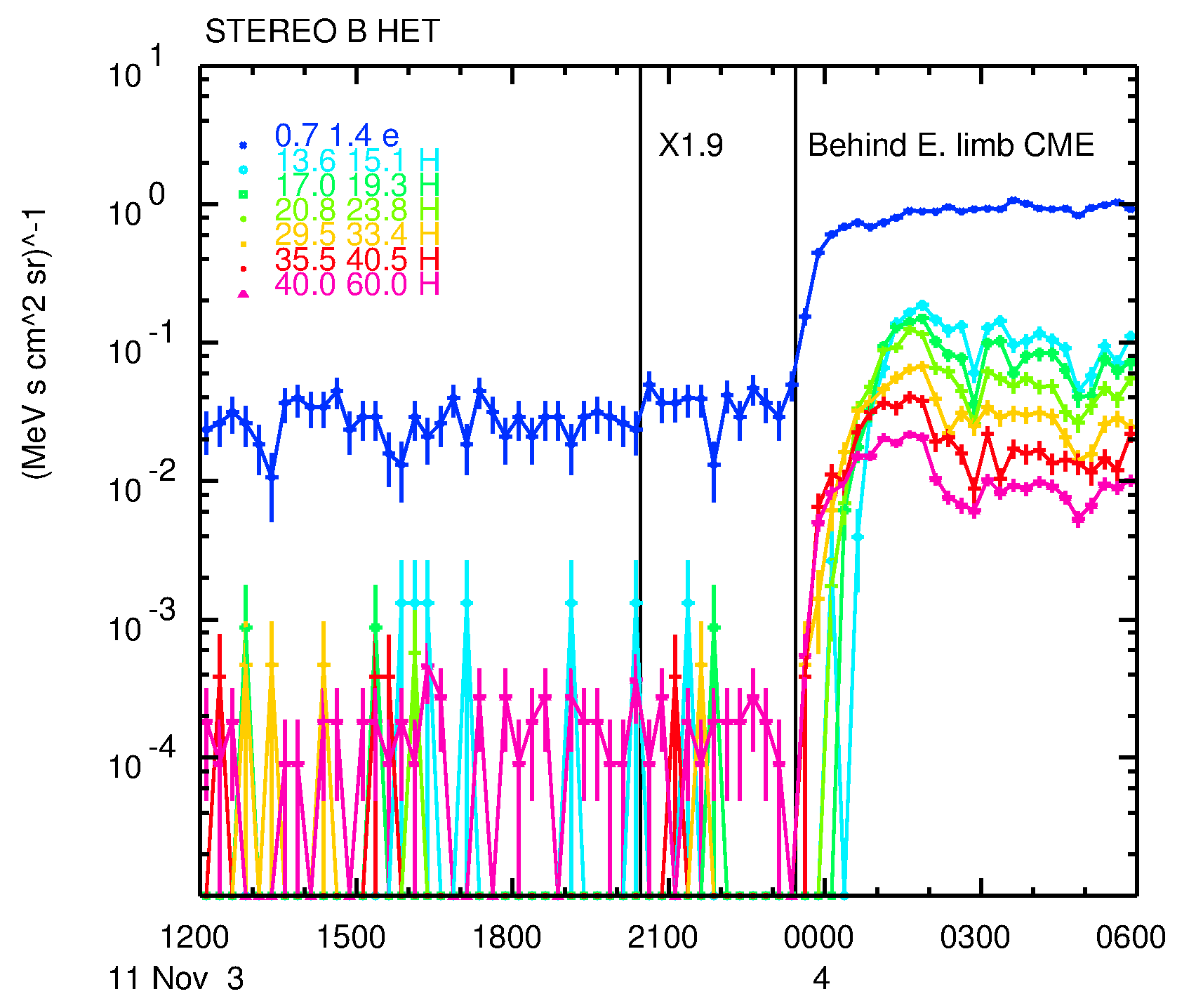}
\centerline{
\includegraphics[width=0.5\linewidth]{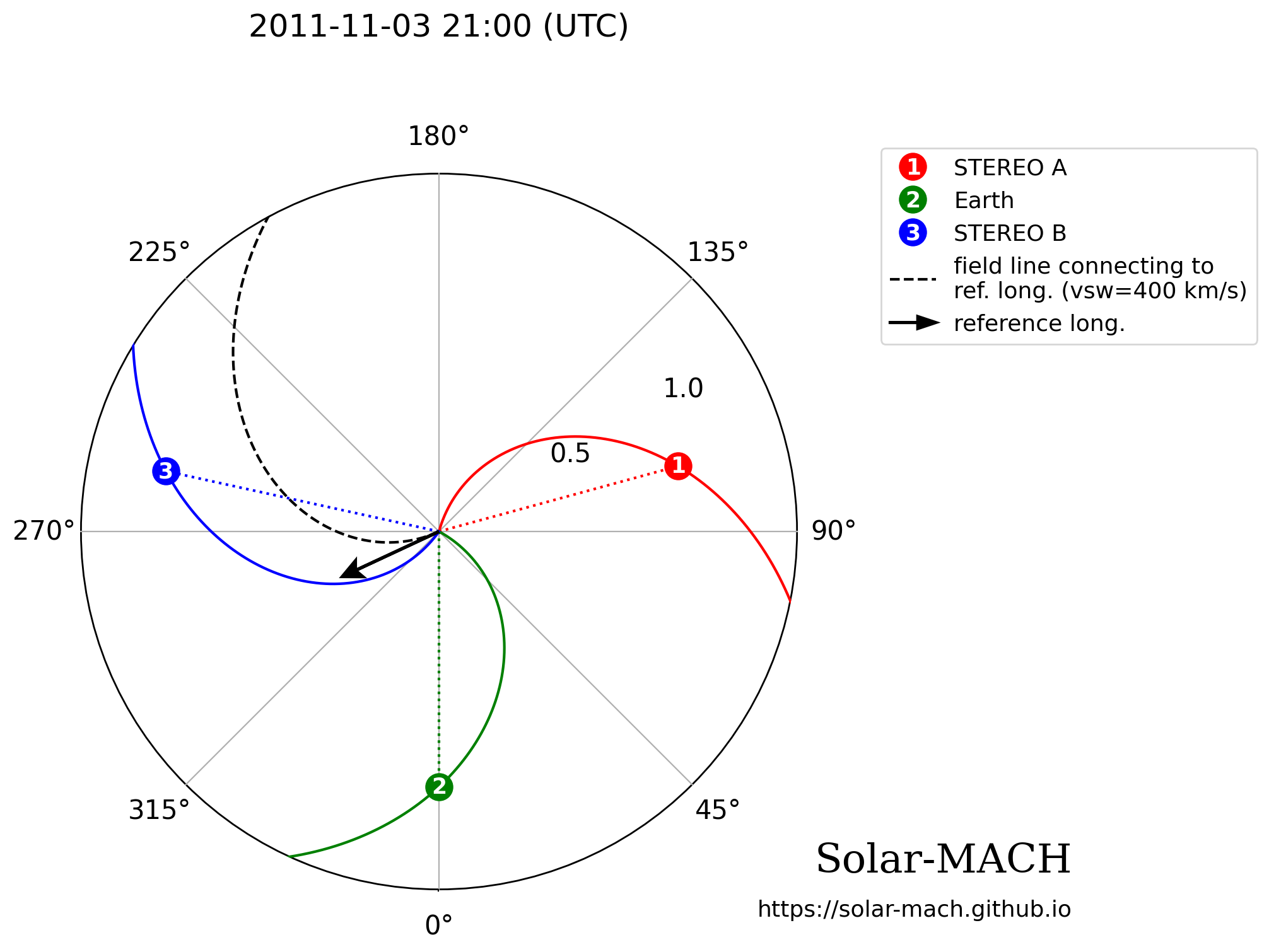}
\includegraphics[width=0.5\linewidth]{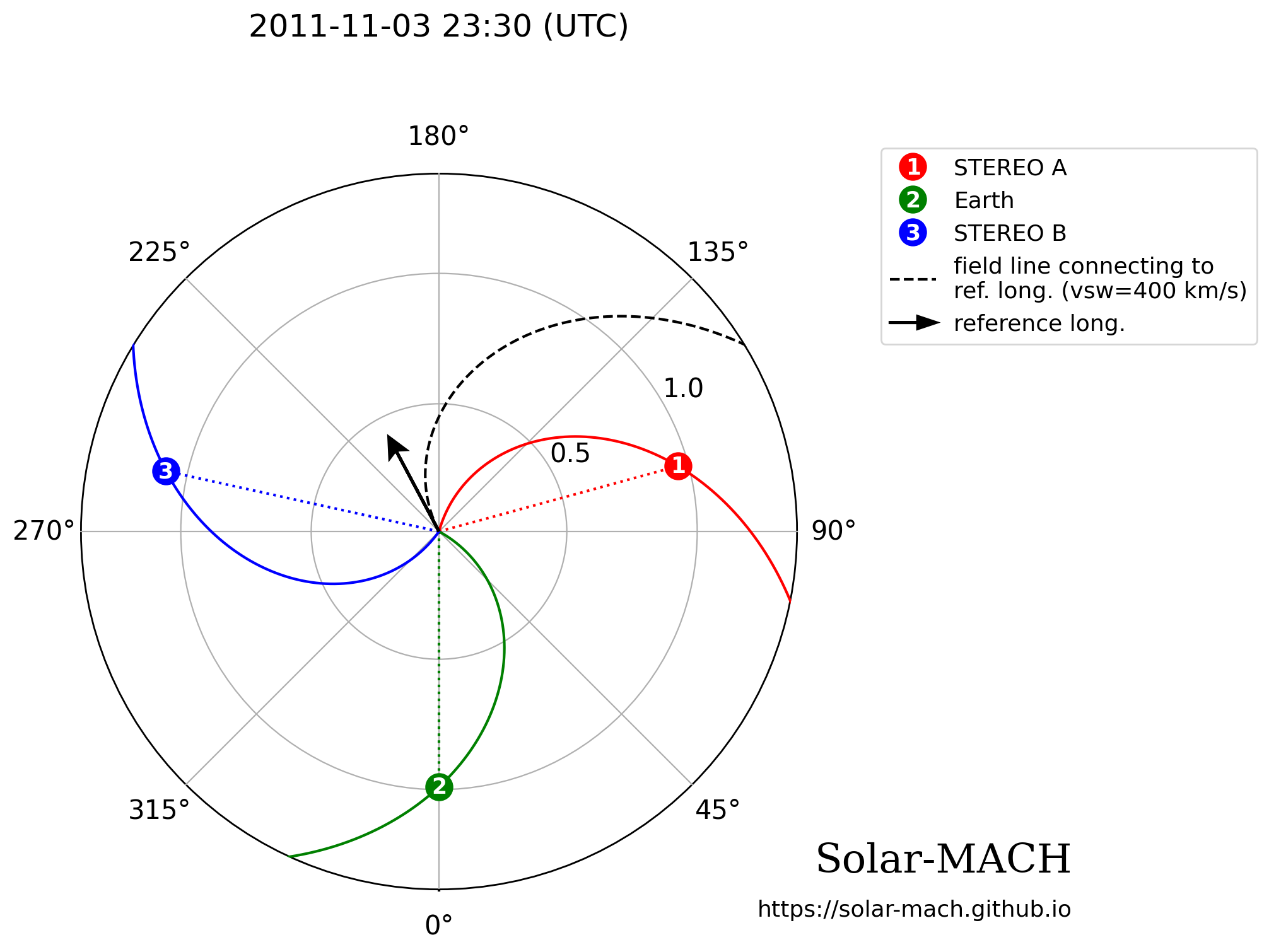}

}
    
    \caption{Top: STEREO~B HET observations on November 3-4, 2011 showing the absence of an SEP event associated with the E65$^{\circ}$ X1.9 flare with peak intensity at 20:27~UT located at W38$^\circ$ relative to the spacecraft (bottom left). Although associated with this flare in some SEP catalogs (see text), the later major SEP event was clearly related with a separate behind-the-limb solar event (bottom right) directly observed by STEREO~B and accompanied by a fast halo CME \citep[e.g.,][]{gomez2015}.}
    \label{fig:20111103HET}
\end{figure}

\begin{figure}
    \centerline{
    \includegraphics[width=0.5\linewidth]{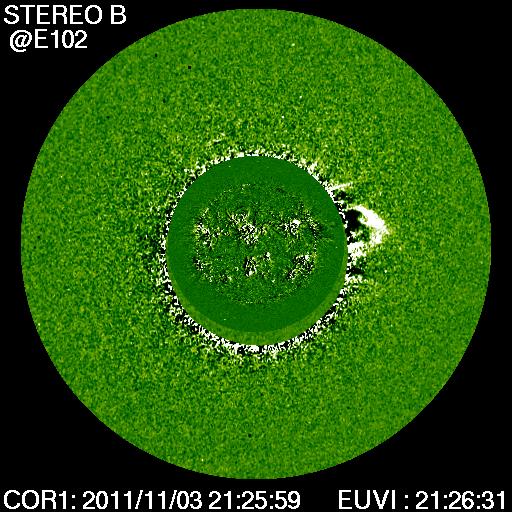}
\includegraphics[width=0.5\linewidth]{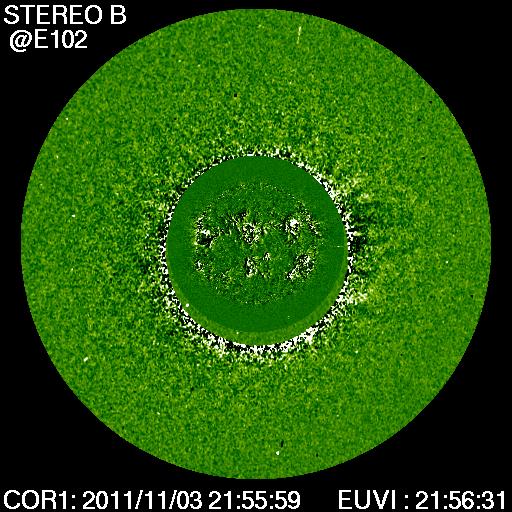}
    }
    \centering
  \includegraphics[width=0.5\linewidth]{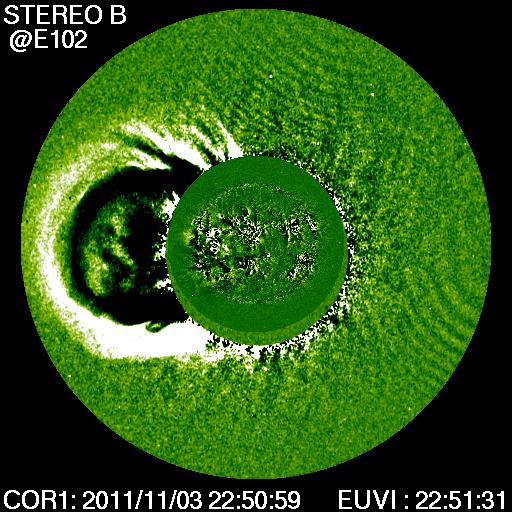}  
    \caption{Top: STEREO~B COR1 observations of the slow ($\sim226$~km/s) narrow CME above the west limb associated with the front side X1.9 flare with peak intensity at 20:27~UT on November 3, 2011 showing the CME falling apart in the COR1 field of view. Bottom: A similar view at 22:51~UT of the faster (991 km/s) wide far side CME associated with the SEP event in Figure~\ref{fig:20111103HET}.  (Images from the CDAW CME catalog.)}
    \label{fig:20111103COR1}
\end{figure}

In the bottom panel of Figure~\ref{fig:flarelongint}, there is one X-class flare without an SEP event despite its well-connected western location relative to the observing spacecraft.  This is the brief X1.9 flare with peak intensity at 20:27~UT on November 3, 2011, located at E65$^\circ$ relative to Earth and W38$^\circ$ relative to STEREO~B (see the bottom left panel of Figure~\ref{fig:20111103HET}) which, even though well-connected to the flare, did not detect any SEP event, as is evident in the STEREO~B HET observations in the top panel of Figure~\ref{fig:20111103HET}. No CME associated with this flare is reported in the CDAW or DONKI catalogs or the ``Dual-Viewpoint CME Catalog from the  SECCHI (Sun Earth Connection Coronal and Heliospheric Investigation)/COR Telescopes” \citep{vourlidas2017}; http://solar.jhuapl.edu/Data-Products/COR-CME-Catalog.php). However, a narrow, slow CME shown in the top panels of Figure~\ref{fig:20111103COR1} was evident in STEREO~B COR1, but fell apart in the field of view.  We estimate the CME leading edge speed in COR1 as $226\pm42$~km/s from a linear fit. Only weak type III emission (not shown) was evident. The absence of any SEP event and weak CME signatures suggests that this X-flare was probably a ``confined" flare; other notable examples were found in the major active region 12192 in October 2014 \citep[e.g.,][]{sun2015}.  This flare was around three hours before a major eruption far behind the east limb (see the bottom right panel of Figure~\ref{fig:20111103HET}), and therefore not associated with a GOES soft X-ray flare, that was directly imaged by STEREO~B and accompanied by a 991 km/s LASCO halo CME, also observed by STEREO~B (Figure~\ref{fig:20111103COR1}), and type II and bright complex type III radio emissions (not shown here).  This eruption gave rise to the widely-studied intense SEP event (Table~\ref{tab2}) evident in Figure~\ref{fig:20111103HET} that rapidly propagated to both STEREO spacecraft and Earth \citep[e.g.,][]{richardson2014,gomez2015,xie2017,lario2017,zhao2018}, although we note that \citet{park2013} and \citet{prise2014} have suggested alternative multi-event interpretations. 

Notwithstanding that these studies have demonstrated that the source of the SEP event on 3 November 2011 is not the front side X1.9 flare, we note that the catalog of \citet{rotti2022} (available at \url{https://doi.org/10.7910/DVN/DZYLHK}) does associate the far side SEP event with the front side X1.9 flare. This incorrect association is also made by the ``Catalog of Solar Proton Events in the 24th Cycle of Solar Activity (2009–2019)" by \citet{logachev2022} and the ``Catalogues of Solar Proton Events in the 20-25 Cycles of Solar Activity" compiled by Moscow State University (\url{https://swx.sinp.msu.ru/apps/sep_events_cat/}). This example illustrates why it is important to consider the possibility that SEP events detected at Earth may originate on the far side of the Sun, as STEREO observations have clearly demonstrated, and should not be assumed to be associated with unrelated front side flares.  The SEP sources listed in Tables~1 to \ref{tab11}, including those on the far side, may help to reduce such errors.

\begin{figure}
    \centering
    \includegraphics[width=0.85\linewidth]{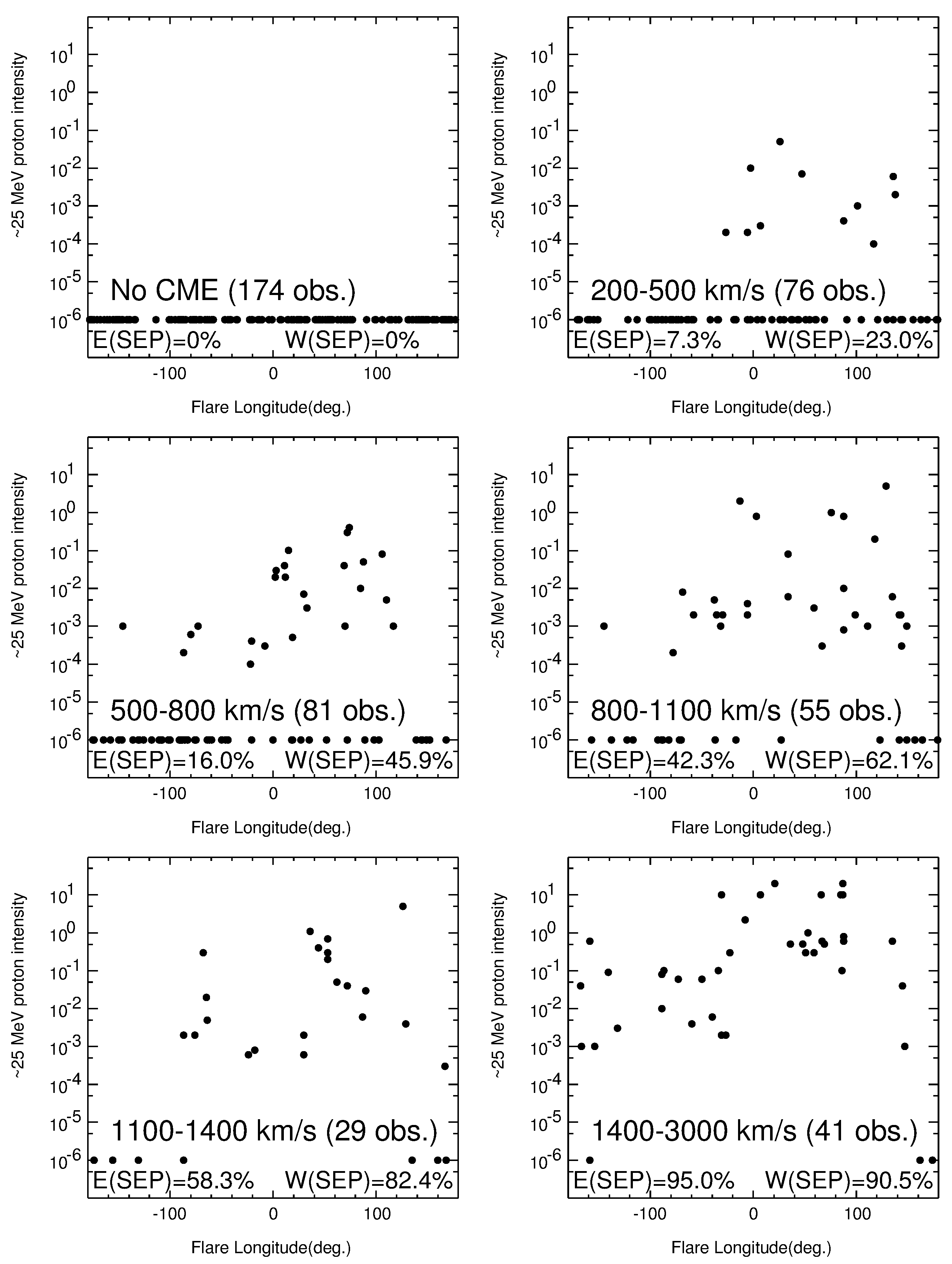}
    \caption{As in Figure~\ref{fig:flarelongint} but for M/X flares associated with no CME (top left), when a $\sim25$~MeV proton event is never detected in any of 174 observations, or associated with CMEs in several speed ranges. Again the probability of an SEP event and the size and extent of the event, increases with CME speed. E(SEP) and W(SEP) indicate the probability that an SEP event is detected for flares east or west of the observing spacecraft, respectively.  }
    \label{fig:cmeintlong}
\end{figure}

We have also examined the dependence on CME speed and flare longitude of SEP occurrence and intensity for our sample of M/X flares. Figure~\ref{fig:cmeintlong}, in a similar format to Figure~\ref{fig:flarelongint}, shows observations of the $\sim25$~MeV proton intensity from STEREO and near-Earth spacecraft plotted against the longitude of the associated flare relative to the observing spacecraft for different speed ranges of the associated CME. Many of the flares are not associated with cataloged CMEs, so we have also examined movies from SOHO/LASCO and STEREO/COR1 and COR2 coronagraphs at the time of each flare to confirm that there is no evidence of an unreported CME or other eruption of material into the corona.  Some 461 independent SEP measurements are available after removing cases with a data gap or high SEP background at a particular spacecraft. (Again, observations where no SEP event is detected are plotted at an intensity of $10^{-6}$ in Figure~\ref{fig:cmeintlong}.) At least for this sample of M/X flares, the 174 SEP observations in the top left panel of Figure~\ref{fig:cmeintlong} show that if an M/X flare is not accompanied by a CME, or evidence of erupting structures in the coronagraph observations, suggesting that the flare is confined, no $\sim25$~MeV proton event is detected.  Thus, based on these results, a necessary requirement for an M/X class flare to be associated with a $\sim25$~MeV proton event appears to be the occurrence of an eruption evident in the corona. 

The remaining panels of Figure~\ref{fig:cmeintlong} show results for flares accompanied by CMEs in increasing speed ranges. They demonstrate that even a well-connected flare (i.e., on the western hemisphere with respect to the spacecraft) accompanied by a CME does not necessarily lead to the detection of a $\sim25$~MeV proton event, in particular when the CME is relatively slow. As in Figure~\ref{fig:flarelongint}, the percentage of observations when the flare is west or east of the observing spacecraft in which an SEP event was detected are given in each panel. As the CME speed progressively increases, the fraction of M/X flares with detected SEP events increases, and these events tend to increase in intensity (with a western bias due to connection) and are detected over a wider longitude range. For example, for flares with relatively slow (200-500 km/s) CMEs, a $\sim25$ MeV proton event was detected for only 7.3\% of observations where the flare was east of the spacecraft compared with 23\% of observations where the flare was to the west.  When an M/X flare was accompanied by a CME with a speed above $\sim1400$~km/s, an SEP event was detected in nearly every spacecraft measurement at all longitudes relative to the flare, though the largest intensities are still associated with flares located at well-connected western longitudes. 

\begin{figure}
    \centering
    \includegraphics[width=1.0\linewidth]{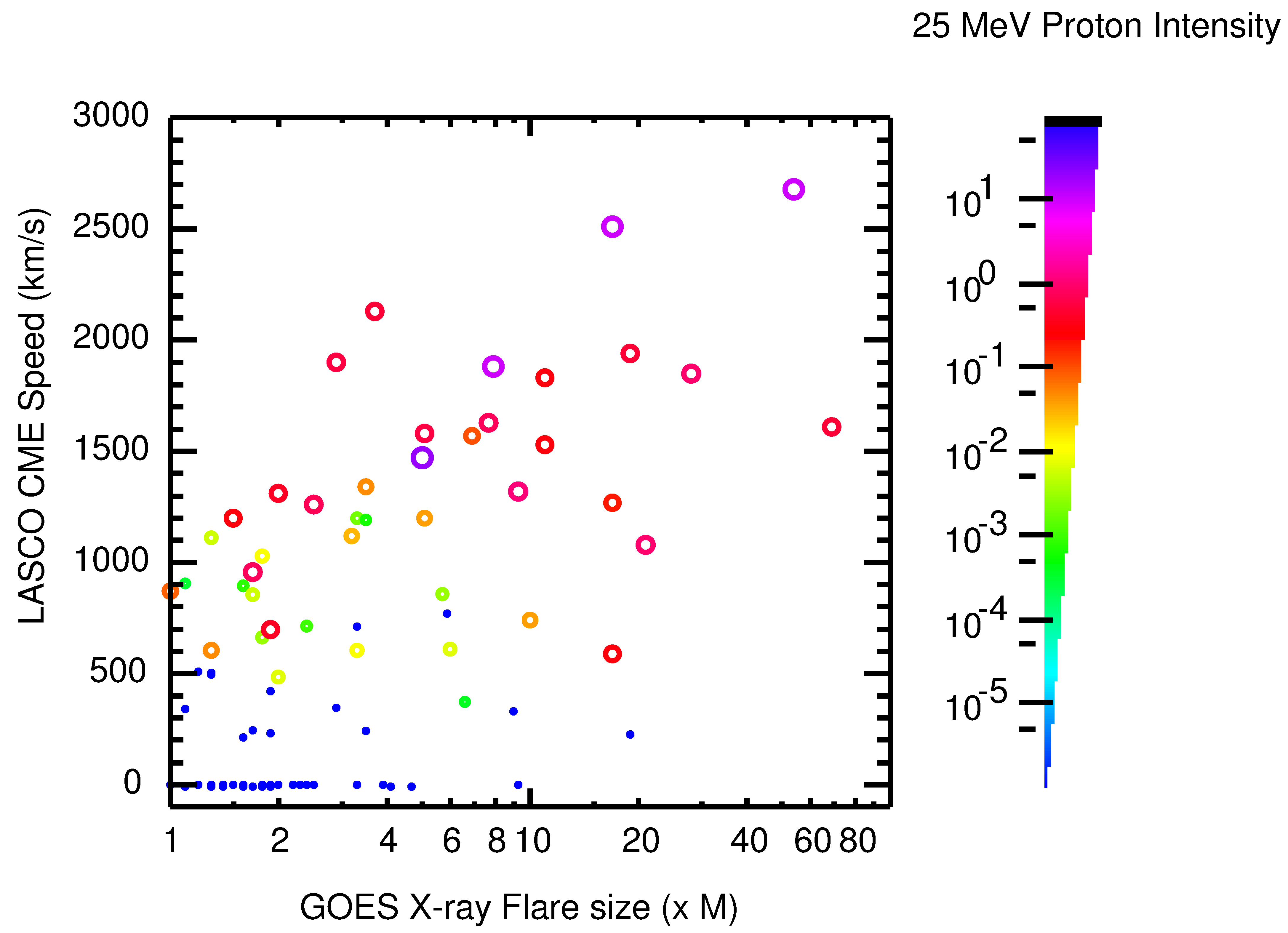}
    \caption{LASCO CME speed vs. GOES X-ray flare size expressed as multiples of M1 for 95 M/X flares at W30-90$^\circ$ with respect to the observing spacecraft. Cases when no CME was observed accompanying the flare are plotted at zero CME speed. The symbol size and color indicate the 25 MeV proton intensity.  The small blue circles indicate cases where no SEP event was detected, which are generally associated with flares below $\sim$M5 and with no CME or a CME below $\sim500$ km/s.}
    \label{fig:cmeflaresep}
\end{figure}

To combine observations of the CME speed, flare intensity and SEP intensity, Figure~\ref{fig:cmeflaresep} shows the LASCO CME speed vs. GOES X-ray flare intensity (in multiples of M1) for the 95 M/X flares in our sample at W30-90$^\circ$ relative to the SEP-observing spacecraft. The symbol size and color indicate the SEP 25 MeV proton intensity. This figure is similar to, and contains a subset of the events in Figure~\ref{fig:cmexrayint} except that here, flares without CMEs (plotted at zero CME speed) or without SEP events (small blue circles) are included. Again, the trend is for larger SEP events to be associated with larger flares and in particular faster CMEs. Flares in this subset of events for which no associated $\sim25$~MeV proton event was detected generally have intensities below $\sim$M5 and either no associated CME, or if present, the CME speed was below $\sim500$~km/s. However, our larger sample of SEP events in Tables 1 to 11 of Appendix~A does include examples associated with weaker flares and slower CMEs, as illustrated in Figure~\ref{fig:cmexrayint}.

These results, starting from a large number of M/X flares and using observations of SEPs from multiple locations with a large dynamic range in SEP intensity (unlike the standard GOES SEP events), clearly demonstrate the relationship between the energy release in a solar eruptive event, as indicated by the soft X-ray intensity or the CME speed, and the peak intensity and extent of the associated SEP event, as well as the probability of an SEP event being produced and detected following the flare. 

\begin{figure}
    \centering
  
     \includegraphics[width=1.0\linewidth]{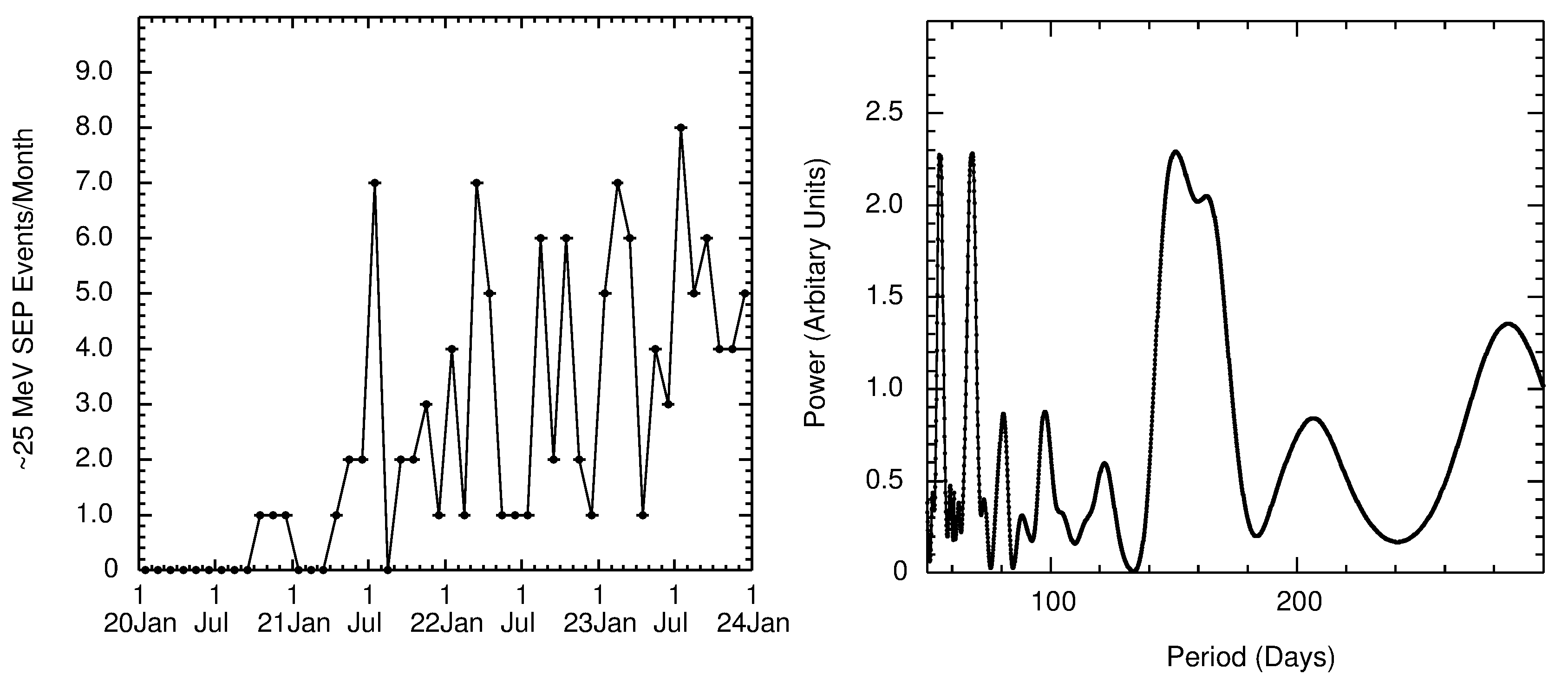}

    \caption{Left: Monthly number of $\sim25$~MeV proton events (from Figure~\ref{fig:3SC}) in 2020-2023 during the rising phase of Solar Cycle~25 showing the tendency for intervals with enhanced numbers of SEP events to be separated by intervals with few events, possibly exhibiting evidence for quasi-periodicity.  Right: Lomb periodogram in the range 50-300 days for the SEP occurrence rate in 2020-2023, showing a broad power enhancement with a peak at 150.8~days and a secondary peak at 163.3 days, consistent with a ``Rieger-like" periodicity.}
    \label{fig:SEPlomb}
\end{figure}

\section{Evidence for Periodicity in the SEP Occurrence Rate 
During the Rising Phase of Cycle 25 (2020-2023)}
\label{s-lomb}
Following the identification of a $\sim154$~day periodicity in the occurrence rate of X-ray flares during Solar Cycle 21 by \citet{Rieger1984},``Rieger-like" quasi-periodicities have been widely reported in various solar phenomena \citep[e.g.,][and references therein]{Lean1990, Lou2003, Lara2008,Lobzin2012, Chowdhury2015, gurgen2021}  including SEP events \citep[e.g.,][]{Bai1990,Cane1998,Dalla2001, Richardson2016}. In particular, \citet{Richardson2016} discussed evidence for $\sim6$ month periodicities in the SEP ($\sim25$ MeV proton) event occurrence rate during the rising
and peak phases of Cycle 24 using the \citet{richardson2014} event catalog. These periodicities were most clearly identified by separating the SEP events into those originating in the northern or southern solar hemispheres, and were shown to be related to variations in the sunspot number or area in the solar hemisphere in which the events originated. Using combined STEREO and near-Earth SEP observations also improved the event statistics significantly over using only near-Earth observations. Here, we point out evidence for a Rieger-like periodicity in the occurrence of SEP events in 2020--2023 during the rising phase of Cycle~25 that might be investigated further (e.g., by considering hemispheric occurence rates) using a longer sequence of Cycle 25 SEP and solar observations.  The left panel of Figure~\ref{fig:SEPlomb} shows the monthly number of $\sim25$~MeV proton events during the rise phase of Cycle~25 in 2020-2023 identified in this study (from Figure~\ref{fig:3SC}),  based on STEREO~A and near-Earth observations. As has been noted during the onsets of some previous solar cycles \citep[e.g.,][]{Dalla2001,Cane1998,Richardson2016}, the SEP rate does not ramp up gradually but shows intervals with large numbers of events (when major active regions are present) separated by periods of reduced activity with fewer SEP events. Inspection suggests that the times of peak SEP rate may occur $\sim$ periodically starting with the first peak late in 2020. To examine this more quantitatively, the right panel shows a Lomb-Scargle \citep{Lomb1976,Scargle1982} periodogram in the range 50-300~days derived from the SEP occurrence rate that includes a broad power enhancement with a peak at 150.8~days and a secondary peak at 163.3~days. These results suggest the presence of a periodicity in the SEP rate which is similar to the Rieger periodicity, though shorter than the six-seven months periodicity found by \cite{Richardson2016} during the rising phase of Cycle 24. \citet{Mcintosh2015}  have discussed the influence of such periodicites, which may originate from Rossby waves in the Solar Tachocline \citep{Dikpati2020}, on solar and heliospheric parameters and their implications for space weather.

\section{Summary} 
      \label{S-Summary} 
\begin{itemize}
    \item Some 450 individual SEP events including $\sim25$~MeV protons were detected by the STEREO~A/B spacecraft and/or by near-Earth spacecraft between STEREO launch in October 2006 and December 2023, encompassing STEREO~A's first orbit of the Sun relative to Earth completed in August 2023.
    \item This period includes the maximum phase of Solar Cycle 24, when the STEREO spacecraft and Earth were widely-separated in longitude, allowing observations of these events to be compared over wide longitudes near 1 AU and demonstrating that SEPs may be transported to extended regions of the inner heliosphere, including those poorly-connected to the observing spacecraft. These observations have also clearly demonstrated that solar events originating on the far side of the Sun can contribute significantly to the SEP population at Earth, with 19\% of the 25~MeV proton events at Earth during our study period originating behind the west limb, and 6.3\% behind the east limb. Unfortunately, contact with STEREO~B was lost in 2014 when on the far side of the Sun.  As a consequence, observations of activity behind the west limb became more limited as STEREO~A moved towards the east limb. Following the loss of contact with STEREO~B, we have used estimates of the direction of the associated CMEs in the DONKI catalog to help infer the general longitude of far side events, in particular those behind the west limb.

\item During the ascending phase of Solar Cycle 25, STEREO A was approaching Earth, and complementary observations from spacecraft in the inner heliosphere, including PSP, SolO and Bepi-Colombo have enabled studies of SEPs using a constellation of spacecraft at varying longitudes and heliocentric distances relative to Earth and STEREO~A. In some cases, SolO was favorably placed to directly observe the origin of an SEP event using remote sensing.  In addition, SEP observations from the inner heliosphere spacecraft may help to identify the onset of an SEP event that was poorly connected to Earth or STEREO A. While such observations from an ever-changing constellation of spacecraft provide opportunities to explore different spatial and temporal variations in SEP events, the STEREO mission has proven the value of observations from spacecraft that are slowly varying in location. In particular, the STEREO spacecraft provided extended (multi-year) near-continuous observations of far side activity that are not available from inner heliosphere missions.  

\item As in \citet{richardson2014}, we have been able to identify the solar event associated with all but a few of the SEP events by assessing the SEP observations together with solar flare, EUV imaging from STEREO or near-Earth spacecraft (SOHO, SDO), coronal mass ejections observed by SOHO/LASCO or STEREO/SECCHI, and solar radio observations from Wind/WAVES and STEREO/SWAVES. Examples of more challenging SEP events are discussed in Appendix~C, as well as a rare small event observed during solar minimum described in Appendix~B.  We have also pointed out that the standard list of SEP events at Earth based on GOES data and defined by thresholds widely used for operations only includes a small subset of the largest SEP events. These do not reflect the rich diversity of SEP events that should be considered, for example, when examining the relationship between solar phenomena and SEP event characteristics that may contribute to the development of models to predict SEP events. 

\item The combined observations from the STEREO and near-Earth spacecraft provide a large sample of $\sim1000$ observations of SEP events associated with solar events at a wide range of longitudes relative to the observing spacecraft that may be useful for statistical studies. For example, we have used this large sample to investigate how the properties of the SEP event depend on those of the solar event, such as the correlation between CME speed and SEP intensity found in many other studies which is also observed at all event longitudes, and a corresponding, though generally weaker, correlation between the flare soft X-ray and SEP intensities.  We have also illustrated how the energetic SEP events discussed by \citet{bruno2018}, and the SEP events associated with long-duration gamma ray flares \citep{bruno2023} lie in the general population at a proton energy of $\sim25$~MeV.  We have also pointed out some biases evident in these data sets, in particular in the longitudinal distributions, due to the locations of the STEREO spacecraft at the times of high SEP activity in Cycles 24 and 25, the loss of contact with STEREO~B, and the restriction of direct X-ray flare observations to front side events (other than a few examples from SolO/STIX).  We have also noted that SEP-associated CDAW LASCO ``non-halo" CMEs tend to be associated with solar events near the limbs, suggesting a bias due to projection, while the ``halo" CMEs are associated with solar events located at all longitudes relative to Earth, consistent with \citet{kwon2014}.  
 
\item We have also summarized the association of $\sim25$~MeV proton events observed by the STEREO or near-Earth spacecraft with a large sample of 397 M/X class GOES X-ray flares in 2010-2014 selected independently of any association with SEP events \citep{Aschwanden2017}. Considering cases with suitable SEP observations (e.g., no data gaps or ongoing SEP event), a $\sim25$~MeV proton event above instrumental background was only detected in 22\% of the observations, demonstrating that even for M/X flares, a majority are not associated with observed SEP events. At least for this set of flares, an SEP event was not detected if no CME or evidence of an eruption accompanied a flare. The likelihood of SEP detection and the longitudinal extent of the SEP event increased as the CME speed increased.  The likelihood of an SEP event being detected and its intensity and longitudinal width also increased with increasing flare size. 

\item As noted above, the observations clearly demonstrate that far side solar events contribute to the SEP population at Earth. We emphasize that, as a consequence, a significant fraction of the SEP events observed at Earth are not associated with front side flares observed, for example in EUV and X-rays, and that care should be taken to avoid erroneously associating far side SEP events with front side flares.  In particular, we highlight the case of the November 3, 2011 event, for which the source far behind the east limb was directly observed by STEREO~B, as several studies have discussed, but has been associated with an unrelated front side X-class flare in several SEP catalogs.  We suspect that further examples of such erroneous associations are also present in older SEP catalogs where SEP events originating on the far side are incorrectly associated with the most conspicuous front side solar event.
\item There is evidence for a ``Rieger-like" periodicity, with peak power at 150.8 days, in the SEP occurrence rate in 2020-2023 during the rising phase of Cycle 25.  This could be explored further using an extended interval of observations and by separating the observations by their solar hemisphere source as in \citet{Richardson2016}.
\end{itemize}

\appendix
\section{SEP Event Tables}

The first column in Tables~1 to \ref{tab11}  gives the date and hour of the likely-related solar event while the second column gives the hour of SEP onset at the spacecraft that first detects the event. We do not provide more detailed estimates of the particle onset times because this requires carefully specifying the instrument/energy considered and for example, possible effects of anisotropies and instrument viewing. Rather the aim is to indicate the solar source associations that may be useful for further detailed studies; several of the other catalogs noted in Section~\ref{S-catalog} include more detailed information on particle onsets that may include inferred solar release times for selected events. 

The next three pairs of columns show the solar event location with respect to the observing spacecraft, and the peak $\sim25$~MeV proton intensity at that spacecraft, for STEREO~B, the Earth, and STEREO~A, respectively. The solar event location, which may be on the far side of the Sun, is inferred from examining movies of solar activity and coronal mass ejections from the STEREO spacecraft, the Solar Dynamics Observatory and SOHO, and cross-checking for consistency with catalogs of solar flares (e.g., NOAA daily solar activity reports (\url{https://www.swpc.noaa.gov/products/solar-and-geophysical-event-reports}, Solarmonitor.org (\url{https://solarmonitor.org}), and the Solarsoft latest
events archive (\url{https://www.lmsal.com/solarsoft/latest\_events\_archive.html}) for front side events, and CMEs (e.g., the CDAW, DONKI (Database Of Notifications, Knowledge, Information; \url{https://kauai.ccmc.gsfc.nasa.gov/DONKI/})   and CACTus (Computer Aided CME Tracking; \url{https://www.sidc.be/cactus/}) catalogs). Occasionally these catalogs do not agree on the solar event properties/location or may include apparent typographical errors, in which case, we have used our judgment guided by observations to make any necessary corrections.  In twelve cases between the launch of SolO in February 2020 and December 2023, the related far side X-ray flare was directly detected by the STIX instrument on SolO. Some of the listed far side events were also directly observed by remote sensing instruments on SolO. ``D" indicates a source longitude based on the CME propagation direction in the DONKI database.  This is mainly used for events that occurred following the loss of STEREO~B observations where the source cannot be viewed directly but appears to be behind the west limb as viewed from Earth (e.g., there is a CME above the west limb with no front side activity) and is also beyond the eastern limb as viewed from STEREO-A. In addition, in some cases, the DONKI database gives an estimate of the behind the limb flare location. The DONKI database typically gives a brief explanation of the rationale for estimating a certain CME direction.  The typical east-west asymmetry of SEP event intensity-time profiles \citep{cane1988, canelar2006} can also often help to corroborate the event location relative to the observing spacecraft.  The solar event locations provided are likely to be accurate only to $\sim10^o$ since the associated activity may cover an extended region, different flare databases often do not agree exactly on the location, and there is uncertainty in the DONKI CME directions (\citet{bruno2021} indicate uncertainties of $\sim10-15^\circ$ depending on the number of spacecraft used in the direction determination).  
 
 The peak $\sim25$~MeV proton intensity at the observing spacecraft generally corresponds to the ``first peak" during the first few ($\sim12$) hours of the event, in (MeV~s~cm$^2$~sr)$^{-1}$. The intensity may be higher in an ``energetic storm particle" enhancement around passage of the associated interplanetary shock, which occasionally can extend above 40~MeV \citep{lario2023} but is typically more prominent at lower energies.  However, here, we choose not to mix intensities from different stages of an SEP event. The proton intensities at $\sim25$~MeV are based on observations from
different spacecraft instruments around this energy. In particular, the STEREO HETs have a channel at 23.8-26.4~MeV,
the SOHO/ERNE export data have a channel at 21-28~MeV. SOHO/EPHIN has
rather wide channels, including 7.8-25~MeV and 25-40.9~MeV that encompass the energy of interest, and Wind/EPACT has a proton channel at 19-28 MeV; some examples of scalings between
different instrument channels are discussed in \citet{richardson2014}. Due to such differences in instrumentation, the $\sim25$~MeV proton intensities are typically only quoted to one significant figure, but are sufficient to indicate the relative sizes of the SEP events at the spacecraft locations. ``BG" indicates that there is a background from an ongoing SEP event that may prevent the SEP event from being observed at a certain location.  ``..." indicates that the SEP event was not detected at $\sim25$~MeV above the instrument background, typically $\sim10^{-4}$ (MeV~s~cm$^2$~sr)$^{-1}$ in all the instruments used (see Figure~\ref{fig:3SC}).

 Column~9 gives the GOES soft X-ray flare size (peak intensity) for front side or just over the limb events, obtained from operational GOES data and reports. 
 Note that before December 18, 2017, the reported intensities are 43\% lower than the actual intensities, due to application of a correction factor, whereas later intensities are from GOES-R with different instrumentation that provides operational data with correctly-calibrated absolute soft X-ray intensities (\url{https://www.goes-r.gov/users/transitionToOperations16.html}; see also the XRS Users Guide, available at \url{https://www.ngdc.noaa.gov/stp/satellite/goes-r.html}). Hence, after this date, the flare intensities listed should be multiplied by 0.7 to be consistent with those of earlier events or alternatively, earlier event intensities should be multiplied by 1.43 to give the absolute intensities. In analyses using the X-ray intensities discussed in Section~\ref{S-dist}, we have chosen to correct the GOES-R data to be consistent with earlier observations since most of the cataloged SEP events occurred before the GOES-R era, while recognizing that these intensities are not absolute. We have also retained the reported X-ray intensities in the catalog, rather than make any corrections, to facilitate cross referencing with other catalogs and studies that typically also quote these intensities.   In a few cases where the related far side flare was detected by the STIX instrument on SolO, the equivalent GOES intensity is obtained from the STIX flare list available at \url{https://datacenter.stix.i4ds.net/} and denoted by `S'.

  Column~10 indicates whether or not (1 or 0, respectively) a type III radio burst \citep{wild1963} observed by the Wind/WAVES or STEREO/SWAVES instruments accompanied the solar event. Observations of type III (fast-drift) radio emissions (usually attributed to flare-accelerated electrons streaming away from the Sun) from the STEREO spacecraft and Wind can help indicate the solar event time and also corroborate the location based on the brightness of the burst at each spacecraft (see \citet{krupar2024} for a recent discussion) and whether there is occultion of the higher frequencies by the limb of the Sun indicating a far side event relative to the observing spacecraft. As discussed for example by \citet{cane2002}, \citet{cane2010}, \citet{laurenza2009}, \citet{macdowall2003}, \citet{macdowall2009}, \citet{winter2015}, \citet{richardson2014}, and \citet{richardson2018}, SEP events are highly (nearly 100\%) associated with type III radio emissions observed by spacecraft, with the larger SEP events tending to be associated with brighter, longer duration type III emissions that extend to low frequencies.  In particular, \citet{richardson2018} used the presence or absence of type III emissions to help distinguish CMEs that were more or less likely to be associated with SEP (specifically, $\sim25$~MeV proton) events. Radio observations made by various spacecraft since November 2015 are summarized at the Coordinated Radiodiagnostics Of CMEs and Solar flares (CROCS) website (\url{https://parker.gsfc.nasa.gov/crocs.html}).
  
  Column~11 indicates whether (1/2) or not (0) a type II (slow drift) radio burst, believed to be evidence for particle acceleration at a CME-driven shock \citep[e.g.,][]{cane1982,nelson1985,wild1963}, was observed by these instruments (from the lists at \url{https://solar-radio.gsfc.nasa.gov/wind/data_products.html} and \url{https://cdaw.gsfc.nasa.gov/CME_list/radio/waves_type2.html}).  If type II emissions extended below 1~MHz, indicative of an ``interplanetary" (IP) type II event \citep[e.g.,][]{canee2005}, this is denoted by `2'. ``..." indicates that this information is not available at the time of writing.  
  
  Columns 12 and 13 give the plane-of-the-sky angular width and speed of the associated CME as observed by SOHO/LASCO.  Values from the CDAW CME catalog (\url{http://cdaw.gsfc.nasa.gov/CME_list/}) are used if available.  Other values are from the CACTus catalog (\url{http://sidc.oma.be/cactus/}, indicated by ``*") while ``A" or ``B" indicates values from the STEREO A or B CACTUS catalog, respectively. 
\begin{landscape}

\newpage

\begin{table}
\label{tab1}
\renewcommand{\arraystretch}{.7}
\setlength{\tabcolsep}{.05in}
\caption{$\sim25$~MeV Proton Events at 1~AU During the STEREO Mission}
\begin{tabular}{llccccccccccc}

\hline
\\
 &
&  
Solar&
25~MeV&
Solar& 
25~MeV&
Solar&
25~MeV&  
&  
&  
&
&
\\
Solar Event&
SEP Event &  
Event--B&
I(B) &
Event--Earth & 
I(Earth) &
Event--A  &
I(A)&  
 GOES&  
Type&  
Type &
CME dA&
CME V  
\\  
(UT) &    (UT)  &  ($^o$)&  & ($^o$)&  & ($^o$) &   &X-Ray   & III &II &($^o$)&km/s    \\ 
\\ 
\hline
2006 Dec~~5,~10 & Dec~~5,~12 &   -68  &   0.04 &   -68   &   0.04 &  -68   &  0.04    & X9.0     &   1  &     2  &DG&DG  \\ 
2006 Dec~~6,~18 & Dec~~6,~19 & -64   & BG   & -64    & BG  & -64 & BG&X6.5& 1 & 2&360&DG\\
2006 Dec~13,~02 & Dec~13,~02 &23 &10&23 &   10 &23 &  10& X3.4  & 1   &    2 &360&1774  \\
\hline
\end{tabular}

\end{table}

\newpage

\begin{table}
\renewcommand{\arraystretch}{.7}
\setlength{\tabcolsep}{.05in}
\caption{$\sim25$~MeV Proton Events at 1~AU During the STEREO Mission}
\label{tab2}

\begin{tabular}{llccccccccccc}

\hline
\\
  &
&  
Solar&
25~MeV&
Solar& 
25~MeV&
Solar&
25~MeV&  
&  
&  
&
&
\\
Solar Event&
SEP Event &  
Event--B&
I(B) &
Event--Earth & 
I(Earth) &
Event--A  &
I(A)&  
 GOES&  
Type&  
Type &
CME dA&
CME V  
\\  
(UT) &    (UT)  &  ($^o$)&  & ($^o$)&  & ($^o$) &   &X-Ray   & III &II &($^o$)&km/s    \\ 
\\ 
\hline
2011 May~~9,~20 & May~10,~00 &    0   &  0.002    & -94    &   ...      & 174   &  ...    &  C5.4  &1      &  2 &292&1318  \\
2011 May~11,~02 & May~11,~04 &  147   &    ... &   53    &   0.004 &  -39  & ... &B8.1        &    1   &  0 &225&745  \\
2011 May~19,~03 & May~19,~06 & 126     &     ...&    32      &  ...    & -61    &   0.0001   & B3.0 &   1     &  0  &132&626 \\
\hline
\end{tabular}

\end{table}

\newpage

\begin{table}
\renewcommand{\arraystretch}{.7}
\setlength{\tabcolsep}{.05in}
\caption{$\sim25$~MeV Proton Events at 1~AU During the STEREO Mission}
\label{tab3}

\begin{tabular}{llccccccccccc}

\hline
\\
  &
&  
Solar&
25~MeV&
Solar& 
25~MeV&
Solar&
25~MeV&  
&  
&  
&
&
\\
Solar Event&
SEP Event &  
Event--B&
I(B) &
Event--Earth & 
I(Earth) &
Event--A  &
I(A)&  
 GOES&  
Type&  
Type &
CME dA&
CME V  
\\  
(UT) &    (UT)  &  ($^o$)&  & ($^o$)&  & ($^o$) &   &X-Ray   & III &II &($^o$)&km/s    \\ 
\\ 
\hline
2011 Dec~~3,~07 & Dec~~3,~08 &  -117  &  ...    &   136   &   DG    &   30   &    0.0003  & ...  &  1   & 0 &155&454 \\
2011 Dec~11,~13 & Dec~11,~14 & -76   &  ...     &    176         &  ...   &  70       & 0.0003    & ...  &  1   & 0&32~A&312~A  \\
2011 Dec~17,~10 & Dec~17,~16 & -115   &  ...     &     136   &   0.0003  &  30 & 0.002     & ... &  1   & 0 & 247&987 \\
\hline
\end{tabular}

\end{table}

\newpage

\begin{table}

\renewcommand{\arraystretch}{.7}
\setlength{\tabcolsep}{.05in}
\caption{$\sim25$~MeV Proton Events at 1~AU During the STEREO Mission}
\label{tab4}
\begin{tabular}{llccccccccccc}

\hline
\\
  &
&  
Solar&
25~MeV&
Solar& 
25~MeV&
Solar&
25~MeV&  
&  
&  
&
&
\\
Solar Event&
SEP Event &  
Event--B&
I(B) &
Event--Earth & 
I(Earth) &
Event--A  &
I(A)&  
 GOES&  
Type&  
Type &
CME dA&
CME V  
\\  
(UT) &    (UT)  &  ($^o$)&  & ($^o$)&  & ($^o$) &   &X-Ray   & III &II &($^o$)&km/s    \\ 
\\ 

\hline
2012 Jul~~2,~08 & Jul~~2,~09 & -10 & 0.001  & -126 & 0.0001 & 115 &0.01    & ... &1  & 2 &360&1074\\ 
2012 Jul~~4,~16 & Jul~~4,~18 & 150 &...  & 34 & 0.003 & -85 & ... & M1.8 & 1 & 1 &360&662 \\
2012 Jul~~5,~21 & Jul~~5,~22  & 161 &  ...  &  46   &  BG   &  -73  &   0.0002  & M1.6    & 1  & 2  & 94  & 980   \\
\hline
\end{tabular}

\end{table}

\newpage

\begin{table}
\renewcommand{\arraystretch}{.7}
\setlength{\tabcolsep}{.05in}
\caption{$\sim25$~MeV Proton Events at 1~AU During the STEREO Mission}
\label{tab5}
\begin{tabular}{llccccccccccc}

\hline
\\
  &
&  
Solar&
25~MeV&
Solar& 
25~MeV&
Solar&
25~MeV&  
&  
&  
&
&
\\
Solar Event&
SEP Event &  
Event--B&
I(B) &
Event--Earth & 
I(Earth) &
Event--A  &
I(A)&  
 GOES&  
Type&  
Type &
CME dA&
CME V  
\\  
(UT) &    (UT)  &  ($^o$)&  & ($^o$)&  & ($^o$) &   &X-Ray   & III &II &($^o$)&km/s    \\ 
\\ 

\hline
2013 May~~1,~02 & May~~1,~04&   37 &  0.009   & -105&  ...    &  120 &  0.007 &   ...  &   1  &  1 &  360 &  762  \\
2013 May~~2,~04 & May~~2,~06&  168 &  BG     & 26  &  0.0005 &  -109&  BG   &   M1.1 &   1  &  0 &  99  &   671 \\
2013 May~~3,~17 & May~~3,~20&   61 &  0.003   & -81 &  ...    &  144 &  0.002 &   M5.7 &   0  &  0 &  274 &  858  \\
\hline
\end{tabular}

\end{table}

\newpage

\begin{table}
\renewcommand{\arraystretch}{.7}
\setlength{\tabcolsep}{.05in}
\caption{$\sim25$~MeV Proton Events at 1~AU During the STEREO Mission}
\label{tab6}
\begin{tabular}{llccccccccccc}

\hline
\\
 &
&  
Solar&
25~MeV&
Solar& 
25~MeV&
Solar&
25~MeV&  
&  
&  
&
&
\\
Solar Event&
SEP Event &  
Event--B&
I(B) &
Event--Earth & 
I(Earth) &
Event--A  &
I(A)&  
 GOES&  
Type&  
Type &
CME dA&
CME V  
\\  
(UT) &    (UT)  &  ($^o$)&  & ($^o$)&  & ($^o$) &   &X-Ray   & III &II &($^o$)&km/s    \\ 
\\ 
\hline
2014 Jan~~4,~18 &Jan~~4,~21  & 113  & ...    &-40  & 0.002 & 169  & ...     & M4.0 & 1  & 2 & 360 & 977    \\
2014 Jan~~6,~07 &Jan~~6,~08  &-97  & 0.003  & 110  &  0.6  &  -41 &0.04     &...   & 1  & 2 & 360     &1402 \\
2014 Jan~~7,~18 &Jan~~7,~19  &164   &0.04    & 11  & 2.2   &-140  &0.6     &X1.2  & 1  & 2 & 360 &1830 \\
\hline
\end{tabular}

\end{table}

\newpage

\begin{table}
\renewcommand{\arraystretch}{.7}
\setlength{\tabcolsep}{.05in}
\caption{$\sim25$~MeV Proton Events at 1~AU During the STEREO Mission}
\label{tab7}
\begin{tabular}{llccccccccccc}

\hline
\\
 &
&  
Solar&
25~MeV&
Solar& 
25~MeV&
Solar&
25~MeV&  
&  
&  
&
&
\\
Solar Event&
SEP Event &  
Event--B&
I(B) &
Event--Earth & 
I(Earth) &
Event--A  &
I(A)&  
 GOES&  
Type&  
Type &
CME dA&
CME V  
\\  
(UT) &    (UT)  &  ($^o$)&  & ($^o$)&  & ($^o$) &   &X-Ray   & III &II &($^o$)&km/s    \\ 
\\ 
\hline
2014 Jul~~8,~16 &Jul~~8,~17  &107  & 0.02    & -56  & 0.0005& 141 & DG     &M6.5 &1     & 0  &360 &773\\
2014 Jul 10,~07 &Jul~10,~08&-81  & BG      & 116D  & 0.001 & -47  &DG     &...  & 1    & 0  & 275 &928 \\
2014 Aug~~8,~16 &Aug~~8,~18  &-44  & 0.001   & 155  & ...  & -10  &0.001   &...  &1     & 0  &360 &1137\\
\hline
\end{tabular}

\end{table}

\newpage

\begin{table}
\renewcommand{\arraystretch}{.7}
\setlength{\tabcolsep}{.05in}
\caption{$\sim25$~MeV Proton Events at 1~AU During the STEREO Mission}
\label{tab8}
\begin{tabular}{llccccccccccc}

\hline
\\
 &
&  
Solar&
25~MeV&
Solar& 
25~MeV&
Solar&
25~MeV&  
&  
&  
&
&
\\
Solar Event&
SEP Event &  
Event--B&
I(B) &
Event--Earth & 
I(Earth) &
Event--A  &
I(A)&  
 GOES&  
Type&  
Type &
CME dA&
CME V  
\\  
(UT) &    (UT)  &  ($^o$)&  & ($^o$)&  & ($^o$) &   &X-Ray   & III &II &($^o$)&km/s    \\ 
\\ 
\hline
2015 Jul~~1,~14&Jul~~1,~15&...           &DG    &133D  &0.03    &...     &DG &...        &1  &2 &360&1435\\
2015 Jul~19,~09&Jul~19,~11&...           &DG    &62&0.0005  &-123     &... &C2.1     &1 &0  &194&782\\
2015 Aug~16,~08 &Aug~16,~09   &...   &DG    &-133  &...      &40&0.0005&...  &1  &0    &114    &544\\
\hline
\end{tabular}

\end{table}

\newpage

\begin{table}
\renewcommand{\arraystretch}{.7}
\setlength{\tabcolsep}{.05in}
\caption{$\sim25$~MeV Proton Events at 1~AU During the STEREO Mission}
\label{tab9}
\begin{tabular}{llccccccccccc}

\hline
 \\
  &
&  
Solar&
25~MeV&
Solar& 
25~MeV&
Solar&
25~MeV&  
&  
&  
&
&
\\
Solar Event&
SEP Event &  
Event--B&
I(B) &
Event--Earth & 
I(Earth) &
Event--A  &
I(A)&  
 GOES&  
Type&  
Type &
CME dA&
CME V  
\\  
(UT) &    (UT)  &  ($^o$)&  & ($^o$)&  & ($^o$) &   &X-Ray   & III &II &($^o$)&km/s    \\ 
\\ 
\hline
\\
2019 Apr~21,~04&Apr~21,~05&...  &DG  &117D         &0.0004       &-147D    &...  &...         &1   &0  &93    &463 \\
2019 May~~3,~23&May~~4,~01&...  &DG  &-88          &...          &7        &0.001 &C1.0     &1   &1  &113   &692\\
\hline
\end{tabular}

\end{table}

\newpage

\begin{table}
\renewcommand{\arraystretch}{.7}
\setlength{\tabcolsep}{.05in}
\caption{$\sim25$~MeV Proton Events at 1~AU During the STEREO Mission}
\label{tab10}
\begin{tabular}{llccccccccccc}

\hline
\\
 &
&  
Solar&
25~MeV&
Solar& 
25~MeV&
Solar&
25~MeV&  
&  
&  
&
&
\\
Solar Event&
SEP Event &  
Event--B&
I(B) &
Event--Earth & 
I(Earth) &
Event--A  &
I(A)&  
 GOES&  
Type&  
Type &
CME dA&
CME V  
\\  
(UT) &    (UT)  &  ($^o$)&  & ($^o$)&  & ($^o$) &   &X-Ray   & III &II &($^o$)&km/s    \\ 
\\ 
\hline
2022 May~11,~18&May~11,~18&...  &DG  &105            &0.1         &135      &0.01    &M2.7    &1   &1  &194   &1100\\
2022 Jun~13,~02&Jun~13,~06&...  &DG  &-45           &0.003       &-17      &0.01     &M3.4    &1   &1  &360   &1150\\
2022 Jul~09,~13&Jul~09,~13&...  &DG  &88            &0.2         &113      &0.009    &C8.4    &1   &1  &148   &1034\\
\hline
\end{tabular}

\end{table}

\newpage

\begin{table}
\renewcommand{\arraystretch}{.7}
\setlength{\tabcolsep}{.05in}
\caption{$\sim25$~MeV Proton Events at 1~AU During the STEREO Mission}
\label{tab11}
\begin{tabular}{llccccccccccc}

\hline
\\
  &
&  
Solar&
25~MeV&
Solar& 
25~MeV&
Solar&
25~MeV&  
&  
&  
&
&
\\
Solar Event&
SEP Event &  
Event--B&
I(B) &
Event--Earth & 
I(Earth) &
Event--A  &
I(A)&  
 GOES&  
Type&  
Type &
CME dA&
CME V  
\\  
(UT) &    (UT)  &  ($^o$)&  & ($^o$)&  & ($^o$) &   &X-Ray   & III &II &($^o$)&km/s    \\ 
\\ 
\hline
2023 Jun~01,~00 &Jun~01,~02&... &DG  &80            &0.0002      &87      &0.0002  &C9.2    &1   &0  &165  &1565\\
2023 Jun~20,~16 &Jun~21,~00&... &DG  &-73           &0.0002      &-68    &0.0006   &X1.1    &1   &1  &360  &1113 \\
2023 Jun~28,~02 &Jun~28,~06&... &DG  &89            &0.0001      &93      &...     &C3.3    &1   &0  &77   &207 \\
\hline

\end{tabular}

\end{table}
\end{landscape}

\clearpage
\section{A $\sim25$~MeV SEP Event During Solar Minimum (May~3, 2018)}

\begin{figure}[tbp] {
\centerline{
  \includegraphics[width=0.44\linewidth]{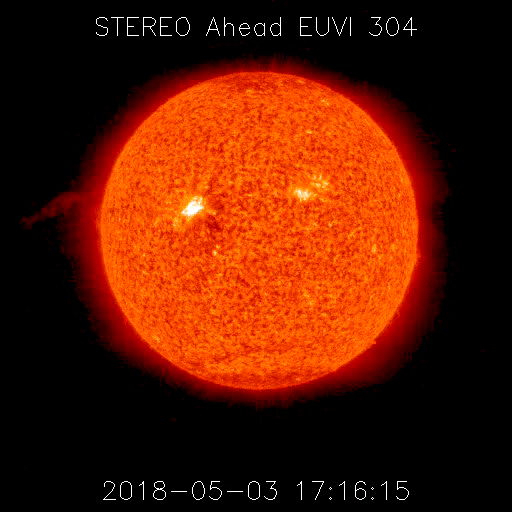}  \includegraphics[width=0.56\linewidth]{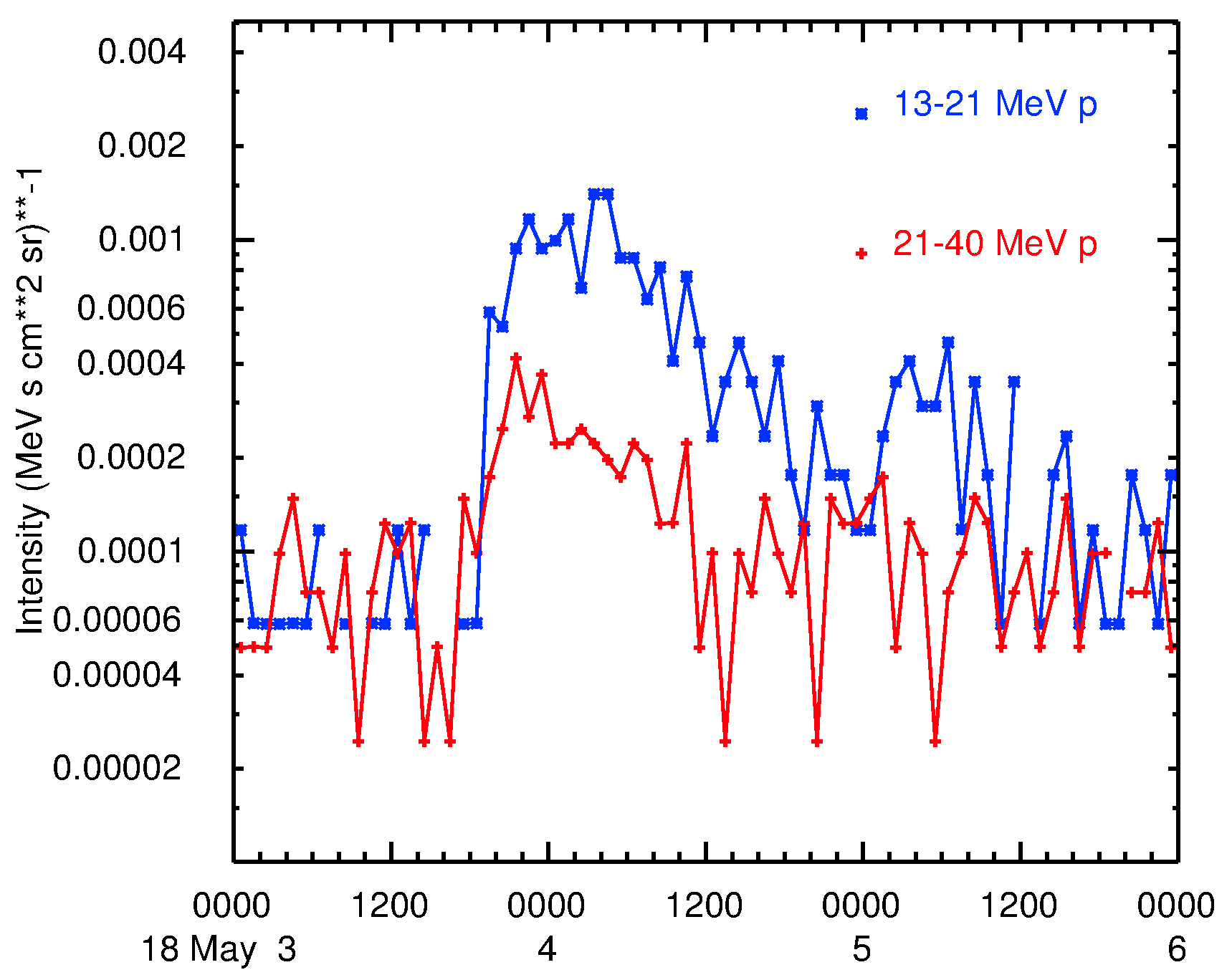}
  }
  \centerline{
      \includegraphics[width=0.6\linewidth]{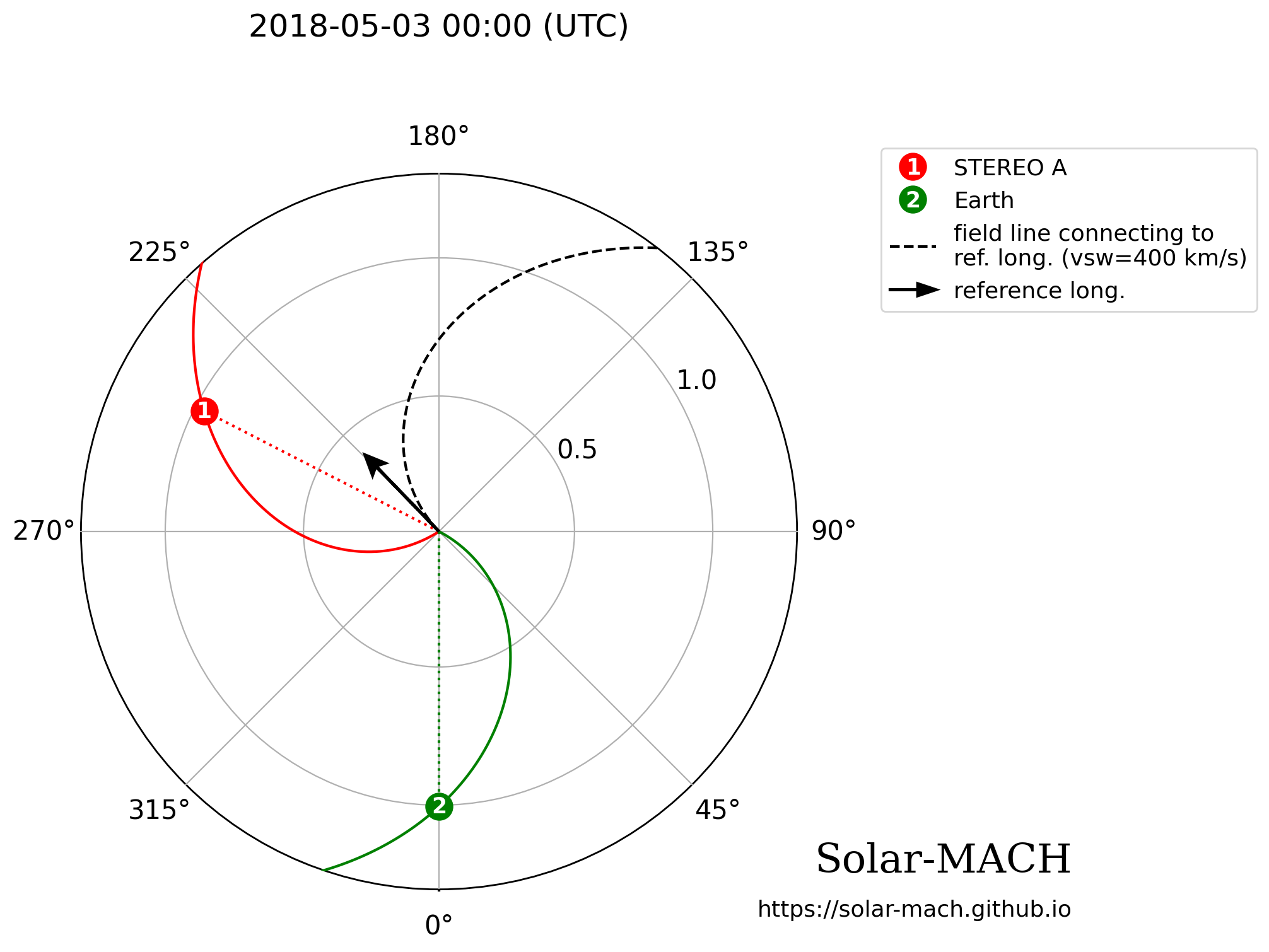}
  }
}
  \caption{One hour averages (top right) of the proton intensity in two energy ranges for a small SEP event on May~3, 2018 detected by the STEREO~A HET associated with flaring at $\sim$E$20^\circ$ relative to STEREO~A ($\sim$E136$^\circ$ relative to Earth) and a filament eruption visible above the east limb observed by EUVI on STEREO~A (top left). The bottom panel shows the spacecraft configuration and longitude of the flaring observed by STEREO~A (black arrow). Interplanetary field lines assume a solar wind speed of 400 km/s. }
  \label{fig:may18}
\end{figure}

\begin{figure}[tbp] {
  \centerline{ \includegraphics[width=0.45\textwidth,keepaspectratio]{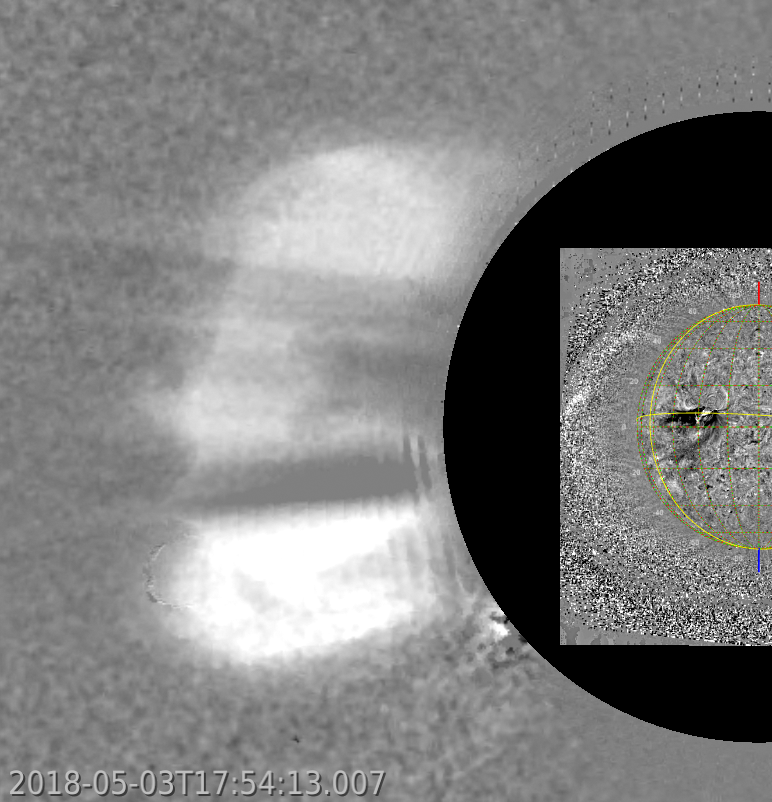}
\includegraphics[width=0.45\textwidth,keepaspectratio]{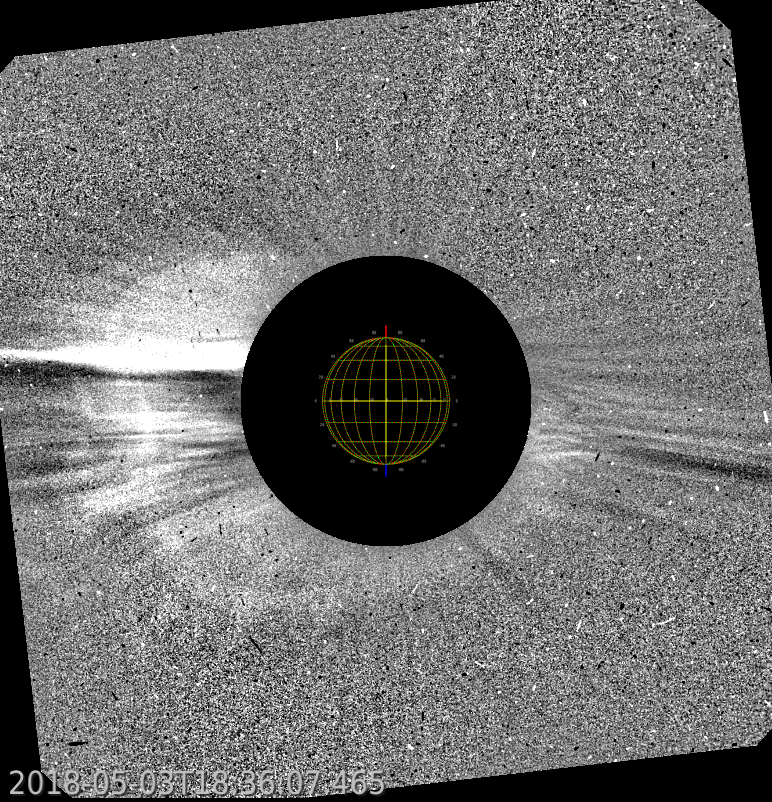}
}
  \caption{The CME associated with the SEP event on May~3, 2018 in Figure~\ref{fig:may18} observed by (left) STEREO~A EUVI 195~\AA  ~at  16:48~UT in the low corona, also showing the dimming associated with the eastern eruption, and by STEREO~A COR2 at 17:54~UT, and  (right) by LASCO~C2 at 18:36~UT. All images are base differences from pre-event images. }
 \label{fig:may18cme}
 }
\end{figure}

Figure~\ref{fig:may18} shows an example of a rare SEP event during the minimum between Solar Cycles 24 and 25 (see Figure~\ref{fig:3SC}) that nevertheless reached at least 25~MeV in protons. This event was observed on May~3, 2018 from $\sim19$~UT and only detected at STEREO~A - the top right panel of Figure~\ref{fig:may18} shows the HET-A proton intensity-time profiles at 13--21 and 21--40~MeV; individual instrument channels (not shown) suggest a maximum energy around 30~MeV.  This was a short ($\sim1$~day duration) event with a rapid rise and slower decay associated with a solar eruption at $\sim$E20$^\circ$ relative to STEREO~A, based on observations by the Extreme Ultraviolet Imager (EUVI) on board STEREO-A \citep[EUVI-A,][]{wuelser2004}, corresponding to $\sim$E136$^\circ$ relative to Earth. The related flaring and erupting filament observed by EUVI-A at 17:46 UT at a wavelength of 304~\AA ~are shown in the top left panel of Figure~\ref{fig:may18}. The bottom panel shows the spacecraft configuration and location of this flaring (black arrow). This small ``impulsive-like" SEP event from a source that was not well-connected to STEREO~A was associated with a CME evident (Figure~\ref{fig:may18cme}) in the STEREO-A COR2 \citep{howard2008} and EUVI-A 195~\AA ~observations in the left panel (also showing the dimming associated with the eruption).  The CME is also evident in observations from  the SOHO/LASCO C2 coronagraph (right panel). Based on the LASCO observations, this CME is classified as a ``partial halo" CME in the CDAW CME catalog with a width of 302$^\circ$ and speed of 533 km/s, though with a deceleration of 22.1~m~s$^{-2}$ in the C2 field of view suggesting that the speed was higher close to the Sun. The automated CACTus COR2 CME catalog gives a plane of the sky width of 102$^\circ$ and speed of 581 km/s.  The DONKI database gives a CME speed of 650 km/s in a direction E155$^\circ$, about $20^\circ$ from, but not inconsistent with, the flare longitude. Type III radio emission (not shown) was detected at STEREO~A but only weakly at Wind, consistent with the far side location relative to Earth.  No type II emission is reported at either spacecraft (e.g., \url{https://cdaw.gsfc.nasa.gov/CME_list/radio/waves_type2.html}). Although the $\sim77^\circ$ separation between the STEREO~A nominal field line footpoint and the solar event does approach the $80^\circ$ criterion used by \citet{Dresing2014} to define a ``widespread" SEP event, the spacecraft configuration is not suitable to  determine if this SEP event was actually detectable over a wide region of the heliosphere.  However, we doubt that this was the case because of the low intensity and short duration of the event at STEREO~A and also because, even though this spacecraft was close to the longitude of the solar event, no shock or ICME was detected in situ during the days following the solar event (\url{https://stereo-ssc.nascom.nasa.gov/data/ins data/impact/level3/STEREO Level3
Shock.pdf}; the associated ``shock arrival" at 10:00~UT on May 7 reported in DONKI is clearly the forward shock of a CIR), and there is no evidence of any extended enhancement of shock-associated particles that might have been expected had this been a widespread event. Nevertheless, this event clearly illustrates that even a poorly-connected spacecraft can detect a such a small SEP event extending to tens of MeV arriving promptly after a modest solar eruption.

\clearpage

\section{Several Problematic Events}

While the likely associated solar event can be identified for nearly every SEP event, there are a few cases that are more challenging for various reasons.  Three are discussed in this appendix.

\subsection{September~30, 2015}

\begin{figure}[tbp] 
  \centerline{
  \includegraphics[width=0.55\textwidth,keepaspectratio]{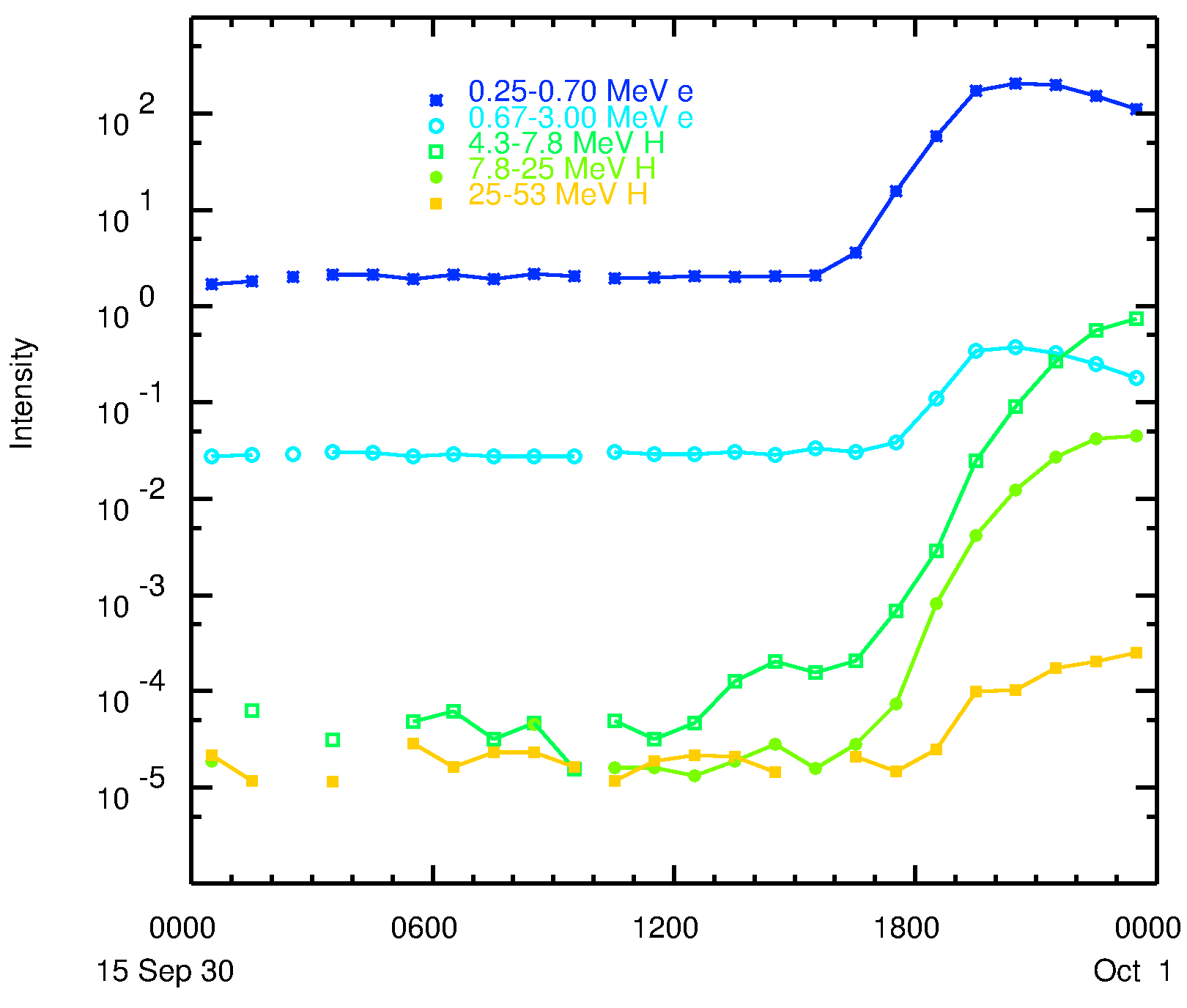}
  \includegraphics[width=0.45\textwidth,keepaspectratio]{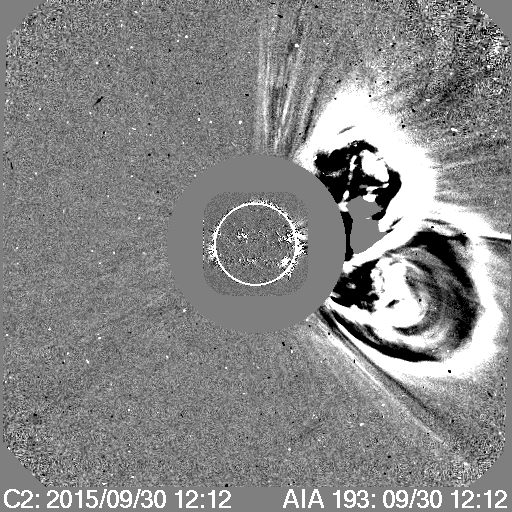}
}
\centerline{
\includegraphics[width=0.45\textwidth]{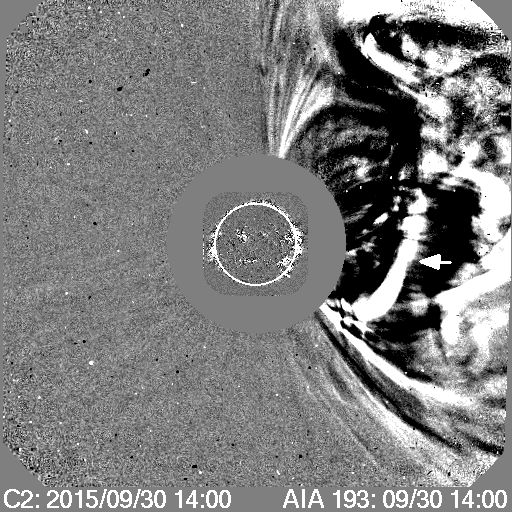}
\includegraphics[width=0.49\textwidth]{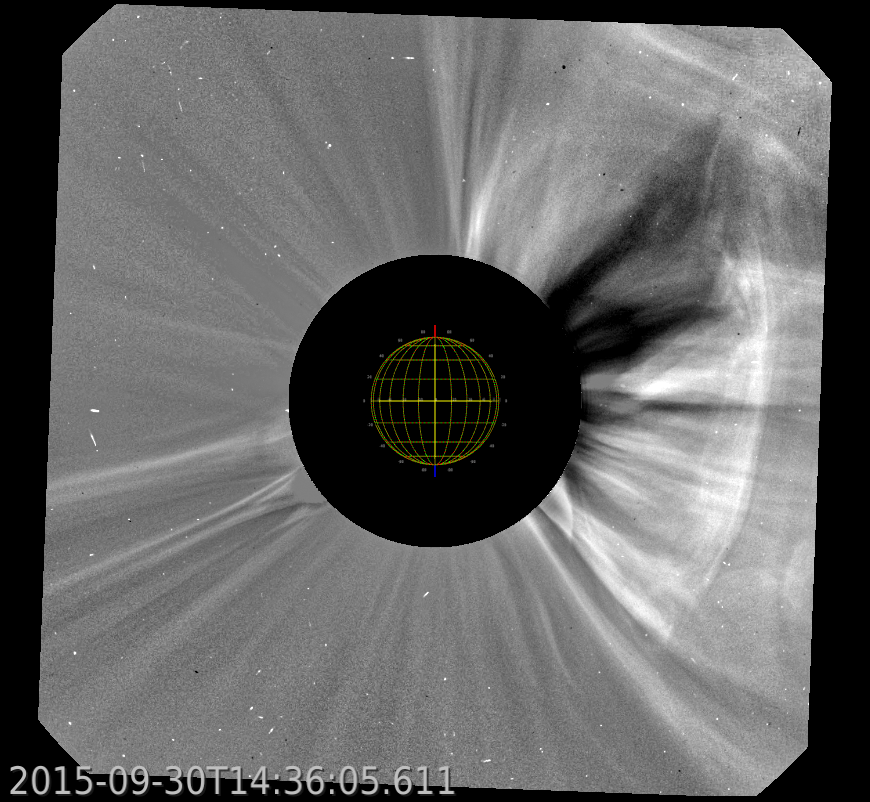}

}
 \centering \includegraphics[width=0.7\textwidth,keepaspectratio]{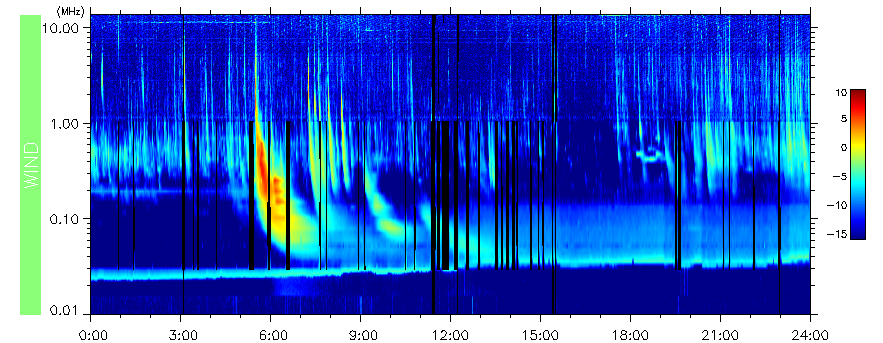}
\centering
  \includegraphics[width=0.6\textwidth,keepaspectratio]{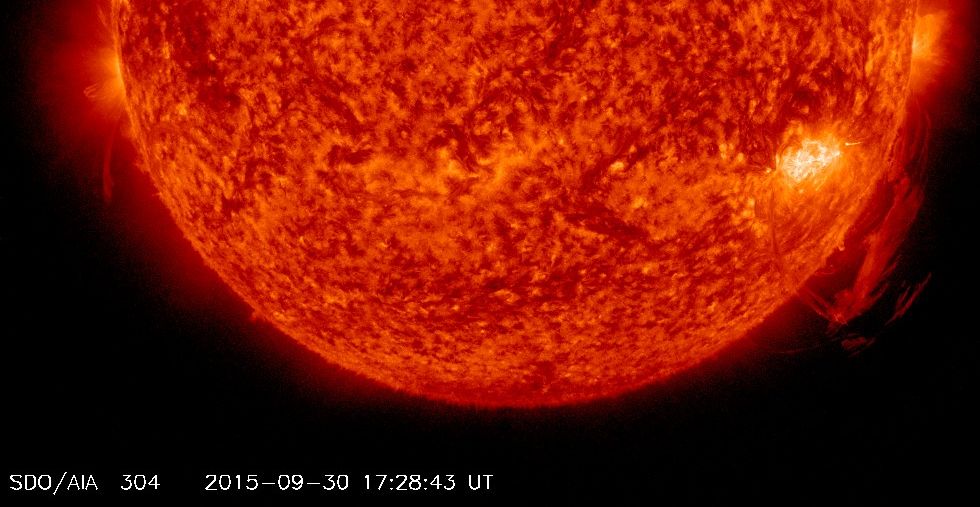}

  \caption{The SEP event of September 30, 2015, $\sim17$~UT (top left; observations from SOHO/EPHIN), for which the associated solar event is uncertain and no STEREO observations are available. The nearest preceding CMEs were a pair at $\sim10$~UT around 7 hours before the SEP event (top right, running difference image from the CDAW catalog) related to the lift off of a large filament (Figure~\ref{fig:20150930fila}). The left figure in the second row (also from the CDAW catalog) shows a likely CME (indicated by the arrow) trailing the first pair of CMEs that is not recorded in the CDAW or DONKI catalogs. This CME is also clearly evident in the right figure showing the base difference between C2 images at 14:36~UT and 09:12 UT.  The Wind/WAVES radio data (third row) show only weak type III emissions at low frequencies early in the day, and no emissions around the time of SEP onset.  There was a slow filament eruption above the southwest limb near the time of SEP onset (bottom figure) - SDO AIA 304~\AA~ observations at 17:29~UT are shown.     }
  \label{fig:20150930}
\end{figure}

\begin{figure}
    \centering
    \includegraphics[width=0.5\linewidth]{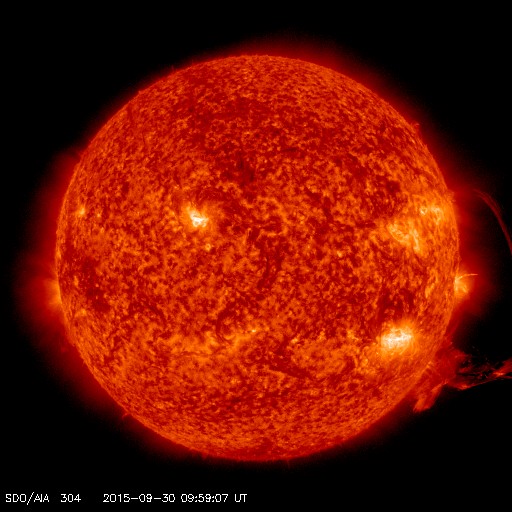}
    \caption{SDO AIA 304~\AA~ observations at 9:59~UT on September~30, 2015, showing a large filament eruption above the west limb.}
    \label{fig:20150930fila}
\end{figure}

As indicated in Table~\ref{tab8} of Appendix~A, the origin of the SEP event on September~30, 2015 (Figure~\ref{fig:20150930}) is not completely clear. One complicating factor is the absence of data from STEREO~A, then located on the far side of the Sun at 176$^\circ$ east of Earth (also, contact had been previously lost with STEREO~B).  This was a modest SEP event at Earth with a fairly prompt electron and proton onset at $\sim17$~UT evident in the SOHO/EPHIN data (top left panel of Figure~\ref{fig:20150930}).  However, no CME is reported at this time in the CDAW catalog (there were no LASCO data gaps on this day).  The closest reported CMEs preceding the SEP onset were at 09:36~UT, going to the southwest with a plane of the sky speed of 586~km/s, and a partial halo CME at 10:00~UT going to the northwest with a speed of 602 km/s that overlap in the LASCO image in the top right panel of Figure~\ref{fig:20150930}. The DONKI database also identifies these two CMEs, with estimated longitudes relative to Earth of W49$^\circ$ and  W116$^\circ$ (just over the west limb), respectively.  Neither of these CMEs was associated with a GOES flare - the ongoing soft X-ray level was around C2. These CMEs were associated with the lift-off of a large filament above the west limb seen as a large erupting prominence in SDO AIA 304~\AA~ 
observations (Figure~\ref{fig:20150930fila}). An M1.3 flare at S20$^\circ$W50$^\circ$ commencing at 10:49~UT was too late to be associated with these CMEs. There was also a brief M1.1 flare commencing at 13:18~UT located at S20$^\circ$W55$^\circ$.  Examination of the CDAW LASCO running-difference movie suggests that this flare was associated with a further CME that is not recorded in the CDAW or DONKI catalogs but is indicated by an arrow in the LASCO C2 running difference image (at 14:00~UT) in the left-hand figure in the second row of Figure~\ref{fig:20150930}, superposed on the complex trailing regions of the previous CMEs. This CME is also evident in the C2 base difference image (14:36~UT minus the pre-event background) in the right-hand figure. The CME leading edge speed, based on a linear fit, is estimated to be $572\pm8$~km/s. There is no clear evidence of any additional CME around the time of the SEP onset.  

Wind/WAVES (third row of Figure~\ref{fig:20150930}) observed brief type III emissions throughout the day. Some brighter emissions, generally weak at high frequencies (and therefore possibly occulted by the solar limb or indicating a lack of particle acceleration close to the Sun), commenced between $\sim5$ and 11~UT, around the time of the two prominent CMEs and the filament liftoff. However, unusually, there were no radio emissions around the onset of the SEP event at $\sim17$~UT. The most notable feature around SEP onset is the slow lift-off of a smaller filament above the southwest limb evident in the SDO observations in the bottom panel of Figure~\ref{fig:20150930}.  Thus, although there were clearly eruptive structures on the western hemisphere of the Sun on September 30, 2015 that might have given rise to the SEP event, the SEP onset does not appear to be unambiguously associated with a specific filament eruption, CME, or X-ray or radio features. 

We have also considered whether the onset of the SEP event on 30 September  2015, though associated with this activity, was delayed due to solar wind structures in the vicinity of Earth. An IMF sector boundary associated with a crossing of the heliospheric current sheet (HCS), was observed at ACE at $\sim19$~UT, which is close to the SEP onset (a similar situation resulting in a delayed SEP onset is discussed by \citet{lario2022}). Thus, it is also possible that the SEPs were largely confined to the region of anti-sunward fields following the sector boundary, i.e., north of the HCS. In the table, we have tentatively indicated the activity beginning at $\sim9-10$~UT associated with the major filament eruption as the origin of this SEP event, but this association is uncertain for the reasons discussed above. 

\clearpage
\subsection{November~26, 2013}

\begin{figure}
    \centerline{
    \includegraphics[width=0.55\linewidth]{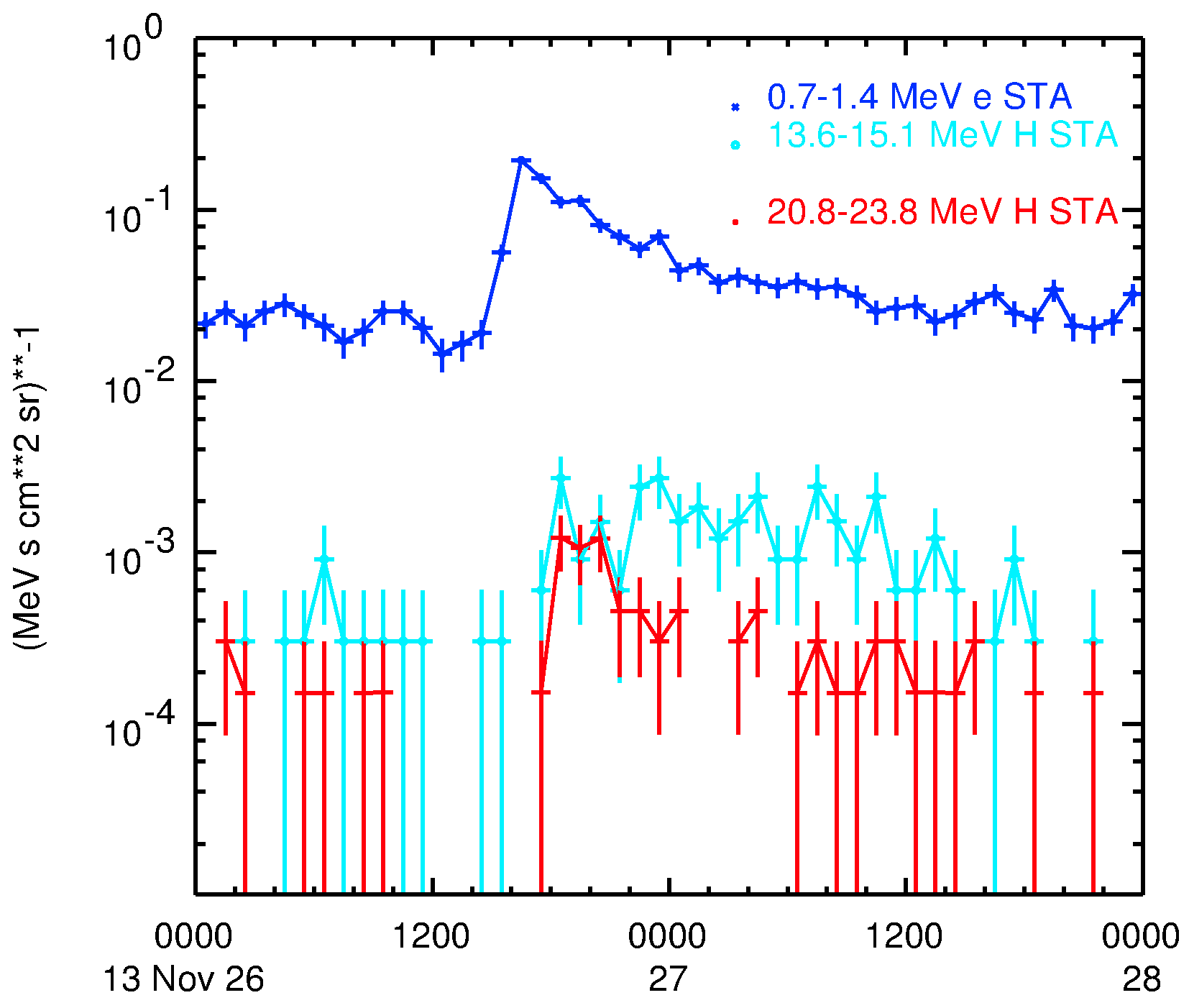}
    \includegraphics[width=0.45\textwidth]{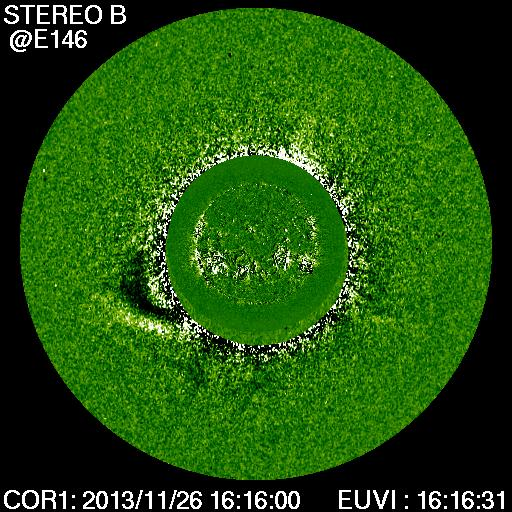}
    }
    \centering
   \includegraphics[width=0.6\textwidth]{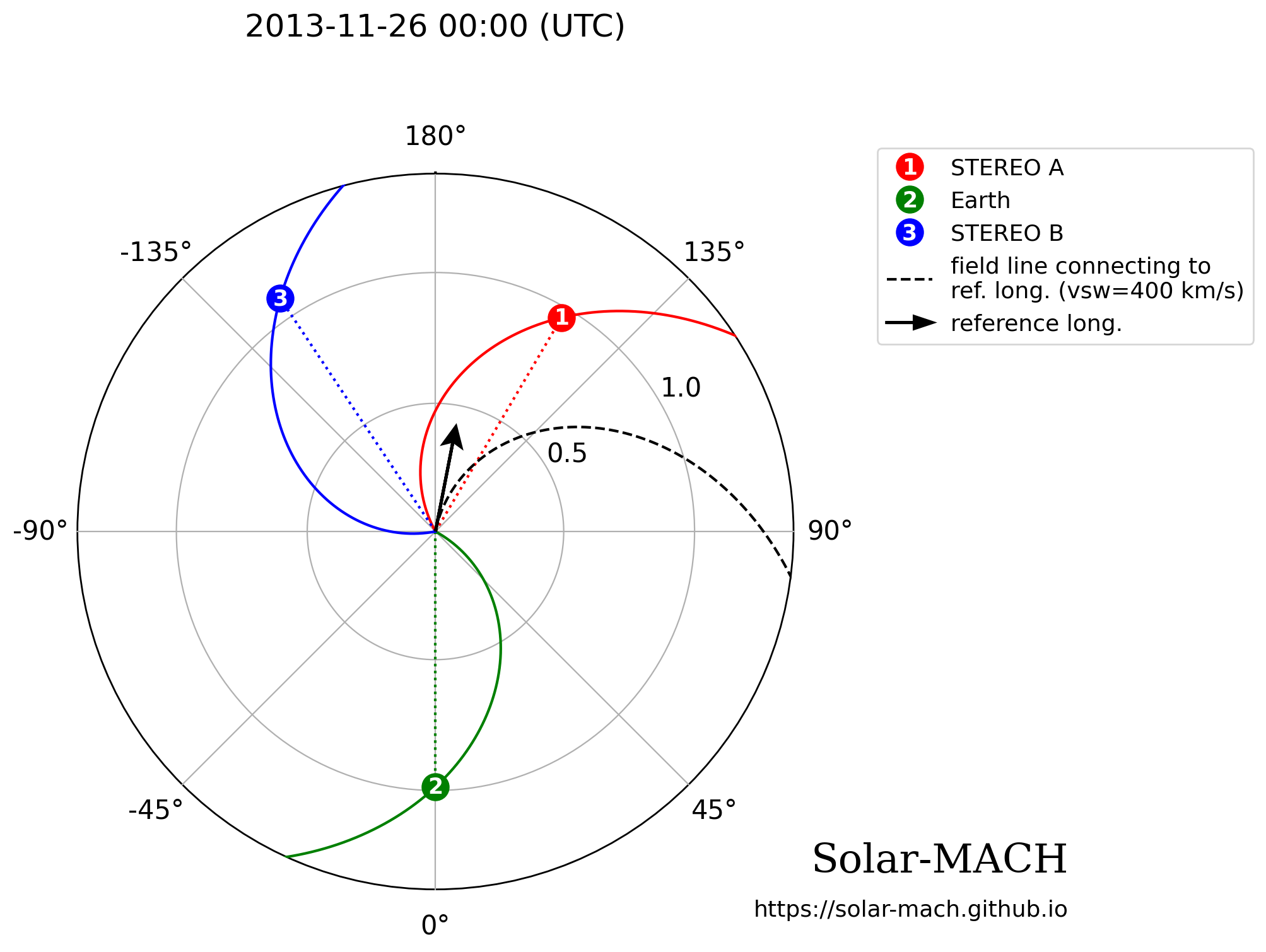}
    \caption{The small prompt SEP event on November~26, 2013 (top left), detected only at STEREO~A (1 in the spacecraft location plot, bottom) that, unusually, was not associated with a reported CME in the LASCO C2 or STEREO A/B COR2 coronagraphs. The arrow in the bottom panel indicates the location of the related solar event observed by the STEREO A/B EUVIs. The top right panel shows that a CME associated with this event was evident in STEREO~B COR1 (figure from the CDAW CME catalog). Although this faded by the edge of the COR1 field of view, close inspection suggests that the CME was observed by COR2 (not shown).}
    \label{fig:20131126}
\end{figure}

Another apparently anomalous SEP event is that on November~26, 2013 (Table~5) which unusually was not accompanied by a CME reported in the CDAW, DONKI or CACTus catalogs based on SOHO/LASCO and/or STEREO~A/B COR2 coronagraph observations. This small prompt SEP event (Figure~\ref{fig:20131126}), just evident at proton energies of $\sim25$~MeV, was observed only by STEREO A, with an electron onset time of 15:46~UT$\pm2$ minutes. The event was associated with a small
region of activity around 20$^\circ$ west of STEREO~A and $\sim44^\circ$ east of STEREO~B seen by EUVI; the spacecraft were nearly $150^\circ$ west
or east of Earth, respectively, at this time (bottom panel of Figure~\ref{fig:20131126}). The SEP event was accompanied by a brief type III burst detected at both STEREO spacecraft (not shown) and also, occulted at high frequencies, at Wind (also not shown), consistent with the far side location. Despite the lack of a reported CME in the above catalogs, examination of the STEREO~A and B COR1 coronagraph running difference movies clearly shows a CME (Figure~\ref{fig:20131126}, top right panel), first evident at STEREO~B at 15:36~UT above the east limb, associated with the aforementioned active region and consistent with the SEP event onset. This CME, which became indistinct within the COR1 field of view, is recorded in the STEREO~A and B COR1 CME lists at \url{https//cor1.gsfc.nasa.gov/catalog}. 
On close inspection of the STEREO A COR2 images, we do see evidence of the associated CME, but this ``falls apart" in the field of view; we estimate the speed as $356\pm10$~km/s from a linear height-time fit. This CME observed at STEREO~A is also noted in the ``Dual-Viewpoint CME Catalog from the SECCHI/COR Telescopes"(\citet{vourlidas2017}; \url{http://solar.jhuapl.edu/Data-Products/COR-CME-Catalog.php}) but no parameters are given and the morphology is described as ``other".  LASCO observations show outflows in the south, but no clear feature associated with this CME.  Thus, this $\sim25$~MeV proton event was associated with a CME observed close to the Sun that was not sufficiently distinct above $\sim2$~R$_s$ to be recorded in the regular CME catalogs. Nevertheless, since there is clearly evidence of a CME in the lower corona, this SEP event is not an anomaly without an associated CME. 

\clearpage
\subsection{September~22, 2023}
\begin{figure}
      \centerline{
      \includegraphics[width=0.5\linewidth]{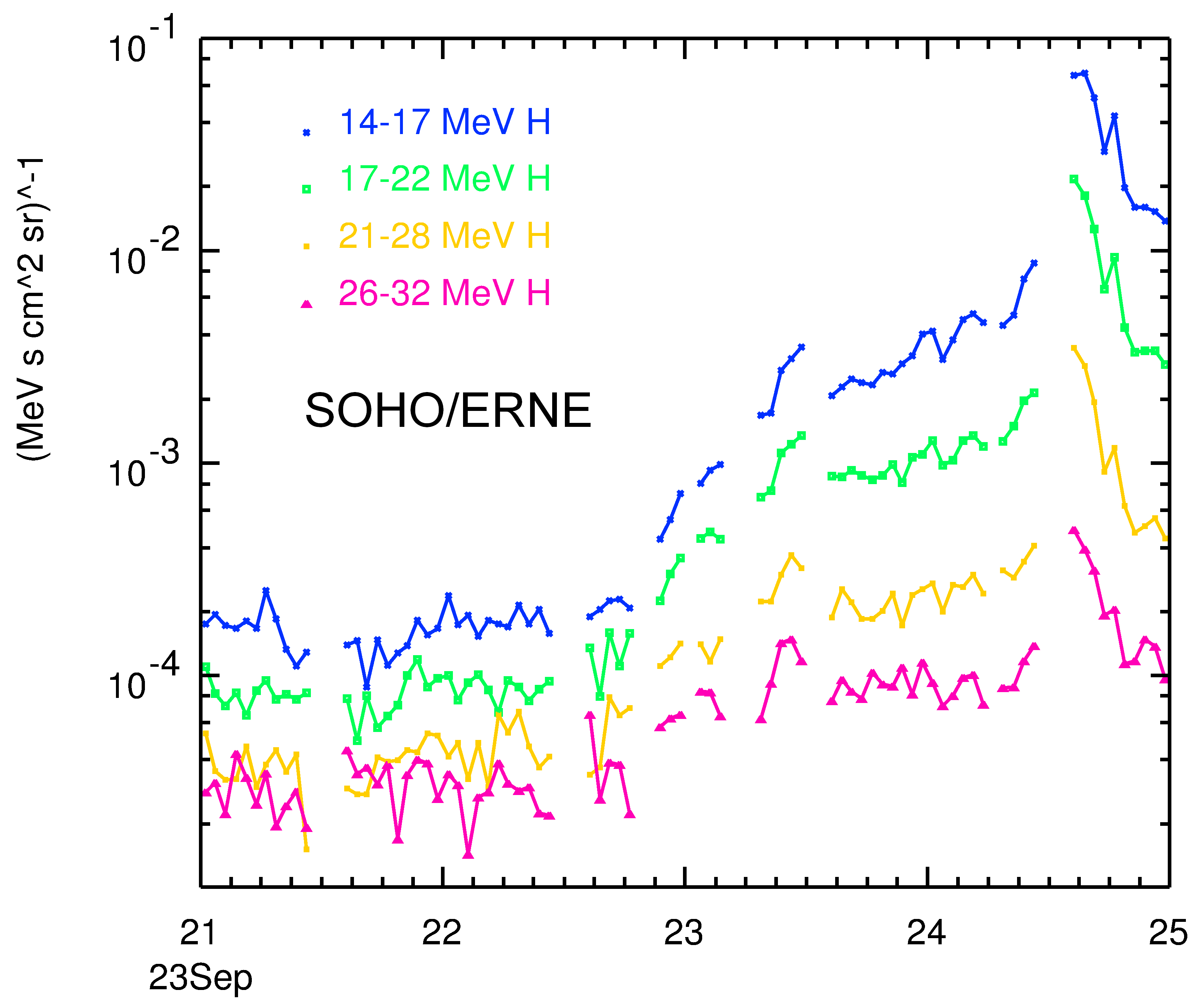} 
    \includegraphics[width=0.5\linewidth]{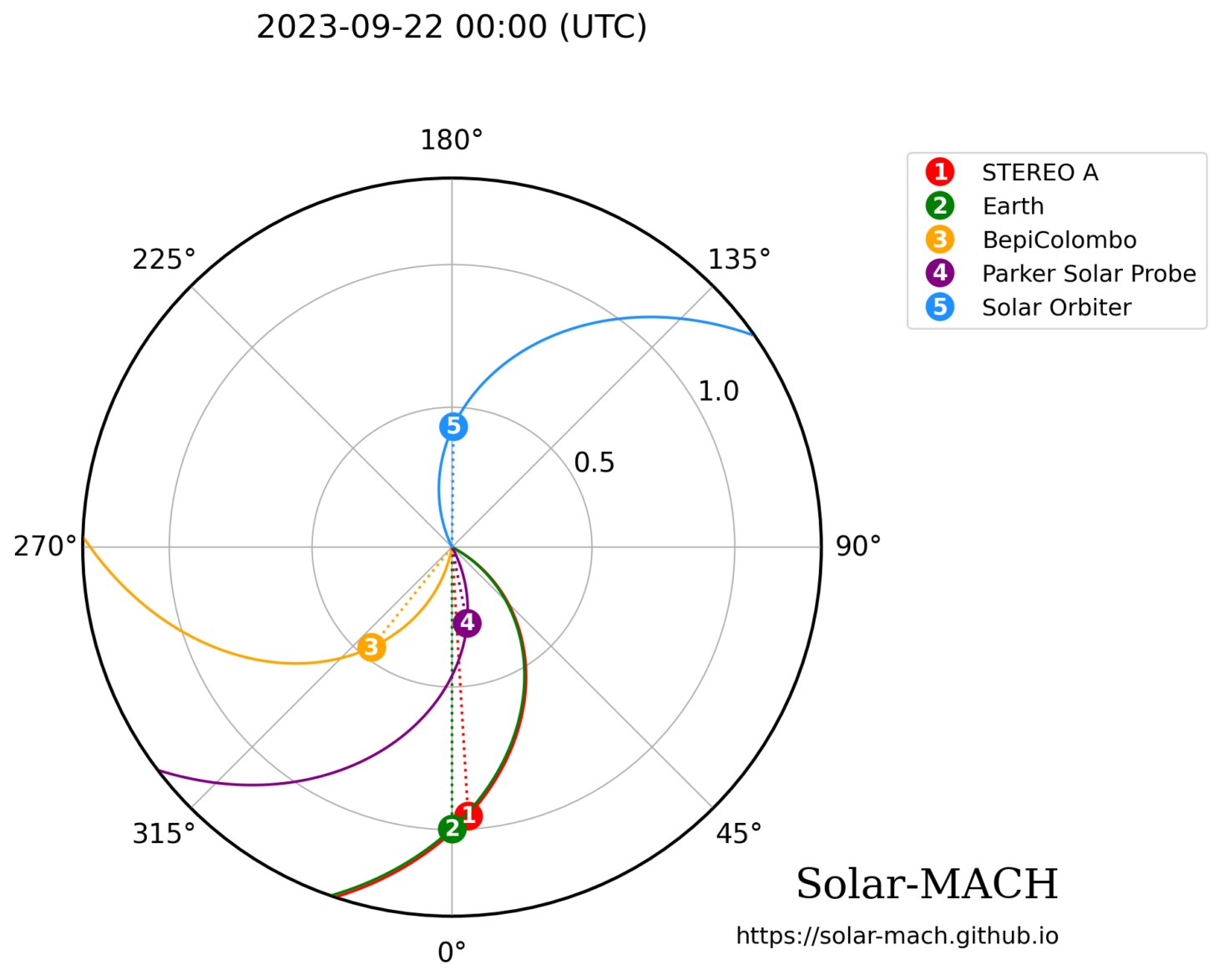}
    }
     \centering
     \includegraphics[width=0.6\linewidth]{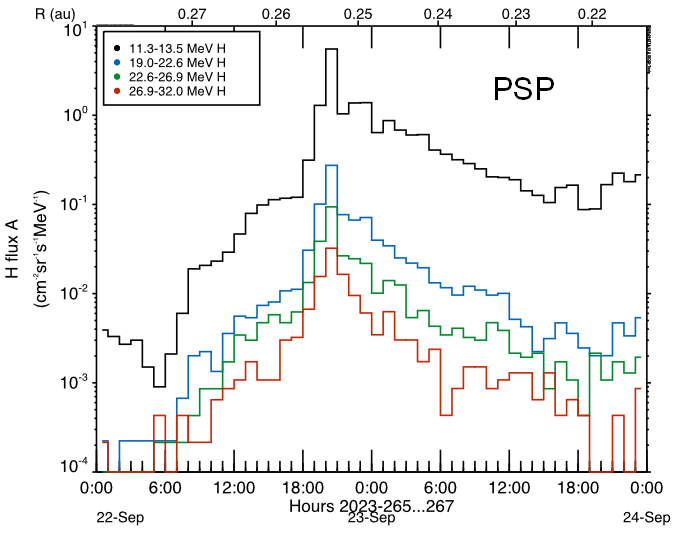}
    \caption{The SEP event observed at L1 (SOHO/ERNE observations are shown in the top left panel) and STEREO~A (not shown)  on September 22, 2023 with an indistinct onset.  At this time, PSP was at $\sim0.27$~AU, $11^\circ$ west of the Sun-Earth line (top right). Hourly-averaged proton intensities from PSP HET A (bottom) show the SEP onset at $\sim7$~UT on September~22. The proton intensity rose to the vicinity of a shock that passed PSP on September 22 and L1 and STEREO~A on September 24.   }
    \label{fig:20230922}
\end{figure}

The associated solar event is  also uncertain for the SEP event at L1 and STEREO~A on September 22, 2023 (Figure~\ref{fig:20230922}) when STEREO~A was separated by just 3$^{\circ}$ in longitude with respect to Earth/L1 (top right panel). This SEP event shows an indistinct onset and rise from $\sim16$~UT (the top left panel shows observations from SOHO/ERNE) that is difficult to associate unambiguously with any of the frequent type III emissions, several overlapping CMEs and significant front side activity on this day; the $\sim25$~MeV proton intensity rose to an ESP event associated with passage of an interplanetary shock on September 24, consistent with a front side origin. The fastest (983 km/s) CDAW CME on September~22 was at 02:24~UT associated with a long duration M1.2 flare at E18$^\circ$, but this is significantly earlier than the slow rise in proton intensity at Earth. At this time, PSP was at $\sim0.27$~AU close to the Sun-Earth line and observed the SEP onset more clearly, at $\sim7$~UT (bottom panel in Figure~\ref{fig:20230922}).
This onset may be consistent with a faint CDAW CME observed at 07:36~UT with a recorded width of 86$^\circ$.  However, inspection of the LASCO observations suggests that it may have been expanding symmetrically around the occultor, though this is uncertain given the other structures in the field of view.  The DONKI database gives a propagation longitude of W04$^\circ$ with a speed of 1508 km/s and an association with a flare at N27W02, though there was only limited brief $\sim$ C-level flaring above the ongoing decay of the earlier long-duration flare. The height-time profile in the CDAW catalog indicates that the CME accelerated from around 500 km/s to 1000 km/s in the LASCO C2 field of view.  We suggest in Table~\ref{tab11} that the SEP event at Earth and STEREO~A was associated with this CME, though it is not possible to completely rule out a different association.  

\clearpage
\begin{acks}
 The authors thank the various researchers who have contributed to the success of the STEREO mission and have made their data available.  We also acknowledge the contribution of data from instruments on near-Earth spacecraft, in particular SOHO EPHIN and ERNE. We thank members of the SERPENTINE collaboration for useful discussions. We particularly acknowledge the dedicated members of the CCMC and Moon to Mars Office at GSFC who have developed, contributed to, and maintain the DONKI database, and also the compilers of the CDAW CME database which is generated and maintained at the CDAW Data Center by NASA and The Catholic University of America in cooperation with the Naval Research Laboratory; SOHO is a project of international cooperation between ESA and NASA.  We appreciate the helpful feedback provided by the reviewer.
\end{acks}



\begin{authorcontribution}
IGR compiled the SEP event list, performed analysis of these events and wrote the manuscript. TvR contributed to the SEP event identifications and analysis. OCStC provided CME analysis and figures, and contributed to the SEP event, solar event and CME associations. DL and JGM contributed to the solar event identifications, and ERC contributed information on the STEREO mission.  All authors read and reviewed the manuscript.
\end{authorcontribution}

\begin{fundinginformation}
IGR and DL acknowledge support from NASA Living With a Star program NNH19ZDA001N-LWS, Heliophysics Guest Investigators program NNH23ZDA001N-HGIO, and the NASA Heliophysics
Space Weather Research Program’s CLEAR SWx Center of Excellence, award
80NSSC23M0191. IGR also acknowledges support from the STEREO mission and NASA Heliophysics Supporting Research program NNH19ZDA001N-HSR. In addition, IGR and OCStC were supported by NASA program NNH17ZDA001N-LWS.
\end{fundinginformation}

\begin{dataavailability}
The energetic particle
data used are available from the NASA Space Physics Data Facility (\url{https://cdaweb.gsfc.nasa.gov/} and \url{https://spdf.gsfc.nasa.gov/research/vepo/}), and the STEREO HET website (\url{https://izw1.caltech.edu/STEREO/Public/HET_public.html}).We have also referred to SolO EPD data, generated and maintained by the EPD team; these data are available at \url{https://espada.uah.es/epd/EPD_data.php}.  A machine readable version of the SEP event list is available at the Harvard Dataverse at \url{https://doi.org/10.7910/DVN/GQPCXZ}.
\end{dataavailability}

\begin{ethics}
\begin{conflict}
The authors declare that they have no conflicts of interest.
\end{conflict}
\end{ethics}


\bibliographystyle{spr-mp-sola}
\bibliography{HETfirst.bib}

\end{document}